\documentclass[fleqn,usenatbib]{mnras}

\usepackage{newtxtext,newtxmath}
\usepackage[T1]{fontenc}

\DeclareRobustCommand{\VAN}[3]{#2}
\let\VANthebibliography\thebibliography
\def\thebibliography{\DeclareRobustCommand{\VAN}[3]{##3}\VANthebibliography}

\usepackage{graphicx}	% Including figure files
\usepackage{amsmath}	% Advanced maths commands
\usepackage{enumerate}
\usepackage{ulem}
\usepackage{booktabs}
\newcommand{\m}{$-$}

\title[]{MIGHTEE: Discovery of a triple-double radio galaxy}

\author[T. F. Rarivoarinoro et al.]{Tombo F. Rarivoarinoro$^{1, 2, 3}$\thanks{e-mail: tombofitahiana@gmail.com},
Zara Randriamanakoto$^{3, 4}$, Russ Taylor$^{1, 2, 5, 6}$, Marisa Brienza$^{7}$, 
\newauthor Fabio Luchsinger$^{ 2, 5}$, Sushant Dutta$^{1, 2}$, Catherine Hale$^{8}$, Jacinta Delhaize$^{1}$, Ndivhuwo Netshiavha$^{1}$,
\newauthor Solohery Randriamampandry$^{4}$, Mattia Vaccari$^{2, 6, 7}$ \\
$^{1}$Department of Astronomy, University of Cape Town, Rondebosch, Cape Town, 7701, South Africa\\
$^{2}$Inter-University Institute for Data Intensive Astronomy (IDIA), Department of Astronomy, University of Cape Town, Rondebosch, 7701, South Africa\\
$^{3}$South African Astronomical Observatory, P.O. Box 9, Observatory 7935, South Africa \\
$^{4}$Department of Physics, University of Antananarivo, PO Box 906, Antananarivo 101, Madagascar\\
$^{5}$Department of Physics and Astronomy, University of the Western Cape, Bellville, Cape Town, 7535 South Africa\\
$^{6}$IDIA, Department of Physics and Astronomy, University of the Western Cape, 7535 Bellville, Cape Town, South Africa\\
$^{7}$ Istituto Nazionale di Astrofisica (INAF) - Istituto di Radioastronomia (IRA), Via Gobetti 101, 40129 Bologna, Italy\\
$^{8}$Institute for Astronomy, University of Edinburgh,  Royal Observatory Edinburgh, Blackford Hill, Edinburgh, EH9 3HJ, UK
\\
}

\date{Accepted XXX. Received YYY; in original form 2025 August nn}

\pubyear{\the\year{}}

\begin{document}
\label{firstpage}
\pagerange{\pageref{firstpage}--\pageref{lastpage}}
\maketitle

\begin{abstract}
Triple-double radio galaxies (TDRGs) are amongst the rarest subpopulations of radio galaxies (RGs). They are characterised by three pairs of radio lobes, where each pair of lobes represents an episode of nuclear activity. Such a feature makes them key objects that can be used to constrain the duty cycle of RGs. In this paper, we report the discovery of J022248\m060934, a new TDRG, hosted by a galaxy at a spectroscopic redshift of $z \approx$  0.94. We have used the MIGHTEE-DR1 data set and MIGHTEE sub-band images as our main data. In total intensity, J022248\m060934 has a bright core and triple-double, edge-brightened-like peaks of radio emission. The polarimetry of the source reveals an inhomogeneous density of the hosting environment which is consistent with the more pronounced bending in its eastern lobes. The spectral index and curvature maps suggest an inverted core and an ultra-steepening of the spectrum towards the outer lobes which reinforce a recurrent nuclear activity. We perform individual spectral age fitting of the components of the source using the JP model and we found a lower limit total age of $\sim$16 Myr. We also derive a short inactive period between the active phases and a rapid duty cycle of 90 per cent for the first cycle of activity. Our spectral ageing analysis suggests that the triple-double structure in TDRGs is not the product of long quiescent periods, as deduced by previous works based on kinematic ages.
\end{abstract}


%%%%%%%%%%%%%%%%%%%%%%%%%%%%%%%%%%%%%%%%%%%%%%%%%%

%%%%%%%%%%%%%%%%% BODY OF PAPER %%%%%%%%%%%%%%%%%%

\section{Introduction}
Double-double radio galaxies (DDRGs, \citealt{2000MNRAS.315..371S}) are a well known subclass of radio galaxies (RGs) often with lobes comparable to those of Fanaroff-Riley Type II (FRII, \citealt{1974MNRAS.167P..31F}). They consist of coaxial double pairs of lobes sharing a common radio nucleus. In general, the outer pair of lobes is devoid of hotspots, tends to be more extended, and is more diffuse than the inner lobes. Their atypical morphology has been interpreted as the morphological evidence of the episodic nuclear activity of radio active galactic nuclei (AGN; \citealt{2000MNRAS.315..371S, 2000MNRAS.315..381K}). The outer lobes are remnants of the earlier phase of activity, which decay once the replenishment of fresh electrons from the core ceased. Thereafter, a new cycle of activity re-ignites, which is typified by the inner pair of lobes. Spectral studies of DDRGs by  \citet{2000MNRAS.315..395S} and \citet{2006MNRAS.372..693K} also support this scenario, i.e. the outermost lobes are steeper than the innermost ones, suggesting increasing spectral age.

In regard to the context above, triple-double radio galaxies (TDRGs, \citealt{2007MNRAS.382.1019B, 2023MNRAS.525L..87C}), which are RGs with three pairs of lobes, are thought to embody sources with three cycles of nuclear activity. They are thus ideal objects to study to constrain the duty cycle of RGs which refers to the different phases of nuclear activity in RG life (i.e. active, dying, and restarted) \citep{2018A&A...618A..45B, 2021Galax...9...88M, 2023Galax..11...74M}. Our understanding of the duty cycle can provide valuable insight into determining jet production and/or evolution mechanisms, the life cycle of AGN and the influence of AGN feedback on their environments. However, there are remaining open questions concerning the duty cycle of RGs. What are the required ingredients for a restarting jet? How often do RGs undergo this cycle? And would the active and/or inactive periods be constant during the RG lifetime? 

Given the analogy between TDRGs and DDRGs, analysing TDRGs will also shed light on the incertitudes concerning DDRGs. The evolution and connection between the outer and inner lobes remain inconclusive. It is unclear how these pairs of lobes develop into a similar system  (i.e. lobes and hotspots), yet they are likely to have evolved under different conditions. The outer pair of lobes are expected to have altered the pressure and/or the density of the medium welcoming the new restarting jets \citep{2000MNRAS.315..381K, 2025A&A...696A..97D}. Simulations confirm that the interaction between successive outbursts can be more complex \citep{1991ApJ...369..308C, 2013MNRAS.433.1453W, 2014MNRAS.439.3969W, 2020MNRAS.497.3638W}.  

Nonetheless, TDRGs are undeniably one of the rarest subpopulations of RGs to date, even in comparison to DDRGs, which are also claimed to be a rare subpopulation of RGs.  In their search for DDRGs, \citet{2019A&A...622A..13M} only found 33 DDRG candidates out of 318542 individual radio sources. Equally, \citet{2020A&A...638A..34J} only identified 3 DDRGs out of 158 radio sources with an angular size larger than 60 arcsec. Though most of the known DDRGs are giant radio galaxies (GRGs, \citealt{1974Natur.250..625W, 2021MNRAS.501.3833D, 2023JApA...44...13D}), which are RGs with a linear extent larger than 700 kpc, giant DDRGs only represent less than 3 per cent of the GRG population. This latter percentage is based on the largest studied sample of DDRGs by \citet{2025A&A...696A..97D}, which consists of  111 sources. Given that, at the time of writing, only six TDRGs have been reported in the literature \citep{2007MNRAS.382.1019B, 2011MNRAS.417L..36H, 2016ApJ...826..132S, 2023MNRAS.525L..87C, 2025A&A...696A..97D, 2025PASA...42..124N}, it suggests that TDRGs are far more scarce.  

In this work, we report the discovery of J022248\m060934, the seventh known TDRG. It was identified during visual inspection of the MeerKAT image of the XMM-LSS field \citep{2025MNRAS.536.2187H}. In previous surveys such as the NRAO VLA Sky Survey (NVSS, \citealt{1998AJ....115.1693C}) or the Faint Images of the Radio Sky at Twenty-cm (FIRST, \citealt{1995ApJ...450..559B}), J022248\m060934 is either completely undetected or its bright core is the only observed component. Indeed, the scarcity of TDRGs might be due to observational limitations such as the low-resolution images of the G4Jy sample \citep{2020PASA...37...17W, 2020PASA...37...18W}. A high spatial resolution image is essential to resolve the triple features of the source, while a high surface brightness sensitivity is needed to recover its large-scale structure. It is therefore plausible that some of the DDRG populations in previous surveys might have a third pair of lobes but their remnant plasma is beyond sensitivity limits (e.g. \citealt{2007MNRAS.382.1019B}).

We give a brief description of the data set used and the host environment in Section \ref{sec: data} and Section \ref{sec: host_env}, respectively. We present the morphological description and the spectral and polarisation properties of J022248\m060934 in Section \ref{sec: radio_properties}. Then we discuss the results we found in Section \ref{sec: disc} before summarising in Section \ref{sec: summary}. Throughout the paper,  we adopt the following cosmological parameters: $H_0$ = 67.8  km s$^{-1} $Mpc$^{-1}, 
	\Omega_m = 0.308$ and $\Omega_\wedge = 1 - \Omega_m$ \citep{2016A&A...594A..13P}. We define the spectral index $\alpha$ as $S_\nu\propto\nu^{\alpha}$, where $S_\nu$ is the flux density at a frequency $\nu$.

\section{Data}\label{sec: data}
 The MeerKAT International GigaHertz Tiered Extragalactic Exploration (MIGHTEE) survey (\citealt{2016mks..confE...6J, 2017MS&E..198a2014T}) is one of the large-scale survey projects of the MeerKAT telescope \citep{2009pra..confE...4J, 2016mks..confE...1J}. Its observations are conducted by using the L-band of MeerKAT centered at 1280 MHz ($880 - 1680$ MHz). Once completed, MIGHTEE will cover 20 deg$^2$ of the sky, distributed across four extragalactic deep fields (E-CDFS, COSMOS, XMM-LSS and ELAIS-S1). In this work, we use the first data release of MIGHTEE (DR1) in the XMM-LSS field. It covers 14.4 deg$^2$ of the field \citep{2025MNRAS.536.2187H}.

  The detailed data reduction processes are reported in \citet{2022MNRAS.509.2150H} and \citet{2025MNRAS.536.2187H}. The final MIGHTEE Stokes I images consist of a first image with higher sensitivity, translating into a thermal noise of $\sim$1.7 $\mu$Jy/beam and a resolution of 8.9 arcsec. It was imaged with the Briggs’ robust weighting parameter of 0.0, while a second image was produced using the value of $-1.2$. The latter has a higher resolution of 5 arcsec but with a lower sensitivity of $\sim$6 $\mu$Jy/beam in thermal noise.   Throughout this work, we will only use the high resolution image (MIGHTEE-hi), it has the effective frequency of $\sim$1211 MHz at the location of our source. 
  
  In addition, we will use sub-band images of MIGHTEE which was imaged with a Briggs’ robust weighting parameter of $-$0.5 and followed by a uv-tapering. These consist of three images at 970 ($880 - 1060$ MHz), 1230 ($1060 - 1400$ MHz) and 1540 MHz ($1400 - 1680$ MHz) with the resolution of 14$\times$11.6 arcsec. Two images at higher resolution of 12.6$\times$11 arcsec and 11.3$\times$9.3 arcsec, respectively at 1230 and 1540 MHz are also shown in this work. The median local rms measured off source range from 10 to 30 $\mu$Jy/beam. The processing and the details about these sub-band images will be given in Luchsinger et al. (in prep).

  The data processing performed to obtain the full polarisation information is similar to MIGHTEE-pol Early science described in \citet{2024MNRAS.528.2511T}. The data set comprises two multiterm multifrequency synthesis (MT-MFS) broadband images in full Stokes and two spectro-polarimetric hypercubes. The MT-MFS broadband \textit{I, Q, U, V} images were obtained with Briggs’ robust
weighting of $-$ 0.5 and $+$ 0.4. The \textit{I, Q, U, V} hypercubes were constructed with Briggs’ robust weighting of 0.0; one with
the native resolution of the associated
frequency for each frequency plane and a second with a common resolution of 18 arcsec.

In addition to these main data sets, we retrieved archival radio continuum image at lower frequencies from LOFAR at 150 MHz which has a median sensitivity of 280 $\mu$Jy/beam and a resolution of 7.5 $ \times $ 8.5 arcsec \citep{2019A&A...622A...4H}. We also made use of an image from the GMRT Legacy at 325 MHz with a median sensitivity of 150 $\mu$Jy/beam and a resolution of 10 $ \times $ 8 arcsec \citep{2014A&A...569A..52S}. We retrieved the Quick Look image at 3 GHz from the first epoch of the Karl G. Jansky Very Large Array Sky Survey as well (VLASS, \citealt{2020PASP..132c5001L}). The image reached a resolution of 2.5 arcsec and a sensitivity of 145 $\mu$Jy/beam.

 \section{Host galaxy}\label{sec: host_env}

\begin{figure*}
		\begin{tabular}{c}
  {\resizebox{1\hsize}{!}{\includegraphics[trim={0.4cm 0 0.1cm 0},clip]{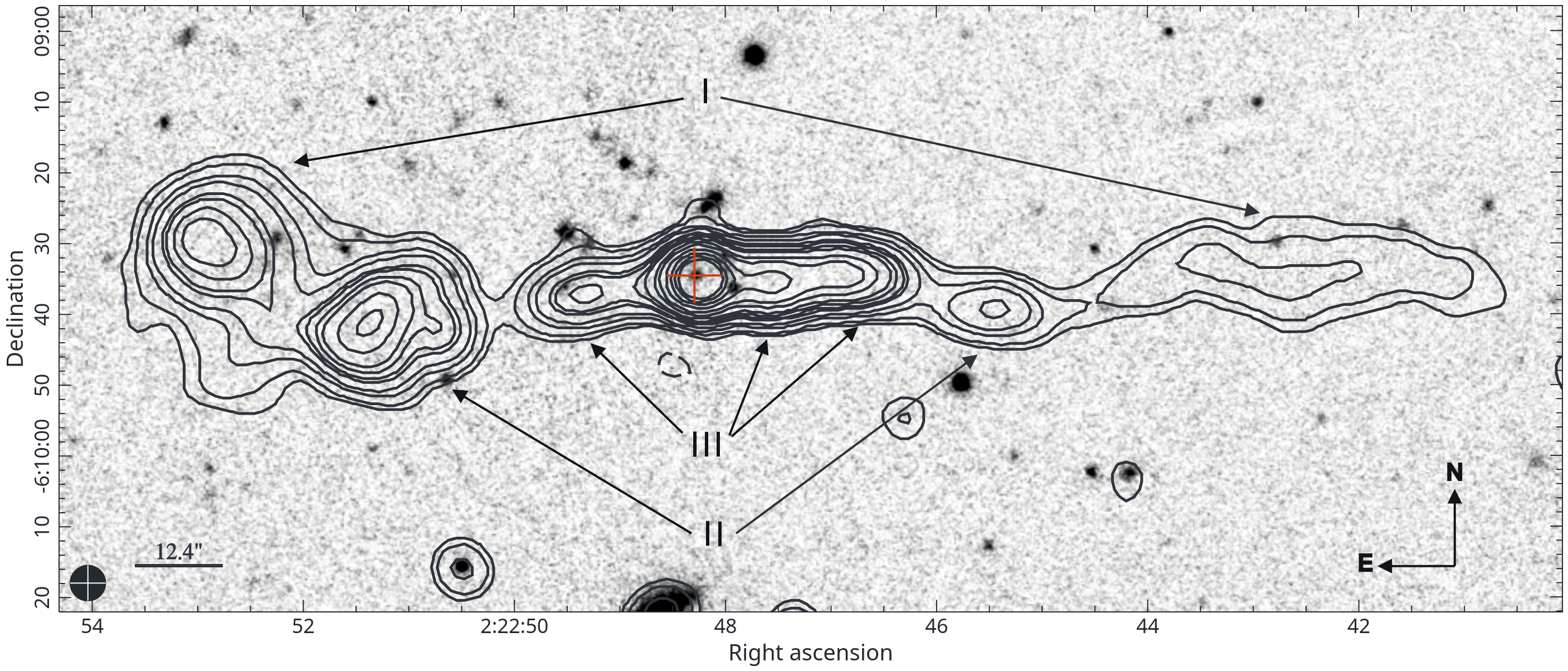}}}
   \end{tabular}
\caption{Radio continuum image of the TDRG showing its morphology as well as the position of its host galaxy as indicated by the red cross. 
\textit{Background}: Pan-STARRS DR1 gri$-$composite optical image; \textit{Contour}: MIGHTEE-hi continuum (levels: $-$3, 3, 5, 10, 15, 25, 30, 35, 40, 50, 75, 100, 150, 200, 250, 300 $\times \sigma_{local}$ = 5.9 $\mu$Jy/beam). The three pairs of lobes are indicated by I (outermost), II (middle) and III (innermost). The scale indicating 100 kpc (12.4  arcsec) and the synthesized beam of MIGHTEE-hi are in the bottom left corner of the image. } 
		
		\label{fig: optical_host}
	\end{figure*}

Figure \ref{fig: optical_host} displays the Pan-STARRS DR1 gri-composite optical
image \citep{2016arXiv161205560C}, which is overlapped by MIGHTEE-hi radio contours (above 3$\sigma_{local}$ = 5.9 $\mu$Jy/beam) showing the extension and the structure of the radio emission from the TDRG. The host galaxy, indicated by the red cross, is located at RA = 02h22m48.29s and DEC = $-$06d09m34.54s, which is 0.85 arcsec away from the centre of the radio core. Using the spectroscopic redshift catalogue of   \citet{vaccari_2025_15176245} based on the Spitzer Data Fusion project \citep{vaccari_2023_8192777}, we found that the TDRG is hosted by a galaxy with a redshift of $z_{spec}$ = 0.9363$\pm$0.0001 \citep{2025arXiv250314745D}, where 1 arcsec corresponds to 8.1 kpc.

To investigate the host environment, we cross-matched the host galaxy location with various galaxy cluster catalogues from different wavelengths (\citealt{2011A&A...535A..65D, 2018A&A...620A..12C, 2018PASJ...70S..20O, 2018MNRAS.475..343W, 2021ApJS..253...56Z, 2022RAA....22f5001Z}) using a search radius of 1 Mpc. We did not find any galaxy cluster within the same redshift range as the TDRG.  Similarly, we used the redshift catalogue of \citet{vaccari_2025_15176245} to search for any neighbouring galaxy within the same redshift. The closest neighbour galaxy with a similar redshift is located at 1.2 Mpc away which suggests that the TDRG seems to
be hosted by a field galaxy. We also perform a meticulous visual inspection of the optical image with the radio contours on top. There are no optical counterparts to any of the hotspots of the TDRG. This is also confirmed by searching for any association using the NASA/IPAC Extragalactic Database (NED\footnote{https://ned.ipac.caltech.edu/}).  

\section{Radio properties}\label{sec: radio_properties}
\subsection{Morphology}
The radio morphology of J022248\m060934 at 5 arcsec from MIGHTEE-hi at $\sim$ 1211 MHz is displayed in Figure \ref{fig: optical_host}. The radio contours show the full extension of the source and its newly resolved multi-components. While higher resolution images are needed to probe the detailed substructures of the source, we interpret these components as “hotspots” following the criteria described in literature.

The common definition of hotspots in literature is that they are associated with jet termination, so they are generally found at the edges of a source (i.e. closer to the end of the jet than the nucleus). They must be the brightest of emissions in the lobes. To further distinguish hotspots from jet knots, one should take into account the discontinuity in jet properties beyond them and whether they are related to major disruptions and/or shocks \citep{1994AJ....108..766B, 2008ASPC..386...46H}. Considering the structure of the source in Figure \ref{fig: optical_host}, J022248\m060934 meets these common criteria. Most of the lobes are characterised by edge-brightened structures which are located further away from the nucleus than the edges of the lobes. The western outer lobe is the exception, which displays fainter features. The observed change of direction and/or the decollimation from one lobe to another and the signs of shocks in them (see Section \ref{sec: polarisation_results} and Section \ref{subsec: disc_morpho_pol}) are also in favour of hotspots rather than jet knots. Furthermore, we have also given consideration to the criteria described in \citet{1997MNRAS.291...20L} on the usage of the terminology “hotspots”. The projected sizes of the brightest peaks in the lobes of J022248\m060934 are less than 10 per cent of the main axis size of the source and all these peaks are also 10 times or more than the rms noise level.

Although unresolved by previous surveys such as FIRST or NVSS, we therefore classify J022248\m060934 as a triple-double radio galaxy based on its resolved structure in the MIGHTEE image at 5 arcsec (see Figure \ref{fig: optical_host} as well as Figures \ref{fig: source_morpho_appendix} and \ref{fig: source_morpho_archival} as comparison). The MIGHTEE-hi contours in Figure \ref{fig: optical_host} show a compact core and triple pairs of lobes hosting hotspots. In optical, the latter components have no counterparts which rules out the possibility of them being independent sources. The total end-to-end projected largest linear size (LLS) of the outermost lobes (I) is 1587 kpc (196 arcsec in largest angular size (LAS)), which classifies J022248\m060934 as a GRG. The angular separations of the middle (II) and inner (III) pairs of lobes are 107 arcsec (866 kpc) and 57 arcsec (461 kpc), respectively (more details in Table \ref{tab: size}). 

    Based on the MIGHTEE high resolution image, the overall morphology of J022248\m060934 is close to an FRII class or an intermediate between FRII/FRI as the outer western lobe slightly resembles an FRI (cf. hybrid sources \citealt{2000A&A...363..507G, 2020MNRAS.491..803H}). There is a faint bridge-like emission connecting the eastern middle-inner lobes and the middle-outer lobes in the western side. We also note that the western inner lobe has an intriguing morphology in which a more edge-brightened part is embedded in a more elongated second part (indicated by the double arrow in Figure \ref{fig: optical_host}). The western lobes also appear more diffuse and elongated with respect to the eastern side. The arm-length ratios (R$_l$) of the eastern to western lobes are R$_{l, out}$ = 0.48, R$_{l, mid}$ = 1.44 and R$_{l, in}$ = 0.73  for the outer, middle and inner lobes, respectively. The eastern components are more compact and brighter than their western counterparts, except for the inner lobes. The outer and middle eastern lobes are respectively $\sim 3$ and $\sim8$ times much brighter than the western sides, i.e. flux density ratios of the eastern to western lobes (R$_f$) of   R$_{f, out}$ = 2.9 and  R$_{f, mid}$ = 7.6 as reported in Table \ref{tab: size} (refer to Table \ref{tab: flux_appendix} as well for flux density measurements at other frequencies). Each pair of lobes in J022248\m060934 therefore presents asymmetry in both projected sizes and flux densities. \\

\indent In terms of orientations, the three generation lobes of J022248\m060934 in the western side seem to be more aligned compared to the opposite side. Measured from the MIGHTEE-hi radio contours, the eastern and western middle lobes respectively bend southward by about 43 and 8 deg from the position angle of the inner lobes. The outer western lobe is almost coaxial with respect to the inner lobes but with a wavy shape. On the other hand, the outer eastern lobe is bending about 43 deg northwards from the inner lobe axis. Assuming the inner double structures to be similar to FRI jets, we use Equations (1) and (2) from \citet{2020MNRAS.499.5719R} to estimate the jet inclination ($\theta_{jet}$) to the line of sight. By measuring a sidedness ratio of 1.4 (the flux density ratio between the jet and the counter jet) and a spectral index of $\alpha = -$ 0.4 (measured from the maps in Figure \ref{fig: alphamap}, see also Section \ref{sec: sepctral_index distribution}), we obtained the inclination of $\theta_{jet} =$ 2.4 deg. \\
\indent At other frequencies (see Figure \ref{fig: source_morpho_archival}), J022248\m06093 is not entirely detected or poorly resolved.  
In the image from LOFAR at 150 MHz, the RG only exhibits a double-double morphology. The outer and middle eastern lobes are detected but marginally resolved as two distinct components. The western side displays three distinct components but the middle one is barely detected at 3$\sigma$ level. The outer lobe has weak evidence of edge-brightening but it is just slightly brighter than the middle lobe. Similarly, at 325 MHz, the TDRG is also barely resolved into three components only. While at 3000 MHz from VLASS, the core is the only component detected.

  \begin{table*}
  
 \caption{Summary of the properties of the source. The columns are as follows: (1) lists the components of the source, (2) and (3) report the LAS and LLS values of J022248\m060934 and its components. The LAS were obtained by measuring the total end-to-end of the source (I, total), the middle and inner pair of lobes (II and III, respectively) and each lobe. The MIGHTEE-hi contours above 3$\sigma_{local}$ were used for the measurements in order to compute the linear arm-length ratios R$_{l}$ in column (4). Columns (5) and (6) list the flux density and radio power measured from MIGHTEE-hi image. The flux density ratios of the eastern to western lobes R$_{f}$ are shown in column (7). }
 \label{tab: size} 
		\centering
		\begin{tabular}{lccccccc}
  
	\hline
 Components & LAS  & LLS & R$_{l}$& $S_{\rm{mightee-hi}}$  & P$_{\rm{mightee-hi}}$ & R$_{f}$ \\
    & [arcsec] & [kpc]& & [mJy] & $\times$10$^{24}$\,[W Hz$^{-1}$]&\\
 (1)& (2)& (3) & (4) &(5)&(6)&(7)\\
    \hline
    \hline
    
    Outer double (I) & 196 &1587&0.48& 3.96 $\pm$ 0.39 &4.6&2.9\\
    Middle double (II) & 107& 866&1.44& 4.02$\pm$ 0.40&6.1&7.6\\
    Inner double (III) & 57& 461& 0.73& 5.36$\pm$ 0.53&10.0&0.2\\
    Core & 9 & 73&-& 4.02 $\pm$ 0.40&7.4&-\\
    Lobe $_{\rm{out (E)}}$ & 27 & 218&-&2.95 $\pm$ 0.29&3.4&-\\
    Lobe $_{\rm{out (W)}}$ & 56& 453&-&1.01 $\pm$ 0.10& 1.2&-\\
    Lobe $_{\rm{mid (E)}}$ & 26& 210&-&3.55 $\pm$ 0.35& 5.4&-\\
    Lobe $_{\rm{mid (W)}}$ & 18& 146&-&0.47 $\pm$ 0.04 &0.7&-\\
    Lobe $_{\rm{in (E)}}$ & 19& 154&-&1.06 $\pm$ 0.10 & 1.7 & - \\
    Lobe $_{\rm{in (W)}}$ & 26 & 210&-&4.30 $\pm$ 0.43 & 7.9& -\\
    \hline
    
		\end{tabular}
        
	\end{table*}
    
\subsection{Spectral index distribution}\label{sec: sepctral_index distribution}
\begin{figure*}
		\centering
		\begin{tabular}{c}		
  {\resizebox{0.9\hsize}{!}{\includegraphics[trim={0.5cm 0 0.5cm 0},clip]{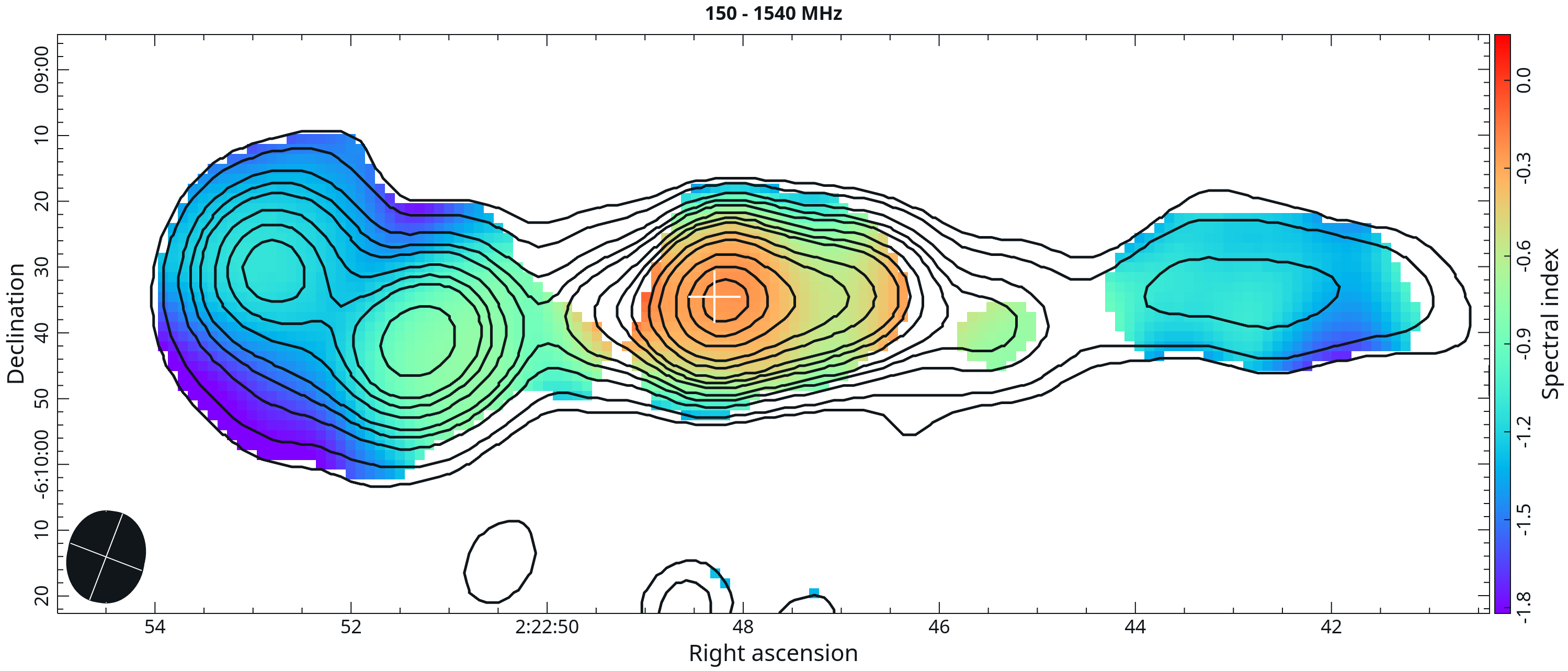}}}\\
  \\
  {\resizebox{0.9\hsize}{!}{\includegraphics[trim={0.5cm 0 0.5cm 0},clip]{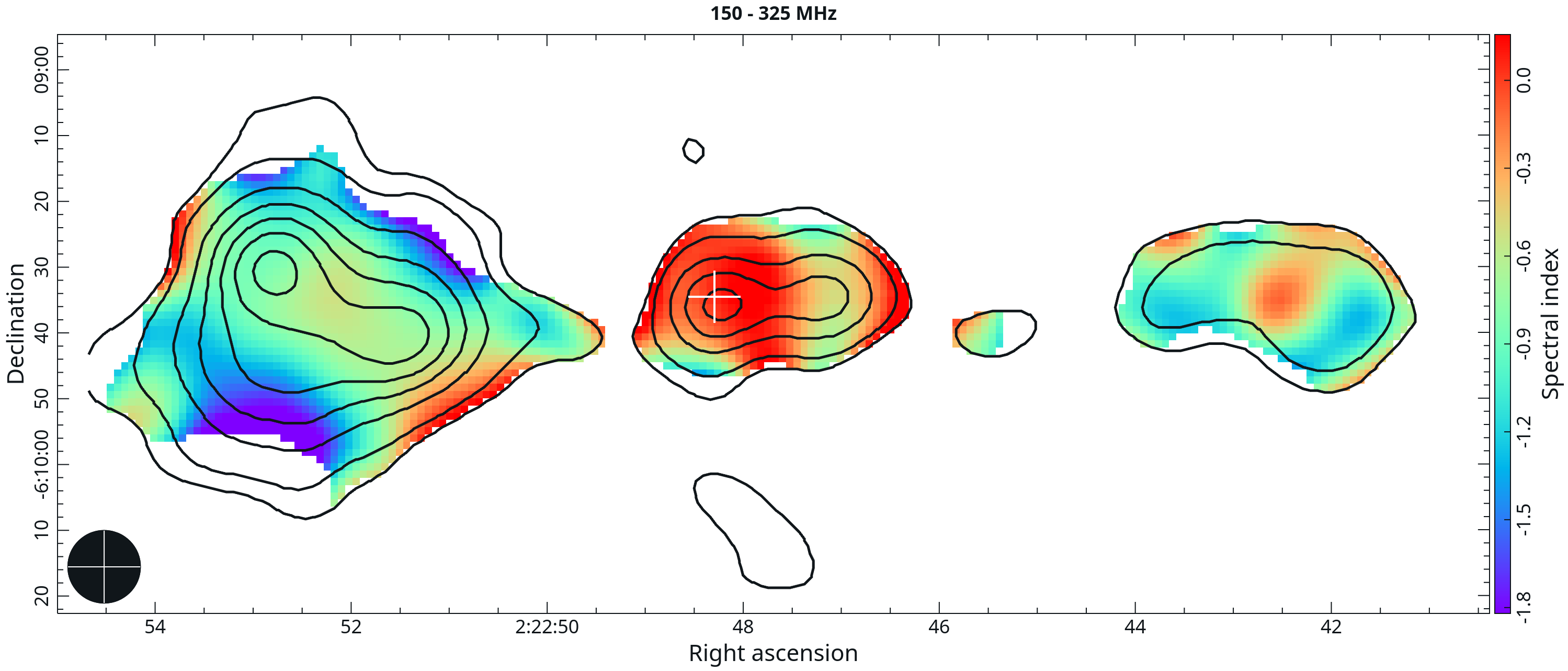}}}\\
  \\
  {\resizebox{0.9\hsize}{!}{\includegraphics[trim={0.5cm 0 0.5cm 0},clip]{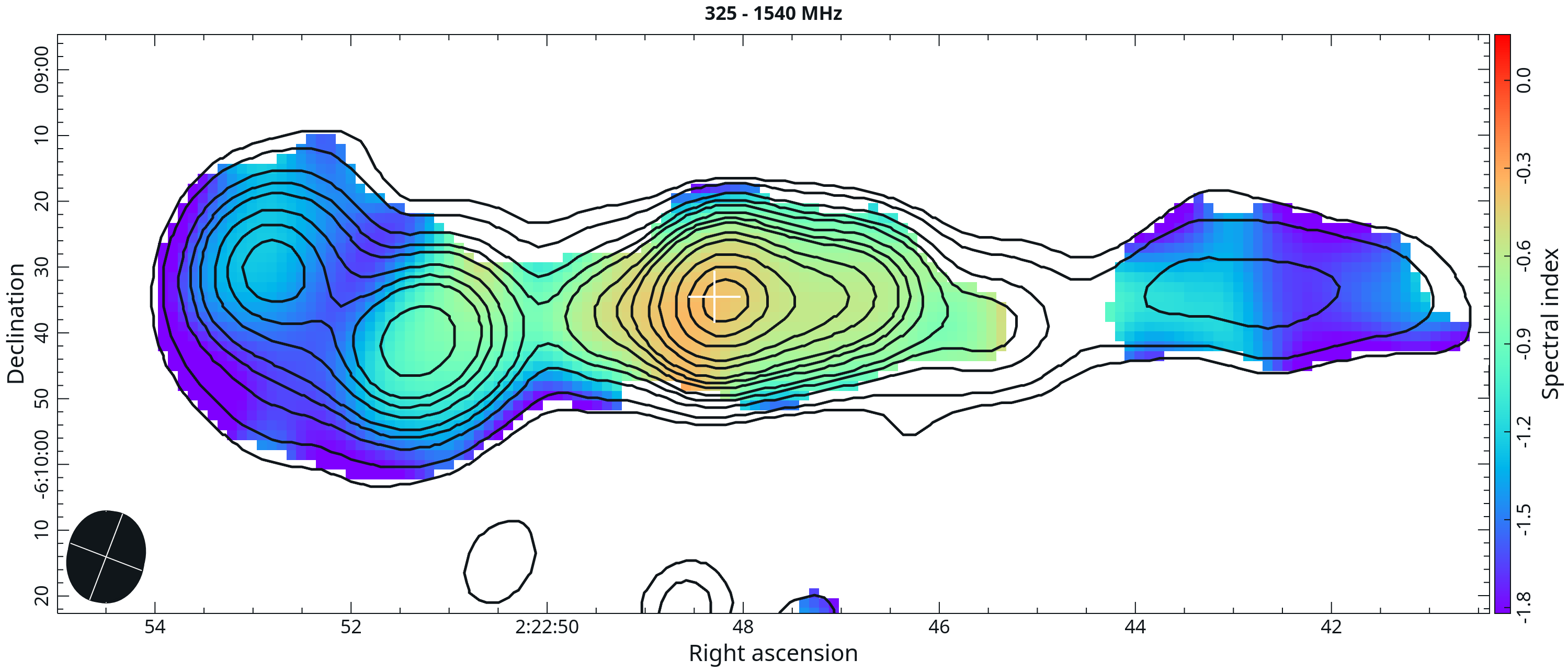}}}\\
   \end{tabular}
\caption{Spectral index maps above 3$\sigma_{local}$ level of the TDRG. \textit{Top panel:} map computed between 150 and 1540 MHz with MIGHTEE-1540 contours (levels: $-$ 3, 3, 5, 10, 15, 20, 30, 40, 50, 75, 100, 150, 200 $\times \sigma_{local}$ = 16.1 $\mu$Jy/beam); \textit{Middle panel:} map computed between 150 and 325 MHz with LOFAR contours  (levels: $-$3, 3, 5, 10, 15, 20, 25, 30, 35 $\times \sigma_{local}$ = 366 $\mu$Jy/beam); \textit{Bottom panel:} map computed between 325 and 1540 MHz with the same MIGHTEE-1540 contours. The solid black circle in the bottom left corner indicates the resolution of each map. The purple and blue colours depict the steepest emissions. The white cross represents the position of the optical host galaxy.}
		
		\label{fig: alphamap}
	\end{figure*}

\indent To further investigate the properties of each component of the TDRG, we compute the spectral index distribution across the source. We use LOFAR, GMRT Legacy and MIGHTEE sub-band images to ensure similar resolution and sensitivity at different spatial scales. Despite any difference between the uv-coverage of the images, we do not expect a major flux density loss as the LAS of the source is below the largest scale recovered by LOFAR, GMRT and MeerKAT.\\
\indent For each map produced, we convolved two images between two different bands to the same synthesized beam which is followed by a regridding. Finally, we computed the spectral index of each pixel above 3$\sigma_{local}$ in both bands. The resolutions of the resulting maps in Figure \ref{fig: alphamap} are 14 $\times$ 11.6 arcsec for both LOFAR versus MIGHTEE-1540 (top panel) and GMRT versus MIGHTEE-1540 (bottom panel) and then 11 arcsec for LOFAR versus GMRT (middle panel). We also computed other spectral index maps using the MIGHTEE-970 and MIGHTEE-1230 images. These additional maps are shown in Figure \ref{fig: alphamap_appendix}.

We notice a general trend in all the spectral index maps from 150 to 1540 MHz. The spectral index values are found to be steeper from the outermost lobes and gradually flatten toward the inner lobes. We can affirm that the outer pair of lobes is the oldest structure of the radio galaxy with spectral index as steep as $\alpha \sim - 2$ at the edges and an average of $\alpha \sim -1.3$. The middle pair of lobes and the inner lobes, the closest to the core, show an average spectral index of $\alpha \sim -1$ and $\alpha \sim -0.5$, %(minimum of $\alpha$ of $ \sim -1.1$)
respectively. The presence of hotspots, especially in the outer and middle eastern lobes, is well traced by flatter indices compared to their surrounding radio plasma. 

In addition, we notice the inversion of the core spectrum. At low frequency, the core shows an inverted spectral index of $\alpha_{150}^{325}=0.2$ which turns into $\alpha_{325}^{1540} = -0.4$ at high frequency. This  spectral behaviour might be caused by synchrotron self-absorption (SSA, \citealt{2009AN....330..120F}) or by free-free absorption (FFA, \citealt{2015ApJ...809..168C, 2015AJ....149...74T}). We plotted the radio spectrum of the core from 150 to 3000 MHz. To limit the contamination from the extended emissions, the integrated peak flux densities of the core were used (see Table \ref{tab: flux_appendix}). The spectrum is then fitted using an homogeneous SSA and a uniform FFA models which are described in more details in \citet{2018MNRAS.477..578C}. Under the assumption of SSA, the spectrum can be modelled by:
\begin{equation}
S_{\nu} = a\left(\frac{\nu}{\nu_m}\right)^{-(\beta-1)/2}\left(\frac{1-e^{-\tau}}{\tau}\right),
\label{SSA_eq}
\end{equation}
\noindent where $a$ is the flux normalisation, $\beta$ is the power-law index of the electron energy distribution, $\tau$ is the optical depth given by $(\nu/\nu_m)^{-(\beta+4)/2}$ and $\nu_m$ is the turnover frequency where $\tau$ = 1. 
On the other hand, by considering FFA, the spectrum can be described as: 
\begin{equation}
S_{\nu} = a\nu^{\alpha}e^{-\tau_{\nu}},
\end{equation} 
where $\alpha$ is the intrinsic synchrotron spectral index, $\tau_{\nu}$ is the free-free optical depth, defined by $\tau_{\nu} = (\nu/\nu_0)^{-2.1}$, where $\nu_0$ is the scale frequency at which $\tau_{\nu} = 1$.

\begin{figure}
		\centering
		\begin{tabular}{c}		
  {\resizebox{1\hsize}{!}{\includegraphics{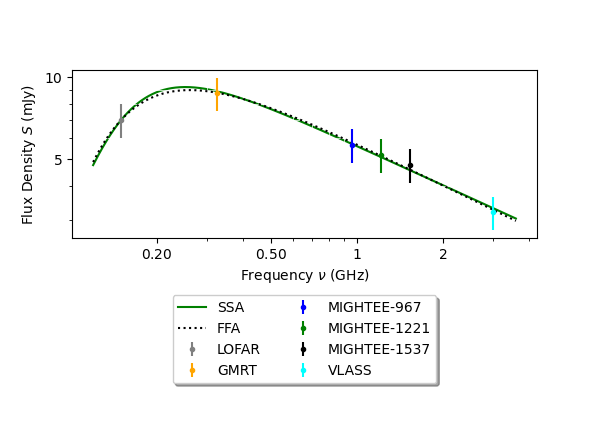}}}
  
  \end{tabular}
\caption{Fitted radio spectrum of the core from 150 to 3000 MHz assuming an homogeneous SSA \textit{(solid green line)} and a uniform FFA \textit{(dotted black line)} models. To limit contamination from extended emissions, the peak flux densities were used. The flux errors were assumed to be 10 per cent for each measurement, except at 3000 MHz where 8 per cent was used instead following the Quick Look users guide. }
%\footnote{https://science.nrao.edu/vlass/data-access/vlass-epoch-1-quick-look-users-guide}	
		\label{fig: core_radio_SED}
	\end{figure}

The results of the fitting are shown in Figure \ref{fig: core_radio_SED}. The respective turnover frequencies are at $\nu_m$ = 0.18$\pm$0.0066 and $\nu_0$ = 0.13$\pm$0.0051 GHz for the SSA and FFA models. We found a difference in Bayesian information criterion ($\Delta$BIC = BIC$_{SSA}$ - BIC$_{FFA}$) of 0.06 and the reduced chi-squared values of $\sim$ 0.06 and $\sim$ 0.04 for the SSA and FFA models, respectively.  Based on these fitting statistics, both models seem to overfit the data and the difference in the output results from them is also negligible. Such results are expected and could arise from various reasons such as possible high uncertainties in our observations and the limited number of data point used, especially at low frequencies, to the absorbed region of the spectrum.  We also lack simultaneous observations at the different frequencies which are required to accurately probe the spectral shape albeit such an effort is beyond the scope of this project.

  Another parameter often used to characterise the spectral properties of a radio emission is spectral curvature, defined as the difference between the spectral index at high frequencies and at low frequencies. Based on the synchrotron spectral ageing model, a curvature above $>$ 0.5 implies no acceleration of fresh particles and therefore it is an indicator of old remnant plasma (e.g. \citealt{2007A&A...470..875P, 2011A&A...526A.148M}). To generate the spectral curvature map from 150 to 1540 MHz, we subtract the difference between two spectral index maps, LOFAR-GMRT at 14 $\times$ 11.6 arcsec and GMRT-MIGHTEE-1540 at 14 $\times$ 11.6 arcsec. The resulting map in Figure \ref{fig: spc_map} shows more curved emission at the edges of the lobes ($\geq$ 0.5), which is consistent with the values of the spectral index in Figure \ref{fig: alphamap}.

%%%%%%%%%%%%%%spectral index map

    \begin{figure*}
		\centering
		\begin{tabular}{c}		
  {\resizebox{0.9\hsize}{!}{\includegraphics[trim={0.5cm 0 0.5cm 0},clip]{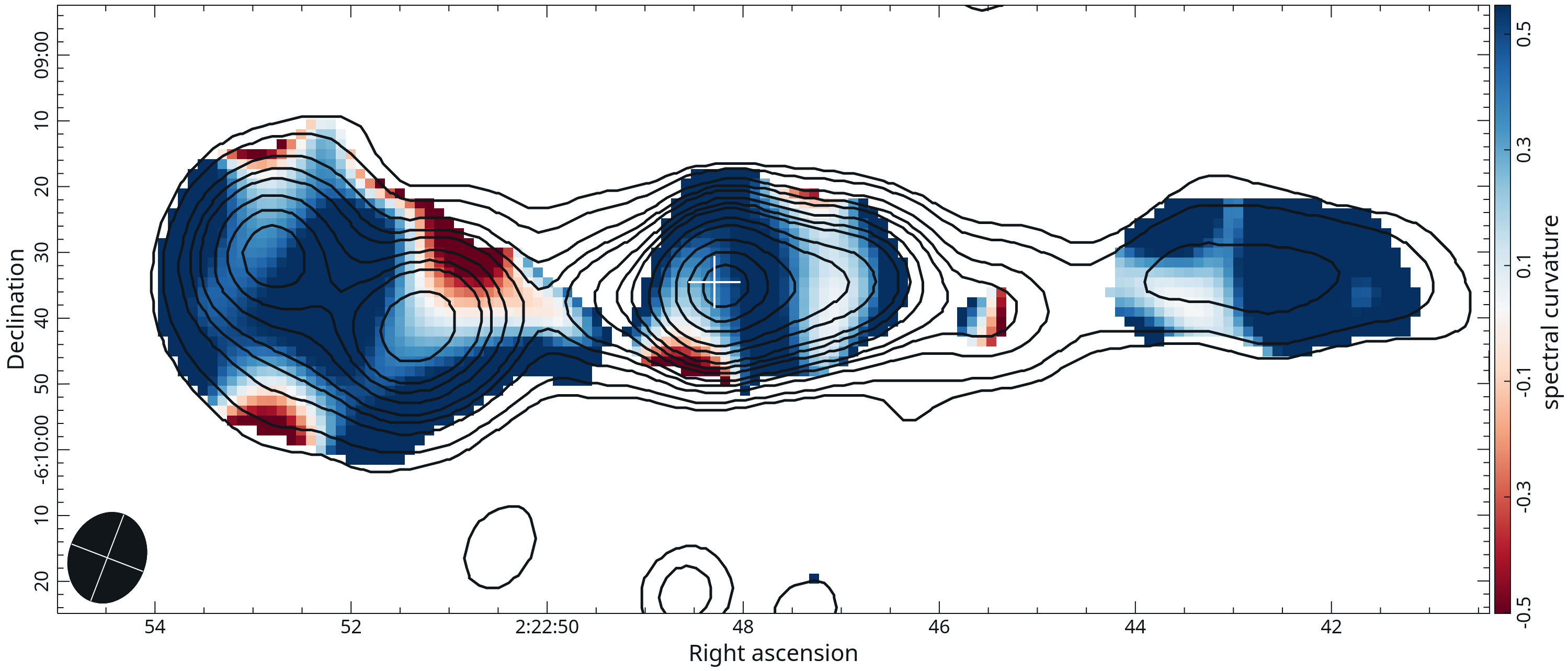}}}
    \end{tabular}
\caption{ Spectral curvature map of the TDRG from 150 to 1540 MHz. The most curved regions are indicated by dark blue colour. The contours on top show the 1540 MHz MIGHTEE radio morphology of the TDRG above 3$\sigma_{local}$ (contour levels: $-$ 3, 3, 5, 10, 15, 20, 30, 40, 50, 75, 100, 150, 200 $\times \sigma_{local}$ = 16.1 $\mu$Jy/beam). The synthesized beam of the map is shown in the bottom left corner.} 
		
		\label{fig: spc_map}
	\end{figure*}

\subsection{Age}\label{sec: results_agemap}
We adopted the method used by \citet{2016ApJ...826..132S} and \citet{2023MNRAS.525L..87C} to compute the kinematic age of J022248\m060934. We assume the average lobe advancement speed of 0.1$c$, 0.05$c$, and 0.01$c$ for the inner, middle and outer lobes, respectively. These are based on the results from the dynamical and spectral ageing of DDRGs studied by \citet{2006MNRAS.372..693K} and \citet{2010A&A...510A..84M}  given that there was no available spectral age studies of TDRGs. By using the projected linear sizes of each pair of lobes in Table \ref{tab: size}, we then compute their kinematic ages. From the innermost pair outward, we found the age estimations of 8 Myr, 28 Myr and 258 Myr. By subtracting the time between two episodes of activity, we can also estimate the time during which the nuclear activity went off. This gives the quiescent period upper limits of 230 Myr between the first and second cycles and 20 Myr between the second cycle and the ongoing activity. These values are similar to what are found in other TDRGs \citep{2016ApJ...826..132S, 2023MNRAS.525L..87C}. However, we strongly caution that these kinematic ages are rough approximations only, considering that the jet advancement speed would not be constant for instance \citep{2012ApJ...760...77A}. Furthermore, none of the morphological asymmetries is taken into account.\\

In contrast to kinematic age, spectral age refers to the time that has passed since the acceleration of electrons, taking into account their energy losses through synchrotron emission and inverse Compton scattering. One way to determine this age consists of modelling the observed radio spectrum with numerical integration models.  
The electrons are accelerated or injected following a power-law energy distribution described as $ N(E) = N_0E^{- 2\alpha_{inj} - 1}$, where $\alpha_{inj }$ is the injection index. \\
\indent We use single injection models if we can assume that the particles from a region within the source have been accelerated in a single event. Among these, we count the JP model \citep{1973A&A....26..423J} and the Tribble model \citep{1991MNRAS.253..147T, 1993MNRAS.261...57T}. The JP model assumes a constant magnetic field, whereas the Tribble model adopts a spatially non-uniform magnetic field described by a Maxwell-Boltzmann distribution. \citet{2013MNRAS.435.3353H, 2015MNRAS.454.3403H} explain that the latter model is more realistic. 
We use the broadband radio astronomy tools software package (BRATS, \citealt{2013MNRAS.435.3353H, 2015MNRAS.454.3403H})\footnote{http://www.askanastronomer.co.uk/brats/} for our spectral age analysis.

To start, we smoothed the radio maps at 150 and 325 MHz to match the resolution of the MIGHTEE sub-band images of 14$\times$11.6 arcsec. Using the formula of \citet{2004IJMPD..13.1549G}, we computed the equipartition magnetic field (B$_{eq}$) of any components of the source that are detected and resolved in the MIGHTEE sub-band. The core was excluded as it is dominated by self-absorption and therefore we could not fit its spectrum with any ageing model. We then determine the injection index value of each component. We first consider the range between 0.5 to 0.8 with a step of 0.1. Then, we continue to constrain the previous best-fit results with smaller steps. The best-fit values of $\alpha_{inj}$ as well as the derived B$_{eq}$ of each component of the source are listed in Table \ref{tab: single_injection}.

\begin{table*}
 \caption{Spectral age fitting results from BRATS using the JP model. Column lists: (1): indicates the fitted component where $^\dagger$ indicates the components for which the spectral age fit was done using the MIGHTEE sub-band data only due to low detection in LOFAR and GMRT, (2): estimated equipartition magnetic field, (3): resulting injection index, (4): average $\chi_{red}^2$ over the fitted components, (5): fit number of degrees of freedom, (6): indicates whether the model can be rejected or not, (7): confidence level at which the model can be rejected or not, (8 $-$ 12): number of pixels with the corresponding level of confidence bin, (13): the minimum age of a pixel within a component and (14): the maximum age of a pixel within a component.} 
 \label{tab: single_injection}
		\centering
        {\resizebox{1\hsize}{!}{
        
		  \begin{tabular}{lccccccccccccc}
          
        \hline
         Component & B$_{eq}$ & $\alpha_{inj}$& Average & dof & Rejected& Confidence & &&Confidence bins&&& Min age& Max age\\
         \cline{8-12}
        & [$\mu$G]& &$\chi_{red}^2$& &&level& $<$68&68$-$90&90$-$95&95$-$99&$>$99& [Myr]& [Myr]\\
        (1)&(2)&(3)&(4)&(5)&(6)&(7)&&&(8 $-$ 12)&&&(13)&(14)\\
        \hline
        \hline
         Lobe$_{out (E)}$&15.9&$-$0.80&0.37&3&No& 68 & 391 & 22 & 0& 0 & 0  & 5.62 $^{+0.55}_{-0.59}$ & 13.18$^{+1.00}_{-1.28}$\\
         &&&&&&&&&&&&\\
         %&Tribble& &&No& 68 & 1626 & 439 & 125& 56 & 171  & 0.00 $^{+}_{-}$ & 26.27$^{+}_{-}$\\
         Lobe$_{mid (E)}$& 11.4&$-$0.70&0.49&3&No& 68 & 272 & 20 & 0& 0 & 0  & 1.76 $^{+1.93}_{-1.76}$ & 12.93$^{+0.85}_{-1.02}$\\
         &&&&&&&&&&&&\\
         $^\dagger$Lobe$_{in (E)}$& 7.4&$-$0.50&0.01&1&No& 68 & 173 & 0 & 0& 0 & 0  & 0.00 $^{+17.12}_{-0.00}$ & 8.78$^{+11.43}_{-8.78}$\\
         &&&&&&&&&&&&\\
        Lobe$_{in (W)}$& 8.8&$-$0.50&0.20&3&No& 68 & 150 & 0 & 0& 0 & 0  & 0.00 $^{+2.87}_{-0.00}$ & 8.82$^{+3.35}_{-6.81}$\\
         &&&&&&&&&&&&&\\
             $^\dagger$Lobe$_{mid (W)}$& 6.45&$-$0.50&0.003&1&No& 68 & 133 & 0 & 0& 0 & 0  & 4.42 $^{+11.77}_{-4.42}$ & 9.59$^{+10.57}_{-9.59}$\\
         &&&&&&&&&&&&\\
        Lobe$_{out (W)}$ & 9.0&$-$0.57&0.19&3&No& 68 & 280 & 0 & 0& 0 & 0  & 10.78 $^{+2.89}_{-4.64}$ & 17.62$^{+1.44}_{-1.79}$\\
        \hline
        
        \end{tabular}{}
        }}
        
\end{table*}
    
\begin{figure*}
		\centering
		\begin{tabular}{cc}		
  {\resizebox{0.5\hsize}{!}{\includegraphics[trim={0.4cm 0 0.25cm 0},clip]{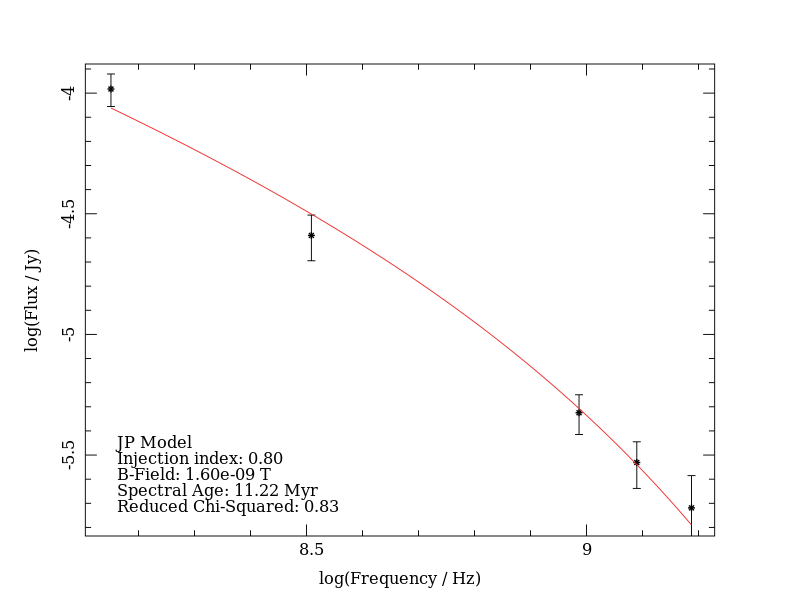}}}&{\resizebox{0.5\hsize}{!}{\includegraphics[trim={0.4cm 0 0.25cm 0},clip]{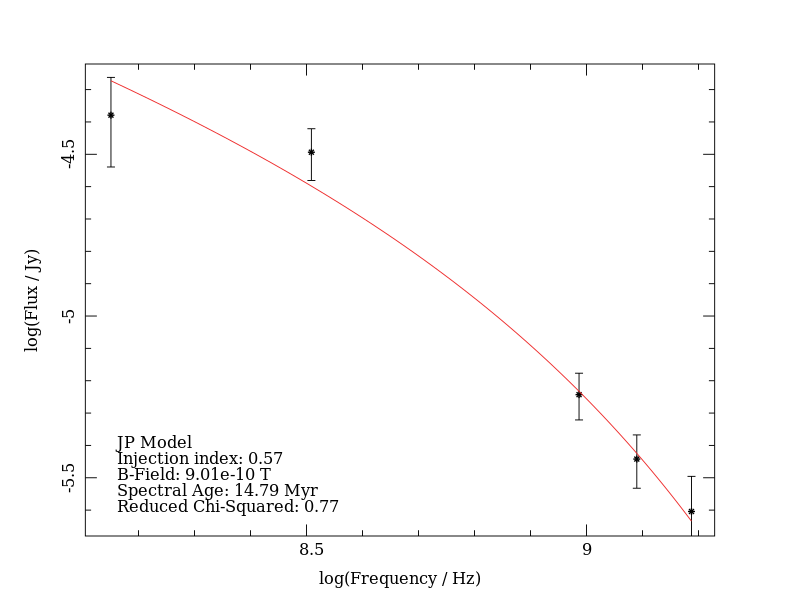}}}\\
  (a) Eastern outer lobe& (b) Western outer lobe\\
  {\resizebox{0.5\hsize}{!}{\includegraphics[trim={0.4cm 0 0.25cm 0},clip]{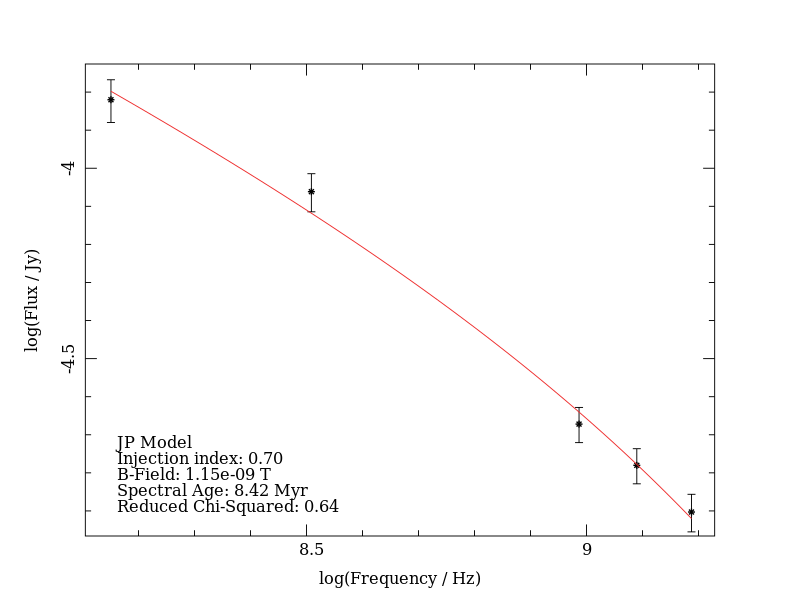}}}&{\resizebox{0.5\hsize}{!}{\includegraphics[trim={0.4cm 0 0.25cm 0},clip]{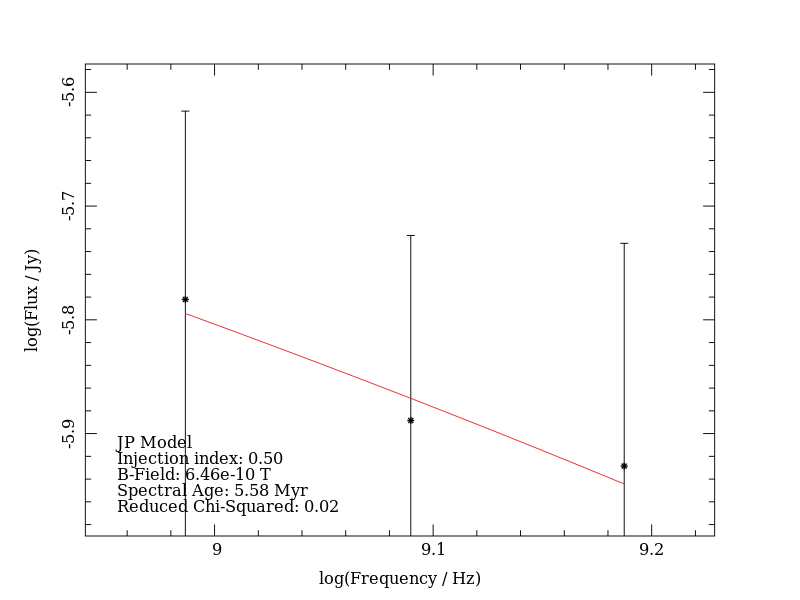}}}\\
  (c) Eastern middle lobe& (d) Western middle lobe\\
  {\resizebox{0.5\hsize}{!}{\includegraphics[trim={0.4cm 0 0.25cm 0},clip]{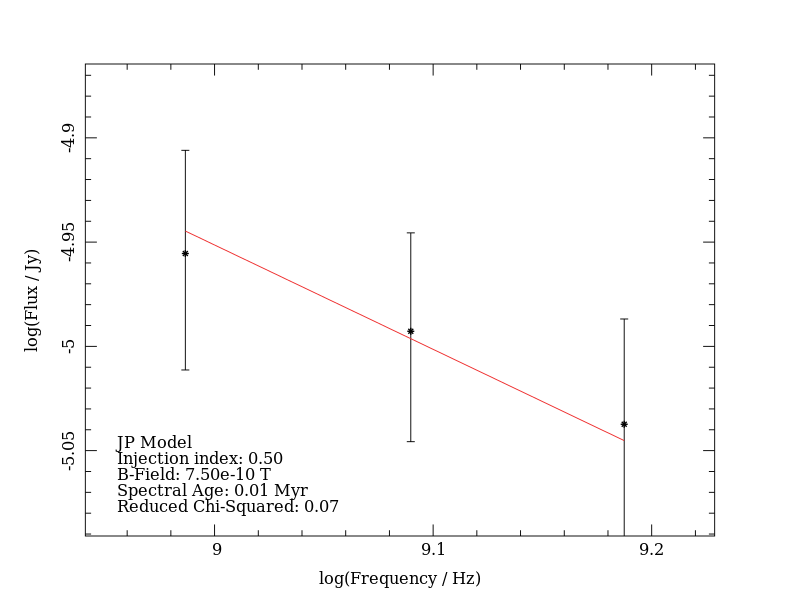}}}&{\resizebox{0.5\hsize}{!}{\includegraphics[trim={0.4cm 0 0.25cm 0},clip]{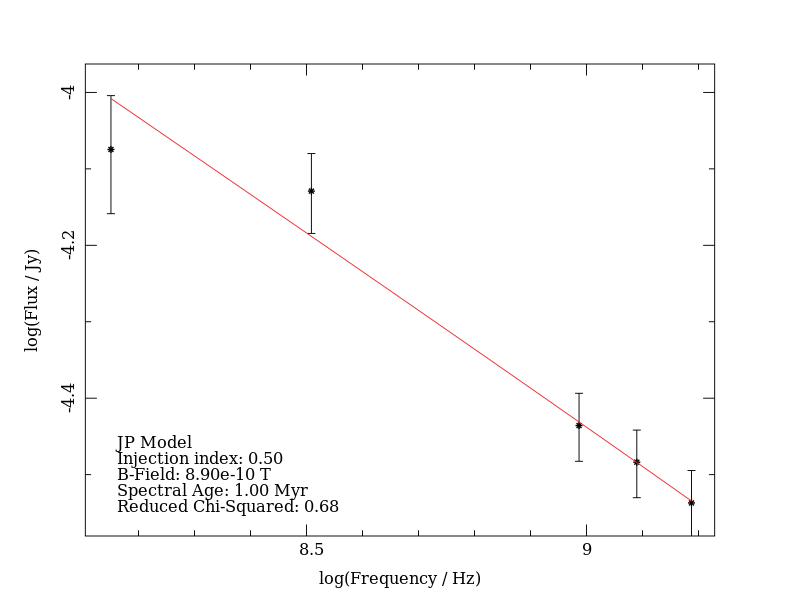}}}\\
(e) Eastern inner lobe& (f) Western inner lobe\\
  
  \end{tabular}
\caption{Representative radio spectrum plots (flux density vs. frequency) of an individual pixel with a good fitting result for each component. The  first column displays pixels from the eastern lobes while the second shows the pixels from the western lobes. The pixels from the outer, middle and inner lobes are respectively displayed from the top to the bottom rows. The observational data are in black points, and the best fit using the JP model is shown with the red solid line. The bottom left corner of the panel lists the parameters used, the ages (in Myr) and the $\chi_{red}^{2}$.}
		
		\label{fig: pixel_bestFit}
        \end{figure*}

We continue with the spectral age fitting using the JP and the Tribble models. For all of the fitted components, neither of the models can be rejected at the 68 per cent confidence level. The spectral ages returned by the Tribble model are slightly higher than those of the JP model. However, the results, including the reduced chi-square ($\chi_{red}^{2}$) of both models are consistent with each other within the margin of error, so we will only discuss the fitting using the JP model throughout the paper. Nevertheless, the results from the Tribble model fitting are reported in Table \ref{tab: Tribble}. The generated spectral age maps, error maps, and the distribution of $\chi^{2}$ are shown in Figure \ref{fig: agemap}. As expected, the oldest emissions are in the outer pair of lobes and the age decreases from the outer to the inner lobes. Based on the fitting, the maximum age of 17.62 $^{+1.44}_{-1.79}$ Myr is found in the western outer lobe (see Section \ref{subsec: disc_duty_cycle} where we discuss the derived age).  More information on the fit results can be found in Table \ref{tab: single_injection}. We show a representative spectral plot (flux density versus frequency) of an individual pixel with good fitting results for each fitting (component) in Figure \ref{fig: pixel_bestFit}.

   \begin{figure*}
		\centering
		\begin{tabular}{ccc}		
  {\resizebox{0.3\hsize}{!}{\includegraphics[trim={0.4cm 0 0.25cm 0},clip]{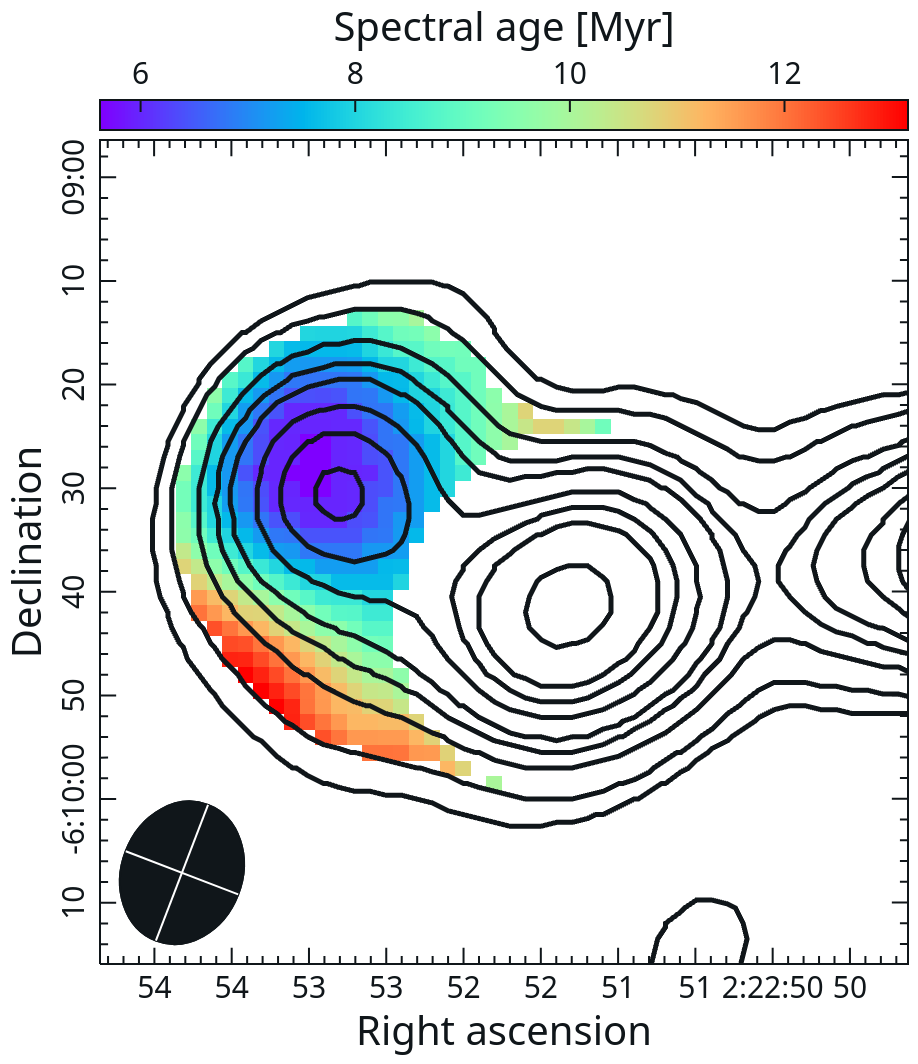}}}&
  {\resizebox{0.3\hsize}{!}{\includegraphics[trim={0.4cm 0 0.25cm 0},clip]{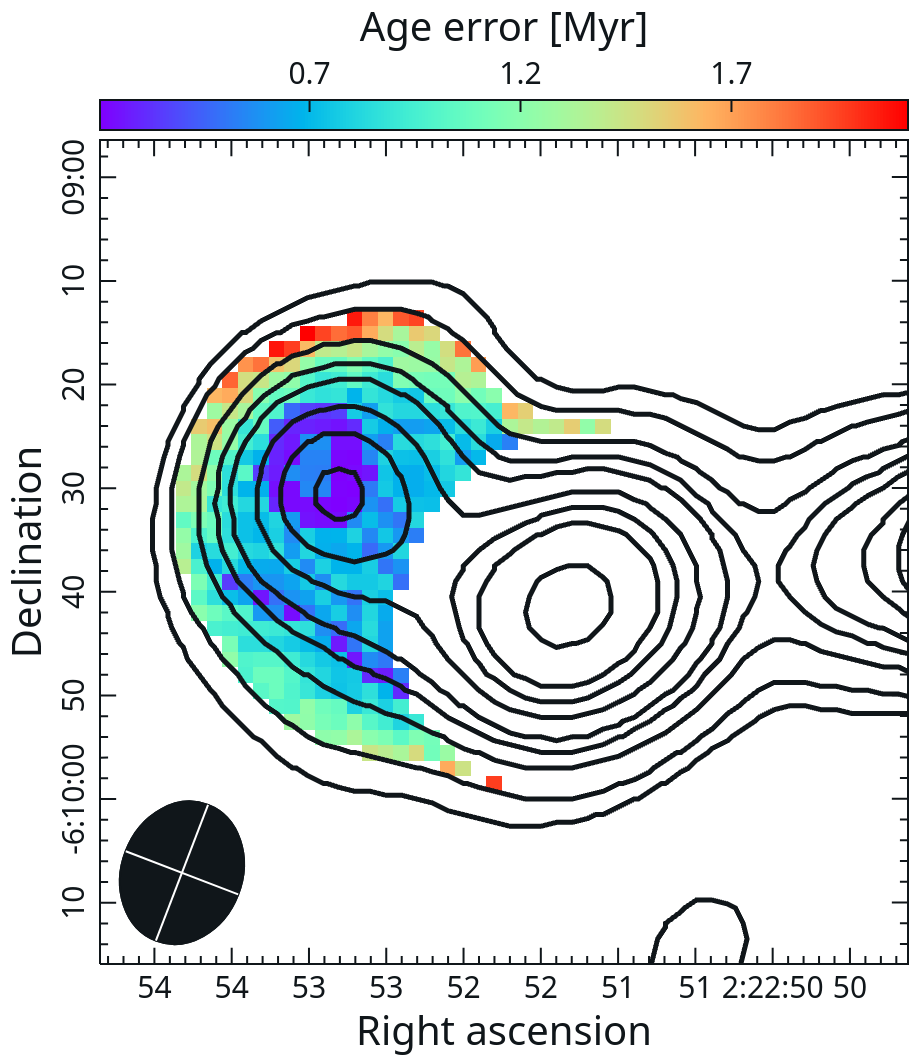}}}&
  {\resizebox{0.3\hsize}{!}{\includegraphics[trim={0.4cm 0 0.25cm 0},clip]{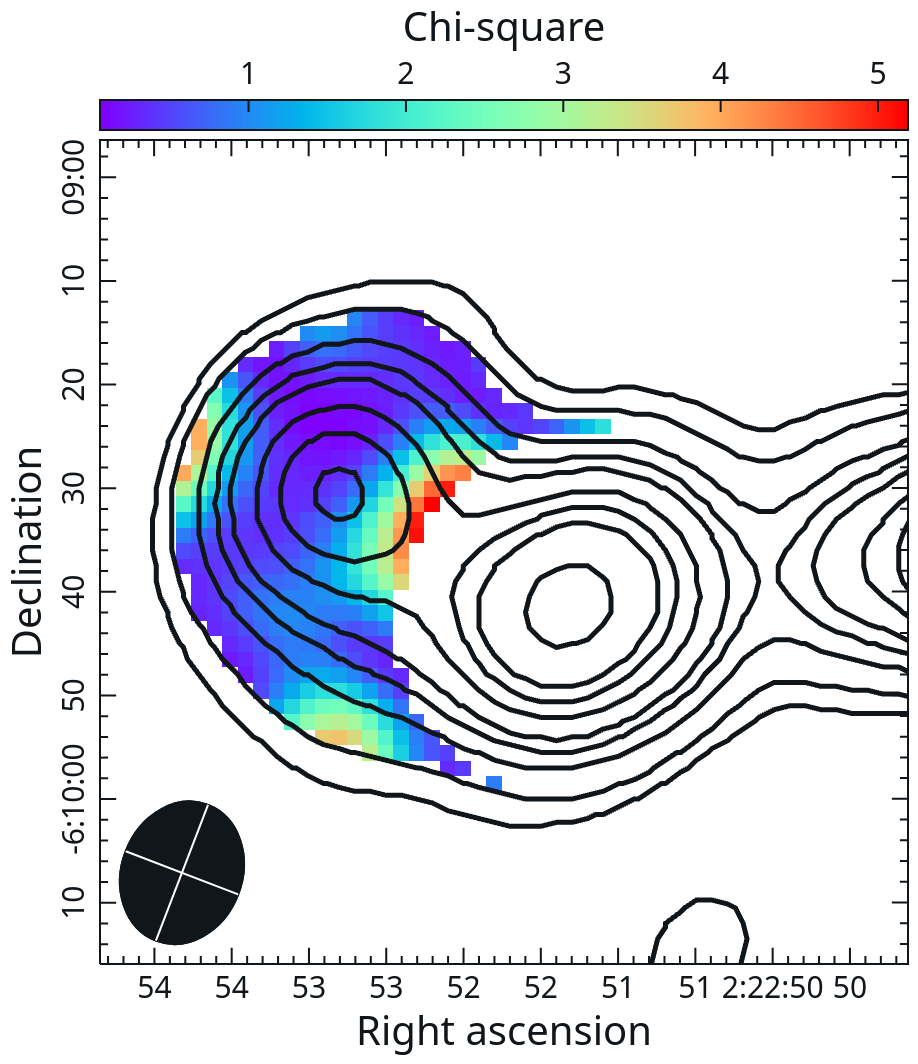}}}\\
  (a)&(b)&(c)\\
  &&\\
  {\resizebox{0.3\hsize}{!}{\includegraphics[trim={0.4cm 0 0.25cm 0},clip]{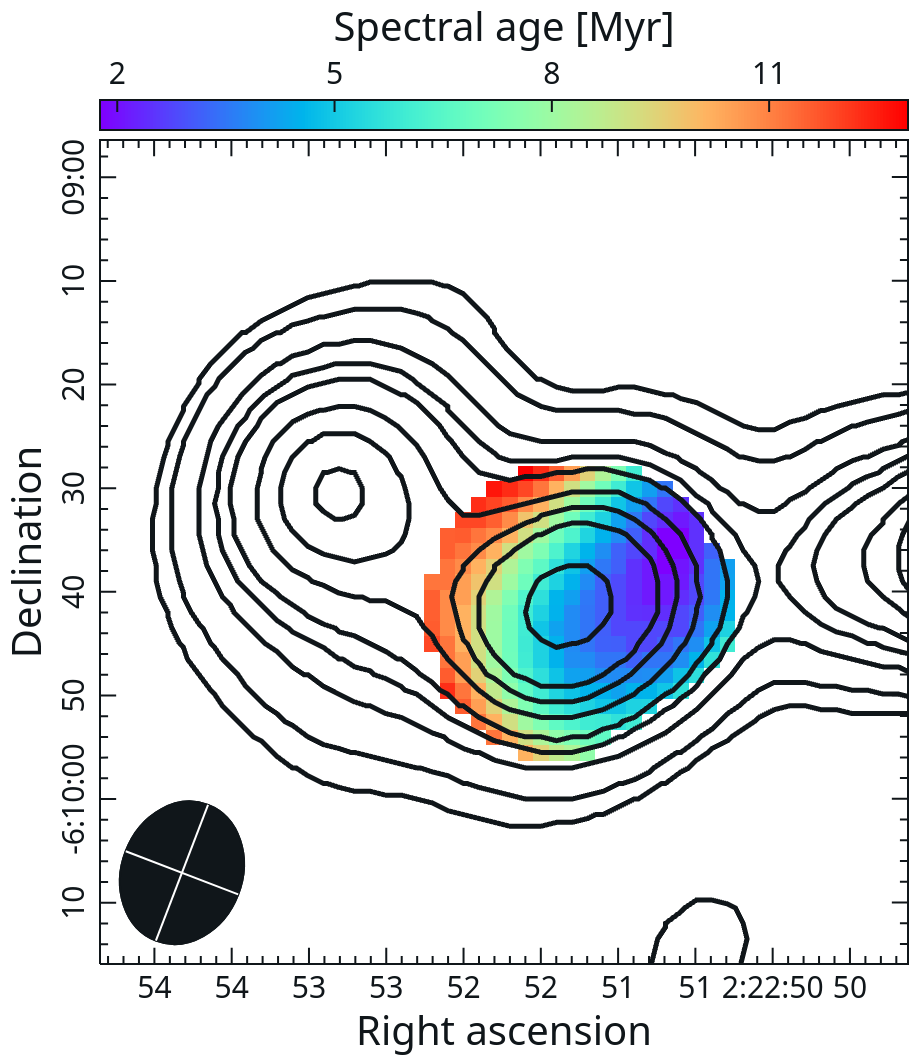}}}&
  {\resizebox{0.3\hsize}{!}{\includegraphics[trim={0.4cm 0 0.25cm 0},clip]{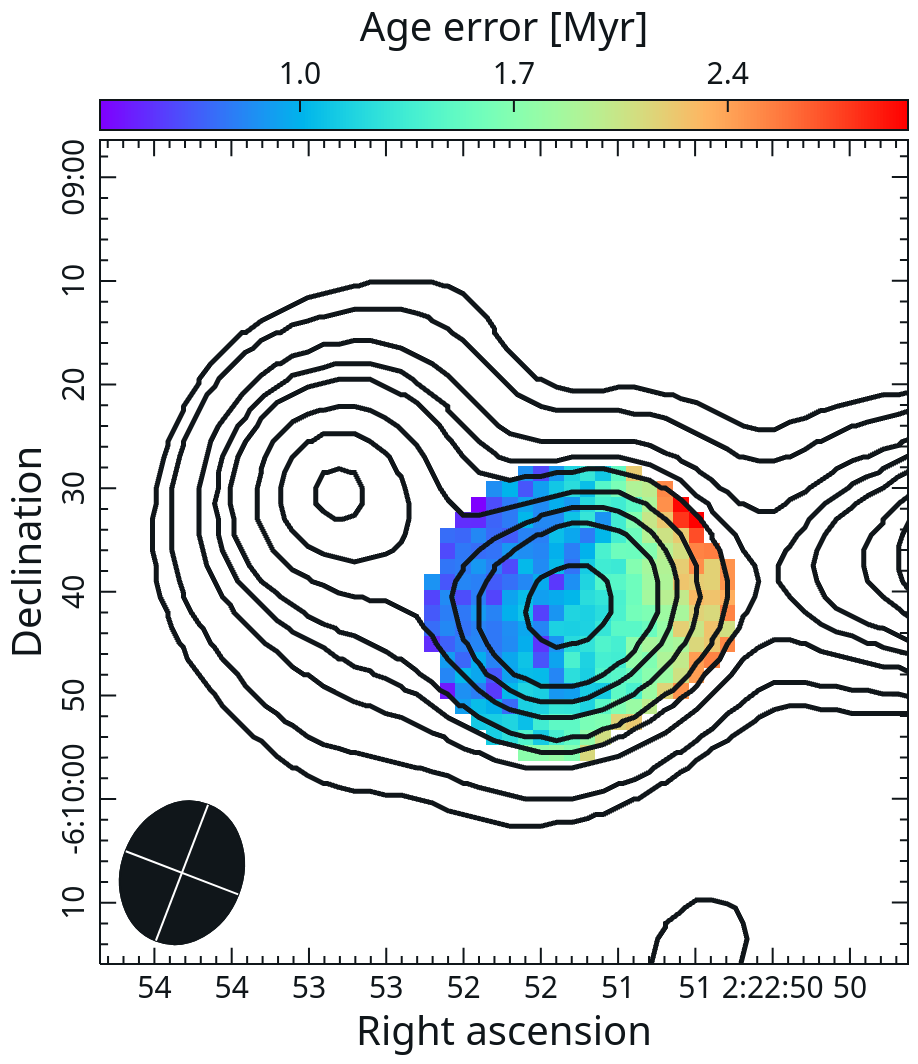}}}&
  {\resizebox{0.3\hsize}{!}{\includegraphics[trim={0.4cm 0 0.25cm 0},clip]{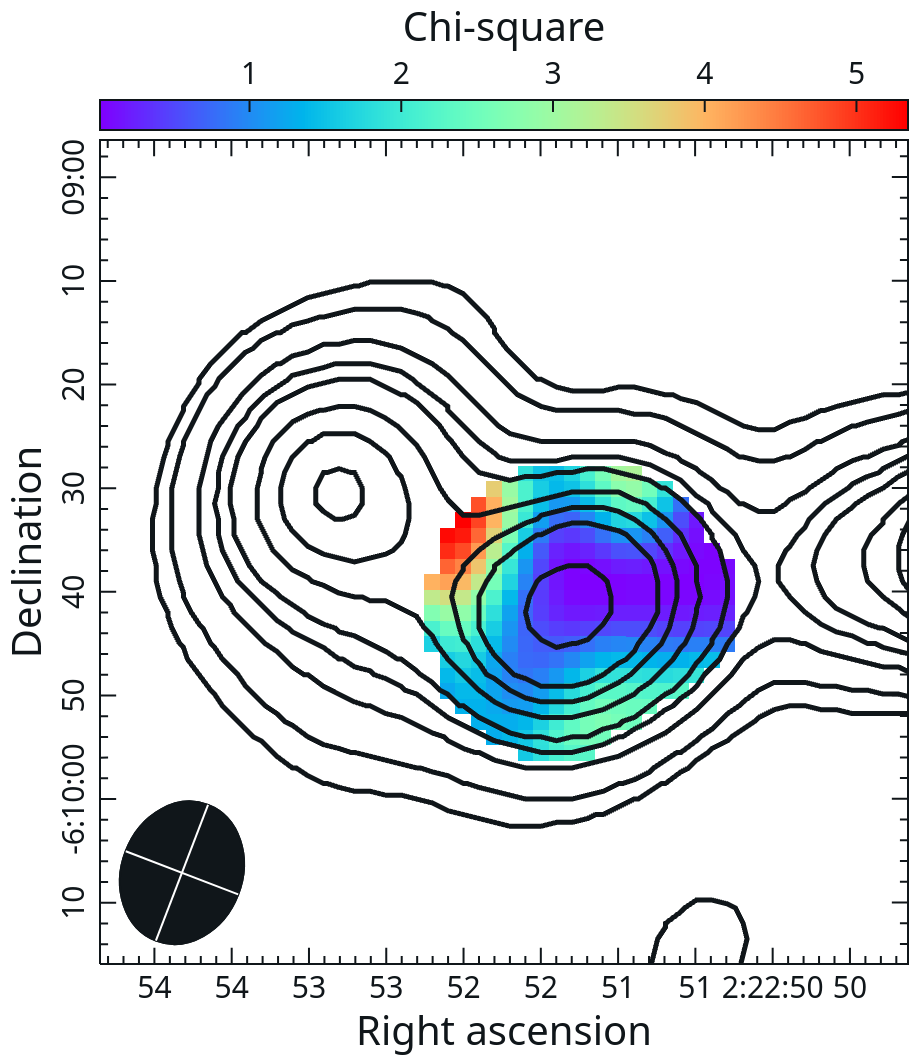}}}\\
  (d)&(e)&(f)\\
  &&\\
  {\resizebox{0.3\hsize}{!}{\includegraphics[trim={0.4cm 0 0.25cm 0},clip]{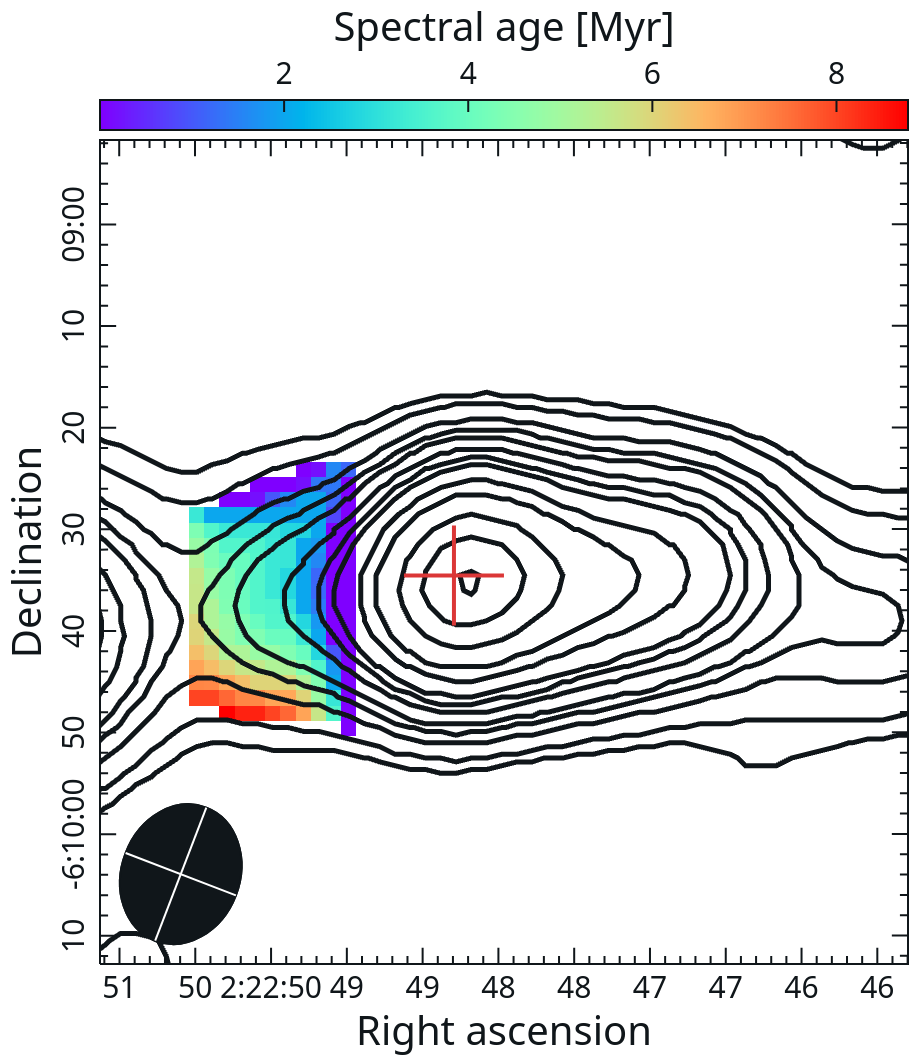}}}&
  {\resizebox{0.3\hsize}{!}{\includegraphics[trim={0.4cm 0 0.25cm 0},clip]{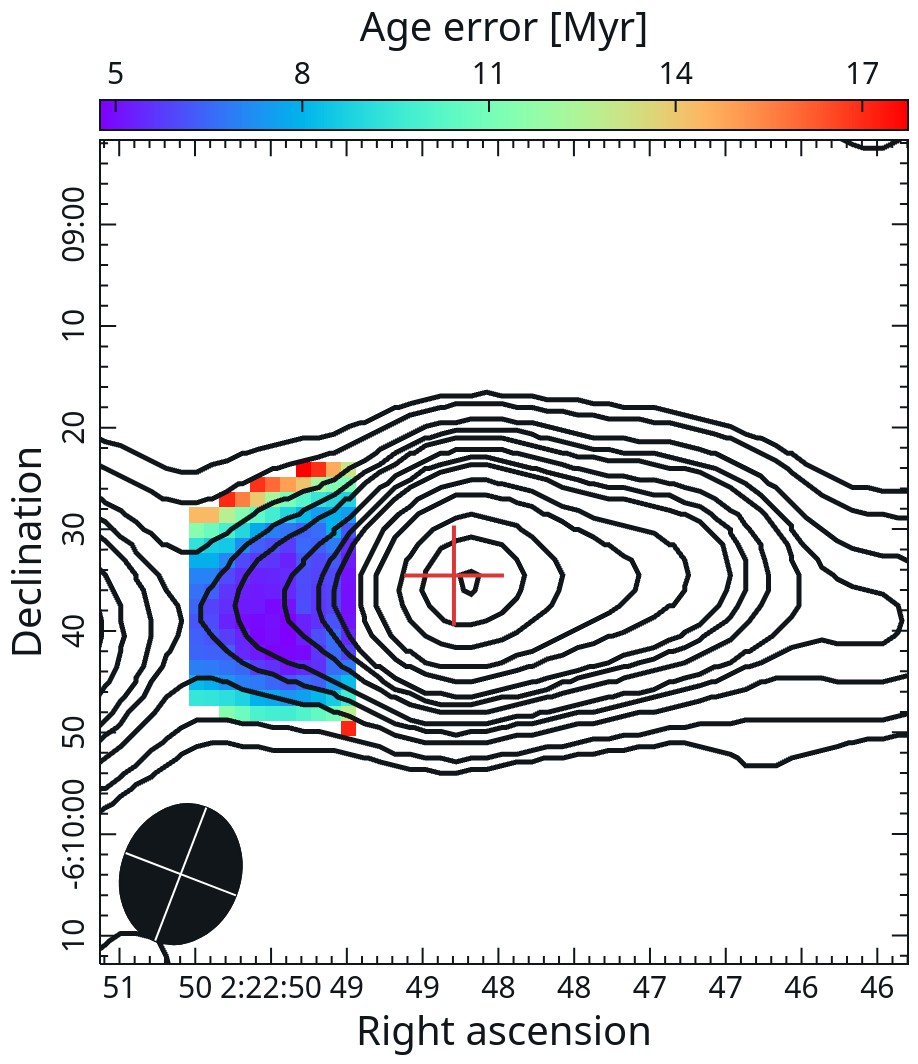}}}&
  {\resizebox{0.3\hsize}{!}{\includegraphics[trim={0.4cm 0 0.25cm 0},clip]{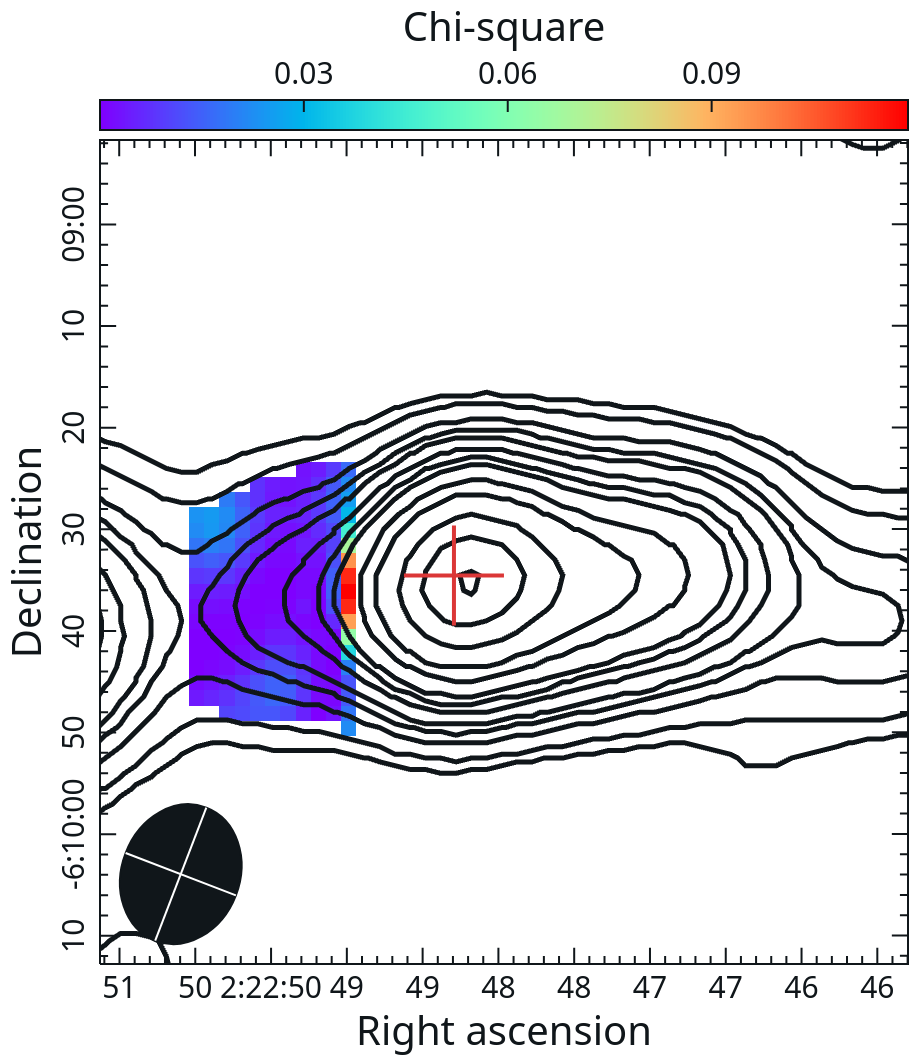}}}\\
   (g)&(h)&(i)\\ 

  \end{tabular}
\caption{Resulting maps of the spectral age fitting using JP model. The outcomes of the fitting for the eastern to the western outer lobes are displayed from top to bottom. The spectral age distributions are shown in the first column, followed by the error maps and the $\chi^2$. The lowest (highest) values are indicated by the blue (red) colours. For the eastern inner lobe and the western middle lobe, the fits were done using the MIGHTEE sub-band data only due to low detection in LOFAR and GMRT. }
		
		\label{fig: agemap}
        \addtocounter{figure}{-1}
	\end{figure*}

	\begin{figure*}
		\centering
		\begin{tabular}{ccc}	
  {\resizebox{0.3\hsize}{!}{\includegraphics[trim={0.4cm 0 0.25cm 0},clip]{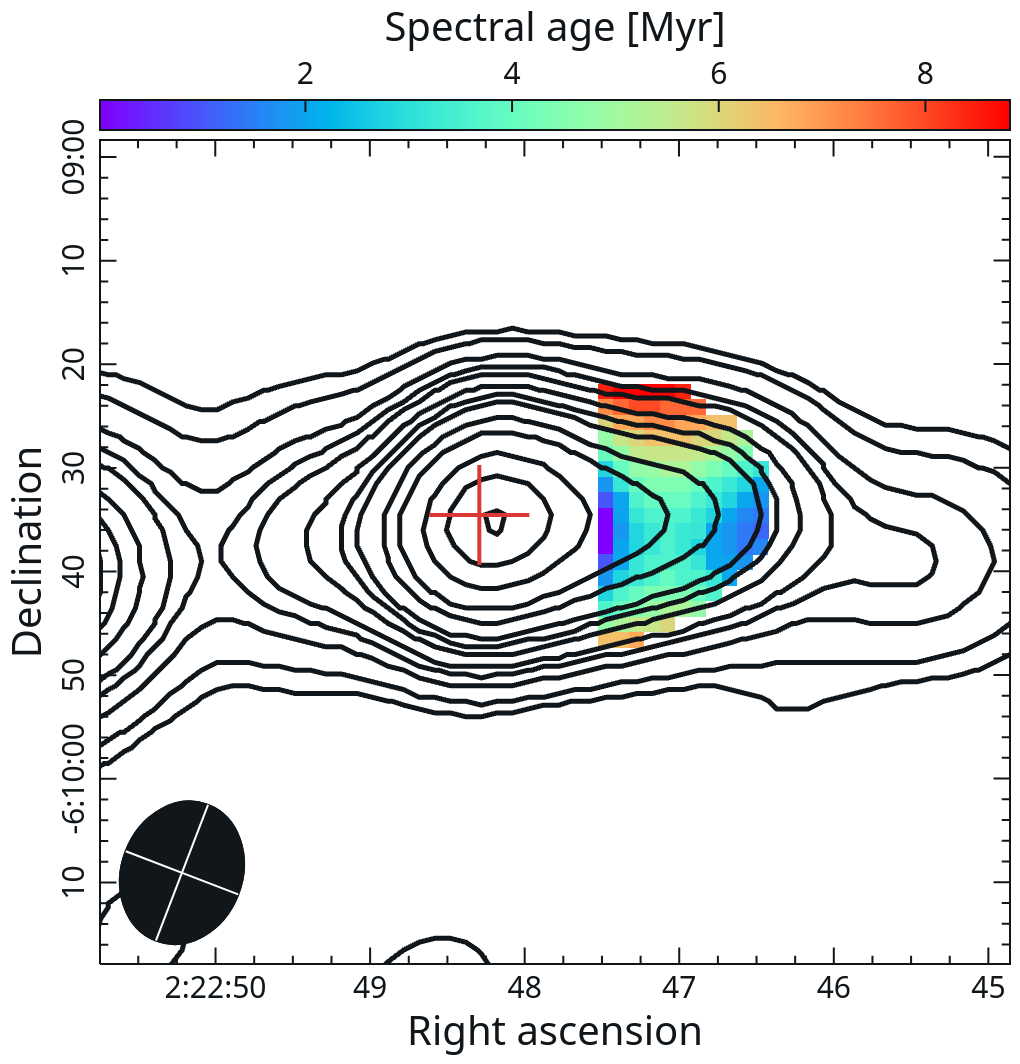}}}&
  {\resizebox{0.3\hsize}{!}{\includegraphics[trim={0.4cm 0 0.25cm 0},clip]{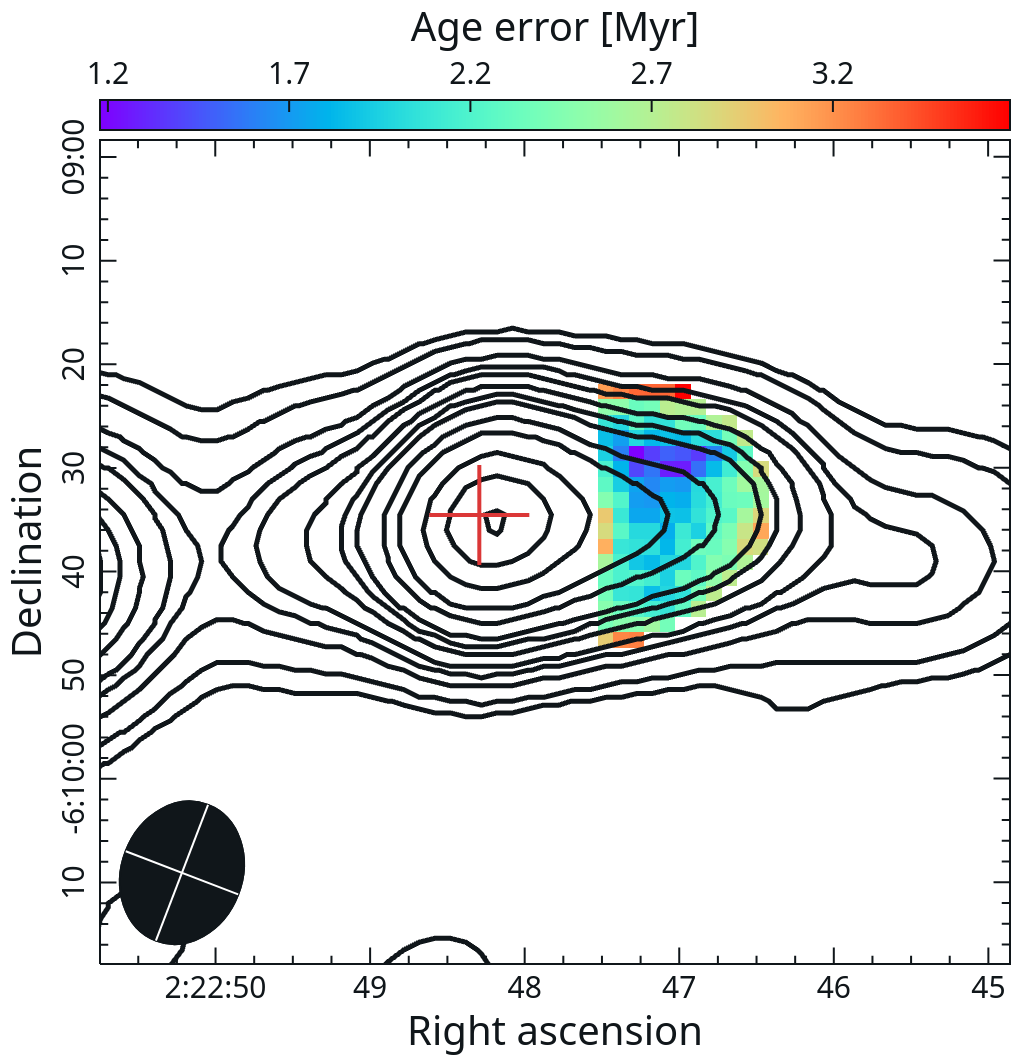}}}&
  {\resizebox{0.3\hsize}{!}{\includegraphics[trim={0.4cm 0 0.25cm 0},clip]{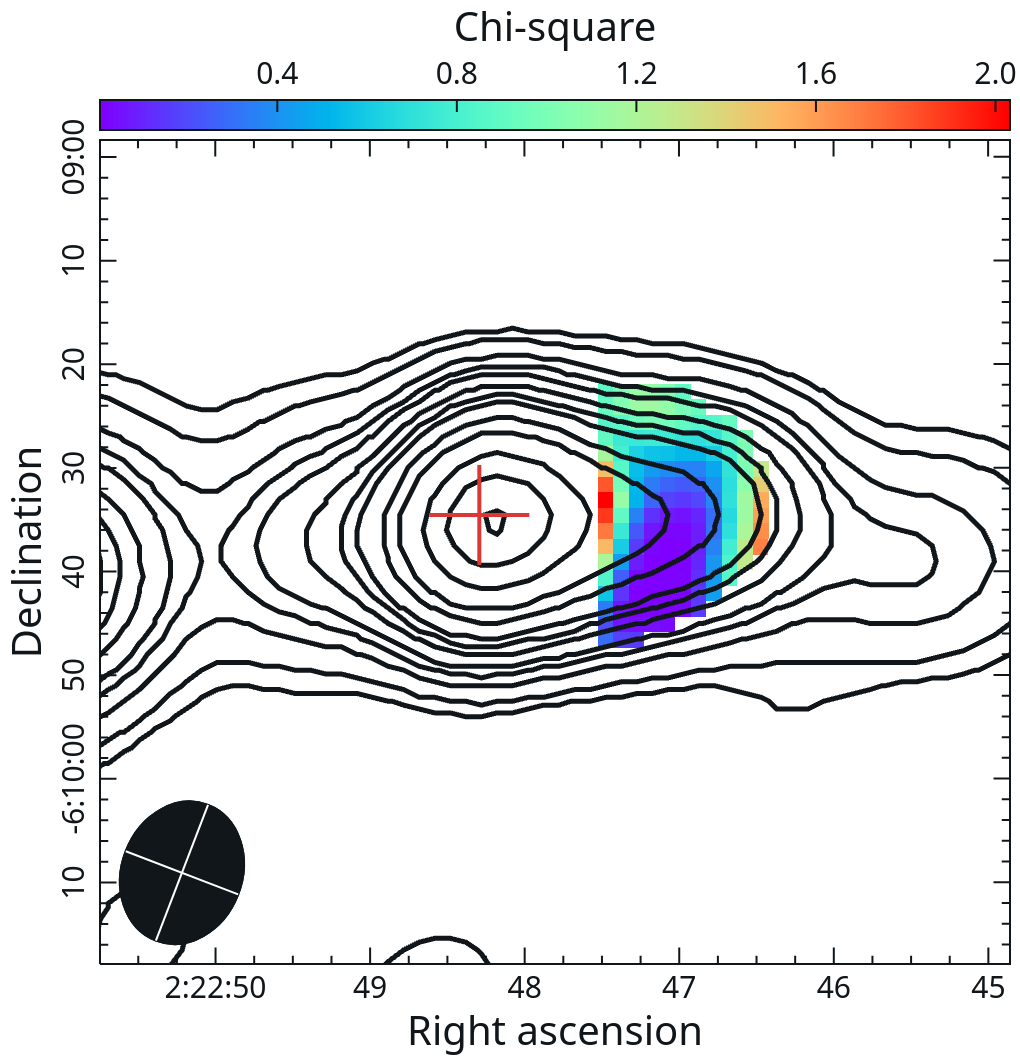}}}\\
  (j)&(k)&(l)\\
  &&\\

  {\resizebox{0.3\hsize}{!}{\includegraphics[trim={0.4cm 0 0.25cm 0},clip]{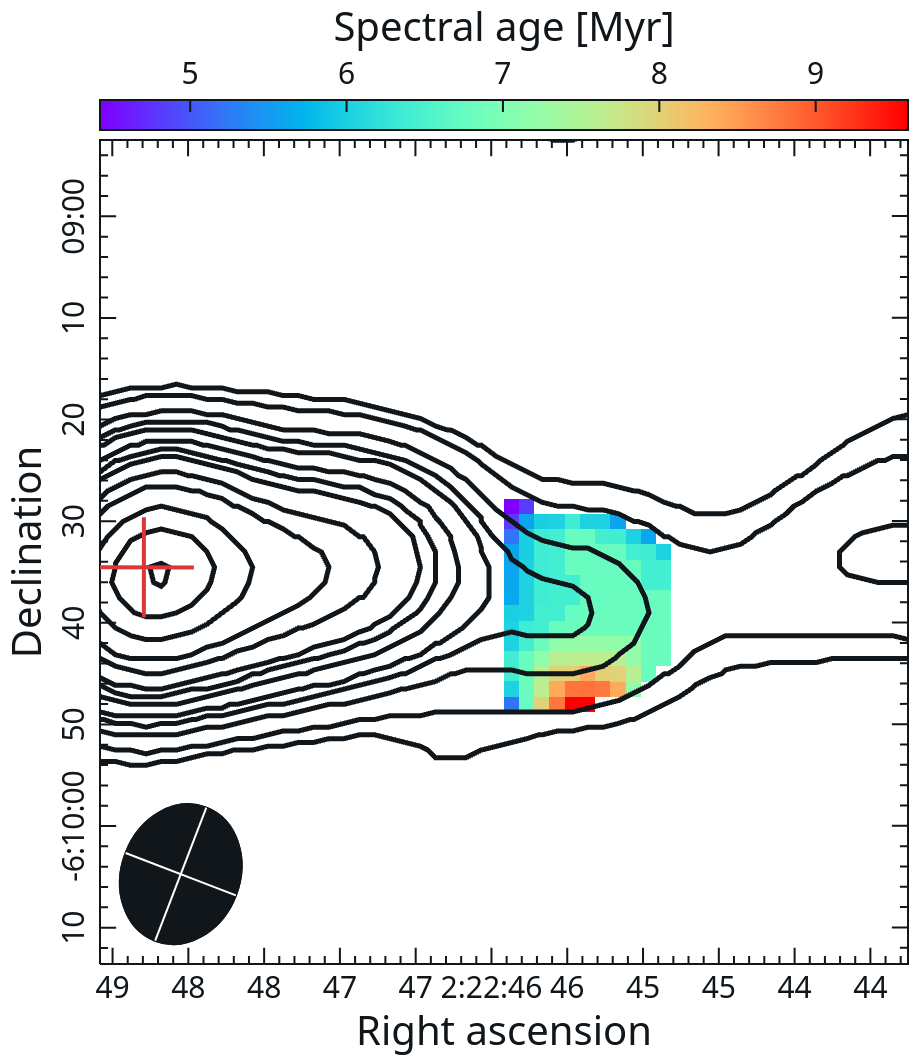}}}&
  {\resizebox{0.3\hsize}{!}{\includegraphics[trim={0.4cm 0 0.25cm 0},clip]{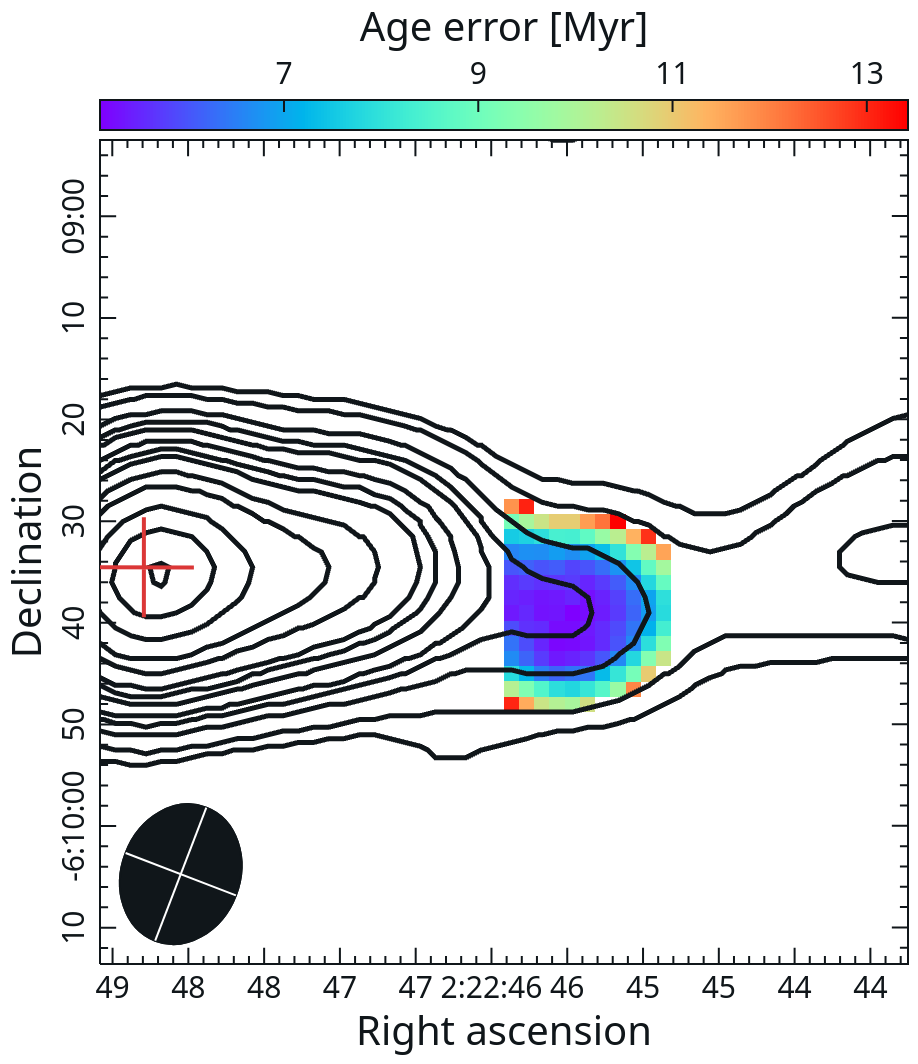}}}&
  {\resizebox{0.3\hsize}{!}{\includegraphics[trim={0.4cm 0 0.25cm 0},clip]{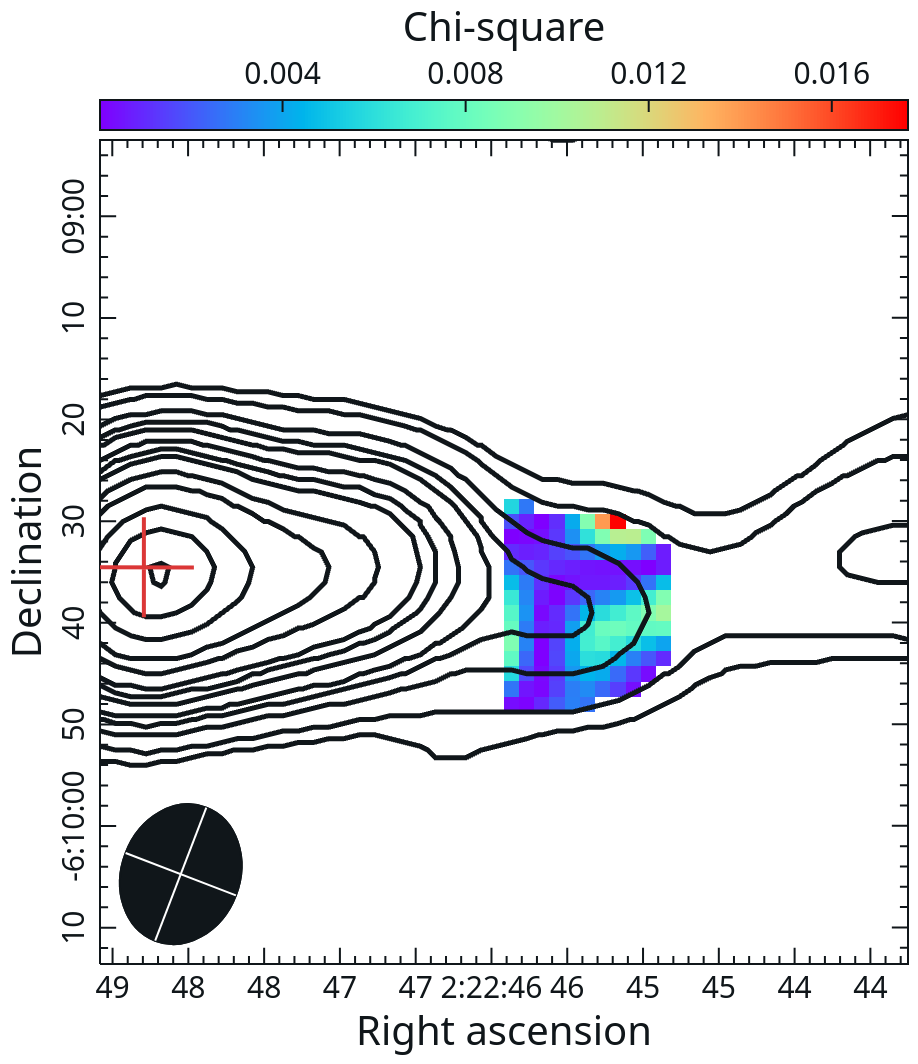}}}\\
  (m)&(n)&(o)\\
  &&\\
  {\resizebox{0.3\hsize}{!}{\includegraphics[trim={0.4cm 0 0.25cm 0},clip]{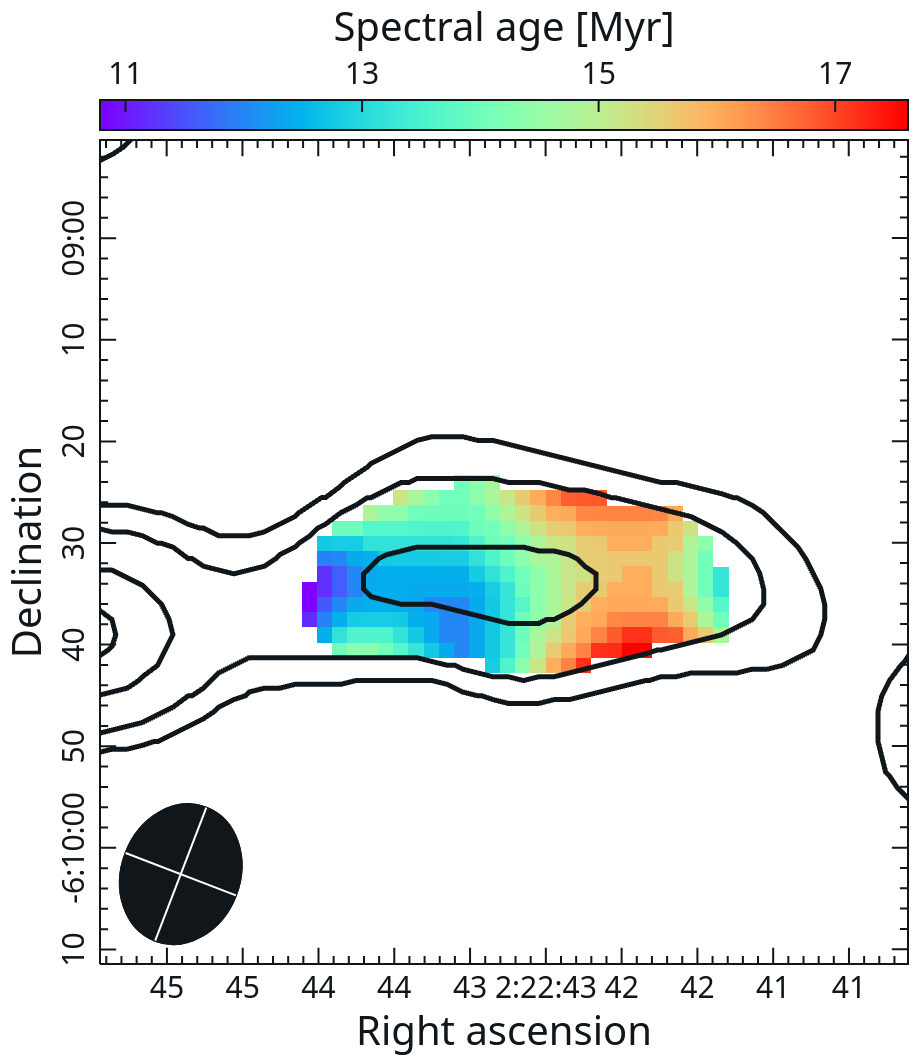}}}&
  {\resizebox{0.3\hsize}{!}{\includegraphics[trim={0.4cm 0 0.25cm 0},clip]{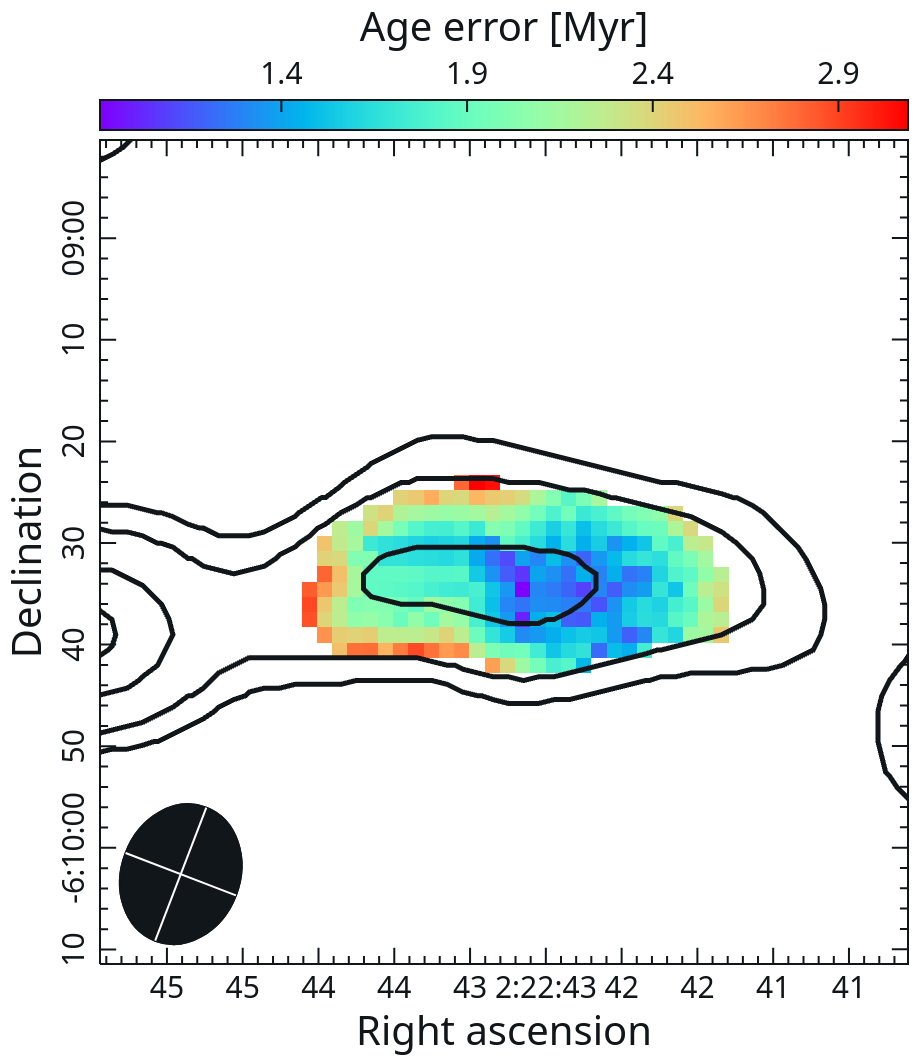}}}&
  {\resizebox{0.3\hsize}{!}{\includegraphics[trim={0.4cm 0 0.25cm 0},clip]{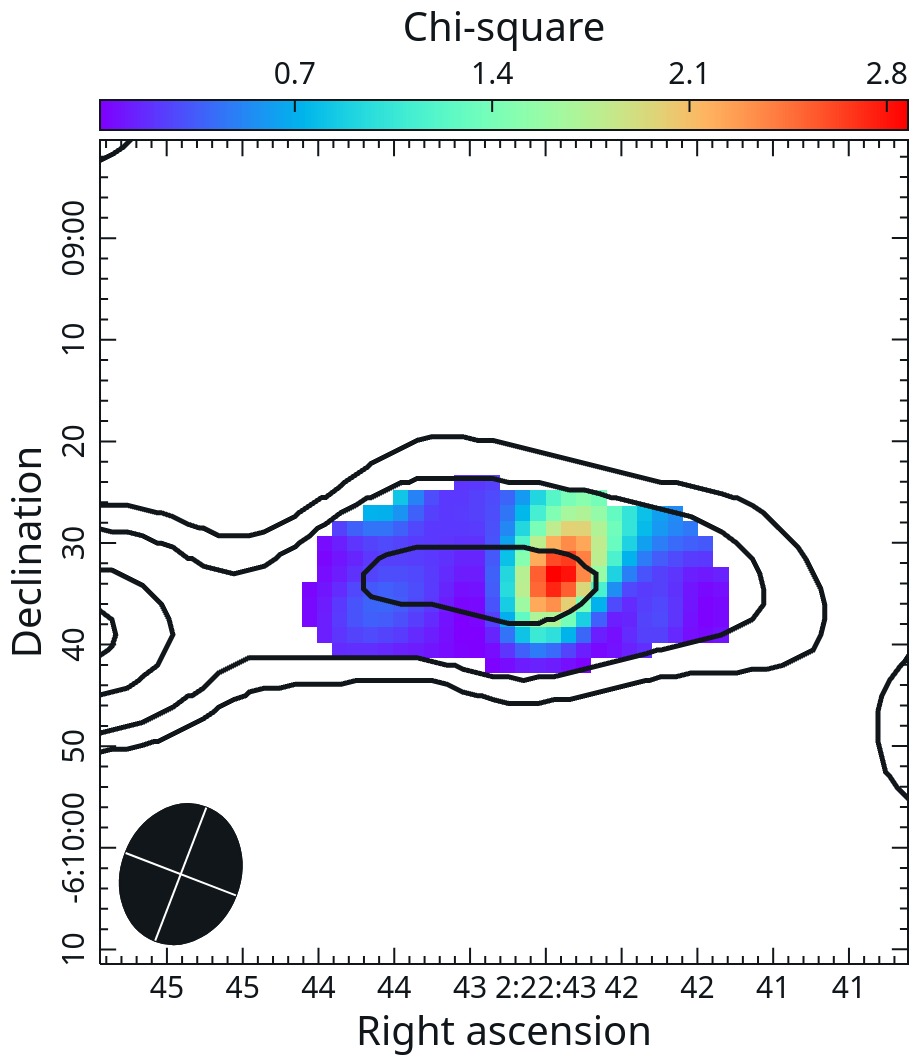}}}\\
  (p)&(q)&(r)\\

   \end{tabular}
\caption{Cont.} 
	\end{figure*}

\indent The level of confidence in the single injection model fittings is relatively low. This might be due to the limitation in resolution and sensitivity at lower frequencies of our data set. We also recognise the limited number of data sets available, as the more wavelengths are covered during the fitting, the more accurate the results will be. 

\subsection{Polarimetry}\label{sec: polarisation_results}
\begin{figure*}
		\centering
		\begin{tabular}{c}		
  {\resizebox{1\hsize}{!}{\includegraphics[trim={0.5cm 0 0.3cm 0},clip]{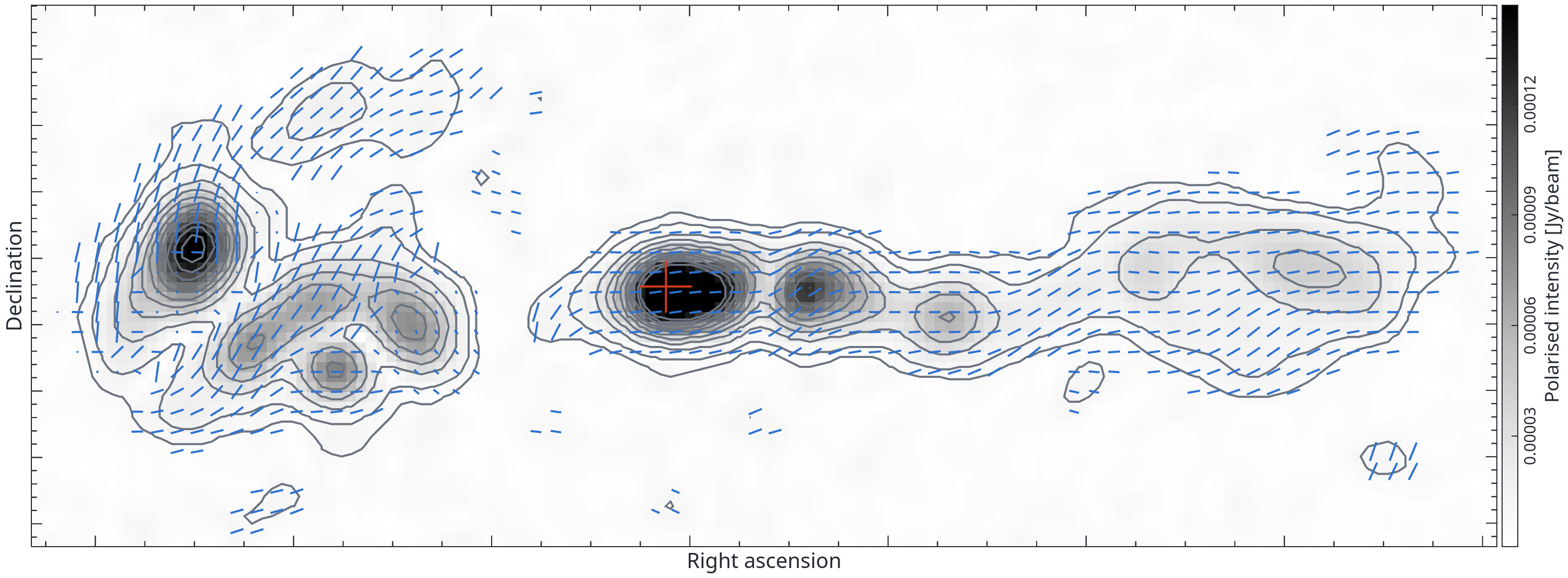}}}\\
  {\resizebox{1\hsize}{!}{\includegraphics[trim={0.5cm 0 0.3cm 0},clip]{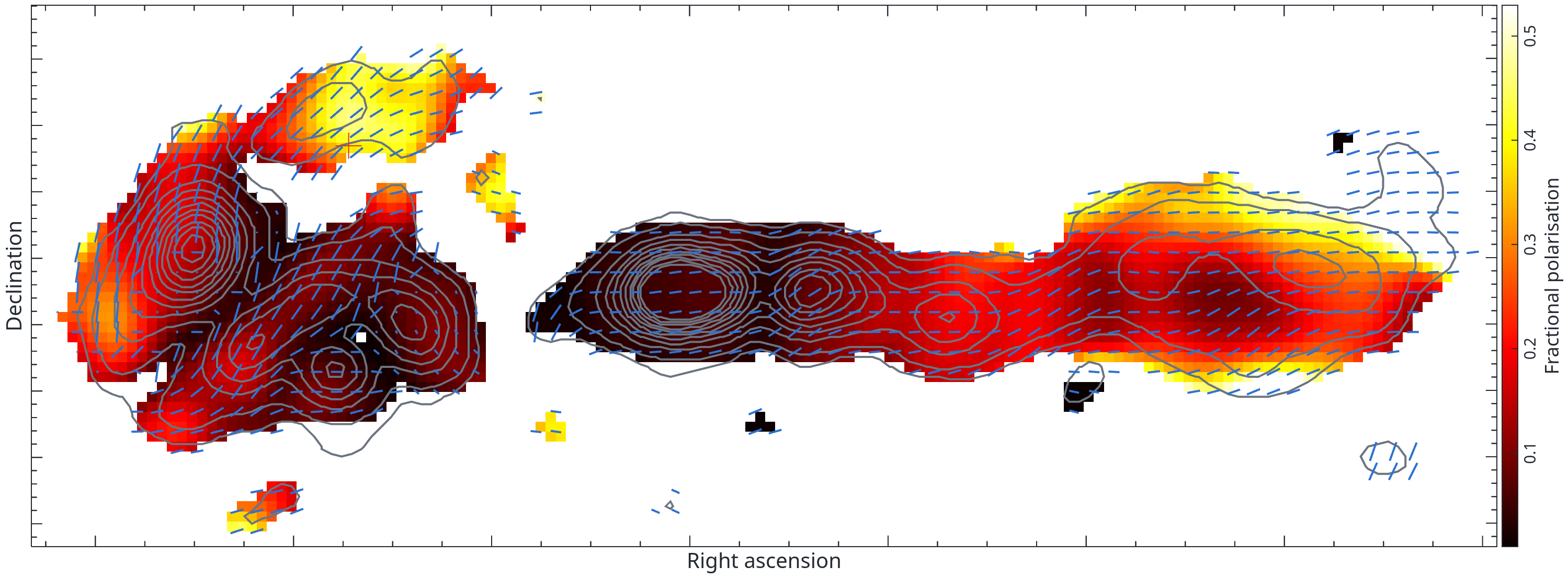}}}\\
 
    \end{tabular}
\caption{ \textit{Top panel:} polarised intensity map and contours overlaid with B vector field. \textit{Bottom panel:} fractional polarisation map overlaid with polarised intensity contours and 
B vector field.} 
		
		\label{fig: PI_pol}
	\end{figure*}

\indent Figure \ref{fig: PI_pol} shows the polarised intensity (top panel) and the fractional polarisation maps (bottom panel). The vectors represent the polarisation angle of the magnetic field. 

At the core, the resolution is good enough as we can see that its magnetic field is oriented along the jet axis as expected. The eastern lobes seem to be more strongly polarised but have a more mixed oriented field. This gives a lower fractional polarisation of 3 to 20 per cent for the outer and middle lobes (dominated by darker colour in Figure \ref{fig: PI_pol} bottom panel). On the opposite side, the lobes seem to be fainter, but with a more ordered field. The fractional polarisation for the western lobes is therefore higher, from 15 to 40 per cent (dominated by brighter colour in Figure \ref{fig: PI_pol} bottom panel).   

It is also interesting that the magnetic field structure at the edges of both outer and middle eastern lobes is orthogonal to the source axis, while their counterparts have a B field parallel to the jet. The overall properties of the magnetic field on each side of the source therefore point towards a non-uniform host environment. The eastern lobes might reside in a denser, clumpier and/or more turbulent medium, unlike the western ones (e.g. \citealt{2007MNRAS.382.1019B}).\\
The rotation measure (RM) map of the source shows uniform values of approximately 4 rad m$^{-2}$ across the source. Since such a value is mainly associated with Galactic foreground contribution \citep{2000MNRAS.315..395S, 2009ApJ...702.1230T}, we did not include the map in the paper.

\section{Discussion}\label{sec: disc}

\indent As already put forward in previous discoveries \citep{2007MNRAS.382.1019B, 2011MNRAS.417L..36H, 2016ApJ...826..132S, 2023MNRAS.525L..87C, 2025A&A...696A..97D, 2025PASA...42..124N}, the intermittent AGN episodes explain the morphology of TDRGs. In addition to the approaches in these works, we will also consider the spectral evidences from our analysis, which has not been done before. We will discuss as well the consistency between the striking morphology of J022248\m060934 in total intensity and its polarimetric properties.

\subsection{Morphological and polarisation properties}\label{subsec: disc_morpho_pol}
The source is not located in an overdense intergalactic medium (see Section \ref{sec: host_env}) but the asymmetry in the observed morphology in MIGHTEE-DR1 (see Figure \ref{fig: optical_host} and Figure \ref{fig: source_morpho_appendix}) suggests a non-uniform host environment. Considering the compact morphology coupled with the bending of the eastern lobes, we assert that its surroundings are more dense than the western side. Such higher density has produced higher shocks which created the prominent hotspots and the bending we observe. On the opposite side, the lobes are more elongated and more relaxed. These hint towards an underdense medium, and consequently a weaker shock that justifies the lack of a distinguishable hotspot or any compact features in the western side.  Furthermore, the advancement speed of the jet in this side should have been higher, which is consistent with the asymmetry in the computed arm-length ratio (i.e the western lobes are more elongated and further away from the core). However, we cannot rule out the possibility of any projection effects \citep{1982ApJ...257..538B, 2020MNRAS.491..803H}. The presence of unobservable bending along the line of sight can explain the differences in the morphologies and length of the lobes.\\

In their analysis of giant DDRGs, \citet{2025A&A...696A..97D} have shown that the asymmetry in arm-length ratio in the inner lobes implies a contamination of the cocoon paved by the outer lobes. On the contrary, symmetric inner lobes are presumed to be from a more homogeneous cocoon. Equally, after modeling B0925+420,  the first discovered TDRG, \citet{2007MNRAS.382.1019B} concluded that the outer and inner lobes are within the same cocoon density but the middle lobes must have grown in a higher density. They thus proposed a gas refill from the dispersion of warm gas clouds.

Disregarding projection effects, in our source, the initial medium in which the first pair of lobes grew was not uniform, as inferred by the asymmetry in the outer lobes. For the middle lobes, the scenario is either that the first activity did not clear up the surroundings, giving a more homogeneous environment or the eastern cocoon has been contaminated. Finally, the inner lobes seem to be in a different condition than their two previous pairs. Though we observe an asymmetry, it is in the opposite sense. The peculiar features in the western inner lobe, indicated by the double arrow in Figure \ref{fig: optical_host}, suggest an asymmetric environment on opposite sides. One of the inner lobes might propagate in an overdense medium. Even if a cluster environment does not determine or foster the occurrence of restarting activity \citep{2019A&A...622A..13M}, we can affirm that the structure of the recurrent jet is influenced by its immediate vicinity.

Comparably, the polarisation properties of any jet of   J022248\m060934 exhibit an asymmetry between the two sides of the source. The alignment of the ordered B field with the jet stream on the western side, compared to the mixed B field direction in the eastern lobes, is consistent with the elongated and relaxed versus compact with pronounced bending morphology observed in total intensity. The B field quasi-perpendicular to the jet axis at the edges of the eastern lobe is an indication of a compression of the field by the lobes or/and an expansion of the lobes (e.g. \citealt{1991ApJ...381...63F, 2007MNRAS.382.1019B}). We also observe a change in angle of the vectors corresponding to the dark patches in the fractional polarisation map (see Figure \ref{fig: PI_pol}). This coincidence suggests that the observed low fractional polarisation in these regions is due to beam depolarisation. As this change in angle is seen in both outer-middle lobe and middle-inner lobe connections, it does not confirm that the density in the inner and/or middle lobes is lower or higher than the density in the outer lobes. However, it suggests a non-uniform density of the medium within one lobe itself. The edges interacting with a younger hotspot (outer lobe case) or a rejuvenated jet (middle lobe case) might be in a more dense/clumpier environment.  

\subsection{Spectral features}\label{subsec: disc_spectral_features}

The variation of the spectral index from low to high frequencies across J022248\m060934 is captured in the maps in Figure \ref{fig: alphamap}. The maps computed at 150 to 325 MHz (Figure \ref{fig: alphamap} middle panel) and 325 to 1540 MHz (Figure \ref{fig: alphamap} bottom panel) highlight flatter values at low frequency than those at high frequency over the entire source. Such spectral behaviour is expected as it signifies that the electrons are ageing. Spectral evolution models state that the spectrum ($S_\nu\propto\nu^{\alpha}$) of an ageing radio source is a modified power law. As the source ages, the slope of their spectrum steepens towards lower frequencies \citep{1973A&A....26..423J}.  \\
    \indent In the $\alpha^{1540}_{150}$ map ( Figure \ref{fig: alphamap} top panel), the outer lobes have values as steep as $\alpha^{1540}_{150} = - 2$. Such ultra-steep values are typical of remnant RGs (\citealt{1970ranp.book.....P, 2011A&A...526A.148M, 2017A&A...606A..98B, 2020MNRAS.496.3381R, 2023ApJ...944..176D}). It is the spectral sign of the shortage of fresh electron replenishment. In the outer lobes, this ultra-steep emission surrounds flatter ($\alpha^{1540}_{150} \geq - 1.1$) emission which indicates the presence of remnant hotspots. Towards the core, the spectrum flattens, the values in the middle lobes vary from $-1.1$ to $-0.7$ in the $\alpha^{1540}_{325}$ map. They imprint a decrease in the age of the emission. 
    For the core, the spectral index jumps from a maximum value of $\alpha_{150}^{325} = 0.2 $ to $\alpha^{1540}_{325} = -0.4$ (see also Figure \ref{fig: core_radio_SED}). This inversion of spectrum is consistent with the peaked-spectrum radio sources. Based on the peak frequency of the spectrum of the core well below 500 MHz in Figure \ref{fig: core_radio_SED}, the core of the TDRG might be a compact steep spectrum source (CSS, \citealt{2021A&ARv..29....3O}). However, a firm distinction between the SSA and FFA as the absorption mechanism at play could not be drawn from our fitting process. Simultaneous observations at the different frequencies are necessary to correctly probe the spectral shape and to infer a robust conclusion.

As similar to all peaked radio sources, CSS sources have also been considered as young unresolved sources \citep{1998PASP..110..493O, 2018MNRAS.477..578C}. Therefore, the detection of such young sources and a diffuse extended radio emission was interpreted to be a possible signature of intermittent nuclear activity. B2 0258+35 is an example of a classified CSS \citep{1995A&A...295..629S, 2005A&A...441...89G} where a surrounding faint diffuse extended emission was discovered which might represent an indication of a previous interrupted active phase \citep{2012A&A...545A..91S, 2018A&A...618A..45B}. 

In the spectral curvature map (see Figure \ref{fig: spc_map}), the core has a mean spectral curvature of $\sim0.44$ (maximum value of $\sim0.66$), although it is not really steep at higher frequencies ($\leq - 0.5$, see Figure \ref{fig: alphamap}). It is in fact, the result of its inverted spectrum and a sign of absorption. On the other hand, the curvature seen in the lobes is consistent with their spectral indices. On the east side, the middle and outer hotspots show lower curvature of 0.25 and 0.5, respectively. In comparison, the emissions at the edges of the outer lobes and in between the outer and middle lobes are extremely curved with a value of $\gtrsim0.7$. On the other side, the western outer lobe closer to the core has a value around $\sim$0.25. Then the rest of the outer lobe is dominated by an ultra-curved spectrum of a curvature of $\sim1$. These spectral shapes with curvature $\gg$0.5 are evidence of old plasma devoid of injection of young materials (e.g. \citealt{2011A&A...526A.148M, 2016A&A...585A..29B}).

In view of these spectral properties, J022248\m060934 is thus most likely hosting a recurrent nuclear activity. Its behaviour is inconsistent with either a typical FRII or a remnant RG. FRII sources are characterised by steeper plasma towards the core due to the backflow of radio plasma re-accelerated by the hotspots. For remnant RGs, they are commonly characterised by old emission without any signs of particle injection or ongoing nuclear activity, which is however the case of the TDRG central region.

\subsection{Duty cycle}\label{subsec: disc_duty_cycle}
Since this work is the first spectral ageing study of TDRGs, we will follow \citet{2017A&A...600A..65S}, \citet{2020A&A...638A..29B} and \citet{2021A&A...650A.170B} in our interpretation of the spectral age maps to reconstruct the duty cycle of the source, though the derived results are only lower limits. The maximum age in the outer lobes will be taken as the total age ($t_{tot}$) of the source. The duration of an active phase ($t_{on}$) can be derived from the difference between the oldest and the youngest particles within the same lobe or pair of lobes. 

For the outer lobes (see Figure \ref{fig: agemap}a and Figure \ref{fig: agemap}p), the oldest region is at the edges of the radio lobes, while the youngest particles are in the hotspots, where the last acceleration took place. From the measurements in the western outer lobe (see Table \ref{tab: single_injection}), the maximum age is about $\sim$17 Myr. However, this value is only from the very few red pixels (see Figure \ref{fig: agemap}p). We will therefore consider the orange emissions of about $\sim$16 Myr. The first active phase then lasted $\sim$10 Myr ($t_{on, 1}$) based on the difference between the source total age of 16 Myr ($t_{tot}$) and the age of the eastern outer hotspot of 6 Myr.

For the second period of activity, we consider the eastern middle lobe which was fitted with more datasets compared to the western middle lobe (see Figure \ref{fig: agemap}d). We estimate the maximum age of $\sim$12 Myr on the eastern side of the lobe. We excluded the northern-east and southern-east edges of this lobe from the fitting as they are likely emissions from the outer lobe. The youngest particles of $\sim$ 5 Myr are then measured from the hotspot. These give the value of $\sim$7 Myr as the duration of the second active phase ($t_{on, 2}$). To obtain a rough estimation of the ongoing activity, we used the maximum age measured in the western inner lobe, which is $\sim$8 Myr ($t_{on, 3}$). It is worth noting that the maximum age measured in both outer and middle lobes are likely lower limits. The oldest emission in FRIIs are in the part of their lobes that is closer to the core due to backflow of radio plasma. However, in our case, the multiple nuclear activities have erased such information. 

For the quiescent period ($t_{off}$), we will consider the difference between the youngest age from the previous (or remnant) outer lobes and the oldest age from the following (or younger) inner lobes. Nevertheless, in our case, the maximum age within the younger lobe is either higher or the same as the youngest age in the previous older lobe. Among the possible reasons for this is the mixing of particles from two successive lobes. The restarting jet might go through the old lobes and lies exactly where we would expect the old plasma from the previous activity (e.g., 3C388 by \citealt{2020A&A...638A..29B}). Another alternative is that the inactive period is very brief. So the outer or previous jet is still travelling when the following phase restarts. The last particle acceleration in the outer hotspot then occurs after the second jet has switched on. Therefore, we will rather compute the upper limit of the quiescent period using the difference between the youngest age from the previous outer lobes and the youngest age from the following inner lobes. We found that the first active phase was followed by an inactive period of $\sim$1 Myr (upper limit of $t_{off, 1}$).
 
The first inactive phase we derived is consistent with the general off-time of a few
Myr to a few tens of Myr observed in other restarted RGs (e.g.   \citealt{2000MNRAS.315..371S, 2006MNRAS.366.1391S, 2008MNRAS.385.2117S,  2012MNRAS.424.1061K, 2015A&A...579A..27S, 2018A&A...618A..45B,2020A&A...638A..29B, 2021A&A...650A.170B}) and the typical timescale derived from two AGN outbursts creating two pairs of X-ray cavities \citep{2014MNRAS.442.3192V}. Furthermore, such a value at the lower end of the range has also been found in other RGs like Hercules A (see e.g., \citealp{2003MNRAS.342..399G}).  

Using the derived jet phases above, we estimated a first-order of the duty cycle defined as $t_{on}/(t_{on}+t_{off})$. We found a very rapid duty cycle of 90 per cent for the first active phase. Such a quick duty cycle was also found in other sources (e.g. \citealt{2008MNRAS.385.2117S, 2020A&A...638A..29B, 2021A&A...650A.170B, 2025A&A...696A.239B}). This implies that the nuclear activity of the source is only interrupted for a short period. Nevertheless, we remind the reader that these derived ages of the jets at different stages are only lower limits. They depend on the reliability of the maximum and minimum spectral ages subtracted from the maps and the possible presence of backflow of radio plasma and the mixing of particles from different phases make such age subtraction challenging. Furthermore, the estimation of the magnetic field is another source of uncertainty as well. For instance, if we consider the minimum magnetic field B$_{min}$ for which the particles have the maximum losses at a given redshift instead of B$_{eq}$, the total age of the source would be $<$ 38 Myr. 

\subsection{Comparison with other TDRGs}\label{subsec: disc_TDRGs_comparision}
The rest of this section will briefly compare the TDRG we report in this work with those from literature \citep{2007MNRAS.382.1019B, 2011MNRAS.417L..36H, 2016ApJ...826..132S, 2023MNRAS.525L..87C, 2025A&A...696A..97D}. Note that this comparison excludes the source reported in \citet{2025PASA...42..124N} as no further details about this source were given. The main properties of the sources are summarised in Table \ref{tab: TDRG_comparison}.

      \begin{table*}
  
 \caption{ Summary of the properties of all TDRGs available in literature. Columns: (1): list of the references reporting a TDRG ((a): \citet{2007MNRAS.382.1019B}, (b): \citet{2011MNRAS.417L..36H}, (c): \citet{2016ApJ...826..132S}, (d): \citet{2023MNRAS.525L..87C}, (e): \citet{2025A&A...696A..97D} and (f) this work), (2): redshift, (3 - 5): projected linear sizes of the outer, middle and inner lobes, respectively, (6 - 8): respective arm-length ratios of the outer, middle and inner lobes, (9 - 11): radio powers of the outer, middle and inner lobes, respectively, where $^{\dagger}$ indicates radio powers at 610 MHz from \citet{2016ApJ...826..132S} and $^{\ddagger}$ values computed at 1.2 GHz from MIGHTEE-hi while the rest are values at 144 MHz and were taken from \citet{2023MNRAS.525L..87C}, (12): host environment of the TDRGs and (13 - 15): respective kinematic ages of the outer, middle and inner lobes.}
 
 \label{tab: TDRG_comparison}
		\centering
        {\resizebox{1\hsize}{!}{
		\begin{tabular}{ccccccccccccccc}
    \hline
     Ref & $z$ & & Size [kpc] & & & R$_{l}$ & & & P$\times$10$^{24}$\,[W Hz$^{-1}$] & & Host & & Kinematic Age [Myr] &\\
     \cmidrule(lr){3-5}\cmidrule(lr){6-8}\cmidrule(lr){9-11}\cmidrule(lr){13-15}      
      &  & (I) & (II) & (III) &  (I) & (II) & (III)  & (I) & (II) & (III) & environment & (I) & (II) & (III)\\
    (1)   & (2) & (3) & (4) & (5) & (6) & (7)  & (8) & (9) & (10) & (11) & (12) & (13) & (14) & (15)\\
    \hline 
    \hline
     (a) &0.36& 2541&732&37&0.63&0.66&0.81&335&168.6&11.2&$-$&415&24&0.60\\
     (b) &0.14& 1433&322&55&0.46&0.98&1.14&51&9.4&1.1&BCG&233&11&1\\
     (c) &0.14& 814&235&95&1.0&1.6&2.0&4.8$^{\dagger}$&0.54$^{\dagger}$&0.35$^{\dagger}$&Group&130&7.6&1.5\\
     (d) & 0.28& 1349&572&118&0.87&0.93&1.44&18.4&11.7&1.6&BCG&220&19&2\\
     (e) & 0.11& 1100&$-$&$-$&$-$&$-$&$-$&$-$&$-$&$-$&BCG&$-$&$-$&$-$\\
     (f) & 0.94 & 1587 & 866 & 461 & 0.48 & 1.44 & 0.73 & 4.6$^{\ddagger}$ & 6.1 $^{\ddagger}$& 10.0 $^{\ddagger}$ & field & 258 &28 & 8\\
	\hline
       
		\end{tabular}
        }}
	\end{table*}

The six known TDRGs are all GRG (size from 0.8 to 2.5 Mpc). This might be due to a selection effect biased towards larger sources, where it is easier to resolve the triple structures of TDRGs, which may explain their rarity as well. Most of the six TDRGs also have features closer to an FRII or intermediate FRII/FRI and they all present a certain form of asymmetry in their morphology. The asymmetries in flux and arm-length ratios of J022248\m060934 middle and inner lobes and in two more TDRGs are in opposite sense \citep{2016ApJ...826..132S, 2023MNRAS.525L..87C}, i.e. the lobes located farther away from the core are also the brightest which supports the possibility of intrinsically asymmetric jets.

In contrast to J022248\m060934 hosted by a field galaxy, most of the TDRGs in literature reside in dense environments. There are three BCGs \citep{2011MNRAS.417L..36H, 2023MNRAS.525L..87C, 2025A&A...696A..97D} and another one in a small group of three galaxies \citep{2016ApJ...826..132S}.  J022248\m060934 also shows an intriguing behaviour in term of radio power as its radio power from the pair of outer, middle and inner lobes increases. Note however that there is no indication from models that the power of jets produced from different outbursts should be the same. Nonetheless, the opposite, decreasing radio power was found in four out of 6 TDRGs (see \citealt{2016ApJ...826..132S, 2023MNRAS.525L..87C}). Given that J022248\m060934 is hosted by a field galaxy, there may be a rapid fading of the emissions from the outer and middle lobes because of the rapid plasma expansion in a low-density environment. 

Regarding the signs of multiple nuclear activity, two of the TDRGs in literature share similar spectral index distribution as J022248\m060934 \citep{2007MNRAS.382.1019B, 2016ApJ...826..132S}. They are characterised by steep/ultra-steep outer lobes coupled with a flattening of the indices towards the core. We also found that by assuming the same advancement speeds, their kinematic ages are within the same range if the sources have similar size \citep{2011MNRAS.417L..36H, 2016ApJ...826..132S, 2023MNRAS.525L..87C}. 

The time between two episodes of activity computed from these kinematic ages also converge in a long  quiescent period, ranging from 120 to 390 Myr and 6 to 23 Myr for the first and second off periods, respectively. However, our results based on spectral age fitting suggest the opposite, short quiescent period and rapid duty cycle. It is worth noting that these latter results are more consistent with the observed spectral indices in J022248\m060934 and two other TDRGs \citep{2007MNRAS.382.1019B, 2016ApJ...826..132S}. These three sources exhibit a nearly gradual steepening spectral index rather than a very sharp spectral discontinuity.

Nevertheless, one should be cautious that the kinematic ages computed in this work as well as in \citet{2016ApJ...826..132S} and \citet{2023MNRAS.525L..87C} are upper limits only. The morphological asymmetries observed in the sources and the probable variation of the advancement speeds were not taken into consideration. Likewise, the spectral ages presented in this work should be considered as lower limit as well. They have their limitations as mentioned earlier and in addition, we might also miss older particles emitting at lower frequencies where we suffer the most from sensitivity limitation.

\section{Summary}\label{sec: summary}
\indent In this paper, we report the discovery of J022248\m060934, the seventh known TDRG, which is a RG characterised by triple pairs of radio lobes. To perform an extensive analysis of the radio source, we used MIGHTEE-DR1 centred at 1.28 GHz, sub-band images of MIGHTEE and additional archival radio data from LOFAR, GMRT and VLASS. The following is a summary of our findings.

\begin{enumerate}[(i)]
    \item The triple-double structure of J022248\m060934 is only well resolved in the 5 arcsec MIGHTEE image. It shows a bright compact core and clear triple peaks of radio emission in both sides (>10$\sigma_{local}$), with evidence of edge-brightened structure except for the western outer lobe. The eastern lobes also show pronounced bending compared to their western counterparts. The total end-to-end projected linear size of the outermost lobes is 1587 kpc (196 arcsec) which classifies the TDRG as a GRG.
   
   \item The orientation of the magnetic field at the core follows the source axis as expected. 
   On the western side, the B field follows the same alignment and it is more ordered. While on the eastern side, there is a certain mixing of B field direction. The variation in the angle, the depolarisation and the orientation of vectors at the edges of the outer lobes suggest the presence of turbulence or shock compression of the magnetic field in these lobes.

  \item The general trend observed in the spectral index and curvature maps from 150 to 1540 MHz is that the outermost lobes are found to be steeper and gradually flatten toward the inner lobes. We can affirm that the outer pair of lobes is the oldest structure of the radio galaxy, while the core has the youngest particles, as supported by its inverted spectrum. 
  
   \item Using the JP model, we performed individual spectral age fitting of the components of the source from 150 to  1540 MHz, depending on their detection at each frequency. The model cannot be rejected at 68 per cent confidence level for all of our fits. We also found that the lower limit of the first outburst is about $\sim$16 Myr. 
   
 \item  Using the spectral age results, we found that the first active phase lasted for $\sim$10 Myr and was followed by $\sim$1 Myr of inactive phase. We also give rough estimations of $\sim$7 Myr and $\sim$8 Myr as the duration of the second period of activity and the ongoing third phase, respectively.  
 
\item We derived a quick duty cycle of about 90 per cent for the first cycle of activity. Therefore, our findings suggest that the triple-double structure in
TDRGs is not the product of a long quiescent period as proposed in previous works based on kinematic age. However, all the spectral age results still need to be interpreted as first order estimates due to observational limitations, though we found similar results using the Tribble model.
\end{enumerate}
   Since there are only seven TDRGs discovered thus far,  we still know little about the mechanisms behind their peculiar features. Nonetheless, this work has shown that an extensive spectral analysis, which has been conducted for the first time to study a TDRG, is essential to determine the duty cycle and the nature of these rare sources. This is largely thanks to the availability of the MIGHTEE survey. However, follow up deeper observations at a much broader frequency range with subarcsec resolution are needed to further investigate the properties and the nuclear activity of this source and all the existing TDRGs. The advent of new radio telescopes such as the Square Kilometre Array Observatory (SKAO) or the upgrade of current telescopes to an unprecedented high sensitivity and resolution will allow systematic search and hopefully the discovery of more TDRGs.  

\section*{Acknowledgements}
We thank the anonymous referee for providing useful comments, which improved our manuscript. TFR acknowledges financial support from the Inter-University Institute for Data Intensive Astronomy (IDIA). IDIA is a partnership of the University of Cape Town, the University of Pretoria and the University of the Western Cape. ZR acknowledges funding from the South African Astronomical Observatory, which is a facilitiy of the National Research Foundation.
The MeerKAT telescope is operated by the
South African Radio Astronomy Observatory, which is a facility
of the National Research Foundation, an agency of the Department
of Science and Innovation. We acknowledge the use of the ilifu cloud computing facility – www.ilifu.ac.za, a partnership between the University of Cape Town, the University of the Western Cape, Stellenbosch University, Sol Plaatje University and the Cape Peninsula University of Technology. The ilifu facility is supported by contributions from the Inter-University Institute for Data Intensive Astronomy (IDIA – a partnership between the University of Cape Town, the University of Pretoria and the University of the Western Cape), the Computational Biology division at UCT and the Data Intensive Research Initiative of South Africa (DIRISA).  The authors acknowledge the Centre for High Performance Computing (CHPC), South Africa, for providing computational resources to this research project. This work made use of the CARTA (Cube Analysis and Rendering Tool for Astronomy) software (DOI 10.5281/zenodo.3377984 –  https://cartavis.github.io). CLH acknowledges support from STFC through grant ST/Y000951/1. MB acknowledges financial support from Next Generation EU funds within the National Recovery and Resilience Plan (PNRR), Mission 4 – Education and Research, Component 2 – From Research to Business (M4C2), Investment Line 3.1 – Strengthening and creation of Research Infrastructures, Project IR0000034 – “STILES – Strengthening the Italian Leadership in ELT and SKA”, from INAF under the Large GO 2024 funding scheme (project "MeerKAT and Euclid Team up: Exploring the galaxy-halo connection at cosmic noon”), the Large Grant 2022 funding scheme (project "MeerKAT and LOFAR Team up: a Unique Radio Window on Galaxy/AGN co-Evolution") and the Mini Grant 2023 funding scheme (project ‘Low radio frequencies as a probe of AGN jetfeedback at low and high redshift’). MV acknowledges financial support from IDIA 
and from the South African Department of Science and Innovation's National Research Foundation under the ISARP RADIOMAP Joint Research Scheme (DSI-NRF Grant Number 150551) and the CPRR HIPPO Project (DSI-NRF Grant Number SRUG22031677).

%%%%%%%%%%%%%%%%%%%%%%%%%%%%%%%%%%%%%%%%%%%%%%%%%%
\section*{Data Availability}

MIGHTEE-DR1 radio images and catalogues are accessible from: https://doi.or
g/10.48479/7msw-r692. The MIGHTEE sub-band images will be available in Luchsinger et al. (in prep).



\bibliographystyle{mnras}
\bibliography{example} % if your bibtex file is called example.bib

@ARTICLE{2020A&A...638A..34J,
       author = {{Jurlin}, N. and {Morganti}, R. and {Brienza}, M. and {Mandal}, S. and {Maddox}, N. and {Duncan}, K.~J. and {Shabala}, S.~S. and {Hardcastle}, M.~J. and {Prandoni}, I. and {R{\"o}ttgering}, H.~J.~A. and {Mahatma}, V. and {Best}, P.~N. and {Mingo}, B. and {Sabater}, J. and {Shimwell}, T.~W. and {Tasse}, C.},
        title = "{The life cycle of radio galaxies in the LOFAR Lockman Hole field}",
      journal = {\aap},
         year = 2020,
        month = jun,
       volume = {638},
          eid = {A34},
        pages = {A34},
          doi = {10.1051/0004-6361/201936955},
archivePrefix = {arXiv},
       eprint = {2004.09118},
 primaryClass = {astro-ph.GA},
       adsurl = {https://ui.adsabs.harvard.edu/abs/2020A&A...638A..34J}
}

@ARTICLE{1974Natur.250..625W,
       author = {{Willis}, A.~G. and {Strom}, R.~G. and {Wilson}, A.~S.},
        title = "{3C236, DA240; the largest radio sources known}",
      journal = {\nat},
         year = 1974,
        month = aug,
       volume = {250},
       number = {5468},
        pages = {625-630},
          doi = {10.1038/250625a0},
       adsurl = {https://ui.adsabs.harvard.edu/abs/1974Natur.250..625W}
}

@ARTICLE{2019A&A...622A...4H,
       author = {{Hale}, C.~L. and {Williams}, W. and {Jarvis}, M.~J. and {Hardcastle}, M.~J. and {Morabito}, L.~K. and {Shimwell}, T.~W. and {Tasse}, C. and {Best}, P.~N. and {Harwood}, J.~J. and {Heywood}, I. and {Prandoni}, I. and {R{\"o}ttgering}, H.~J.~A. and {Sabater}, J. and {Smith}, D.~J.~B. and {van Weeren}, R.~J.},
        title = "{LOFAR observations of the XMM-LSS field}",
      journal = {\aap},
         year = 2019,
        month = feb,
       volume = {622},
          eid = {A4},
        pages = {A4},
          doi = {10.1051/0004-6361/201833906},
archivePrefix = {arXiv},
       eprint = {1811.07942},
 primaryClass = {astro-ph.GA},
       adsurl = {https://ui.adsabs.harvard.edu/abs/2019A&A...622A...4H}
}

@ARTICLE{2022MNRAS.509.2150H,
       author = {{Heywood}, I. and {Jarvis}, M.~J. and {Hale}, C.~L. and {Whittam}, I.~H. and {Bester}, H.~L. and {Hugo}, B. and {Kenyon}, J.~S. and {Prescott}, M. and {Smirnov}, O.~M. and {Tasse}, C. and {Afonso}, J.~M. and {Best}, P.~N. and {Collier}, J.~D. and {Deane}, R.~P. and {Frank}, B.~S. and {Hardcastle}, M.~J. and {Knowles}, K. and {Maddox}, N. and {Murphy}, E.~J. and {Prandoni}, I. and {Randriamampandry}, S.~M. and {Santos}, M.~G. and {Sekhar}, S. and {Tabatabaei}, F. and {Taylor}, A.~R. and {Thorat}, K.},
        title = "{MIGHTEE: total intensity radio continuum imaging and the COSMOS/XMM-LSS Early Science fields}",
      journal = {\mnras},
         year = 2022,
        month = jan,
       volume = {509},
       number = {2},
        pages = {2150-2168},
          doi = {10.1093/mnras/stab3021},
archivePrefix = {arXiv},
       eprint = {2110.00347},
 primaryClass = {astro-ph.GA},
       adsurl = {https://ui.adsabs.harvard.edu/abs/2022MNRAS.509.2150H}
}

@INPROCEEDINGS{2016mks..confE...6J,
       author = {{Jarvis}, M. and {Taylor}, R. and {Agudo}, I. and {Allison}, J.~R. and {Deane}, R.~P. and {Frank}, B. and {Gupta}, N. and {Heywood}, I. and {Maddox}, N. and {McAlpine}, K. and {Santos}, M. and {Scaife}, A.~M.~M. and {Vaccari}, M. and {Zwart}, J.~T.~L. and {Adams}, E. and {Bacon}, D.~J. and {Baker}, A.~J. and {Bassett}, B.~A. and {Best}, P.~N. and {Beswick}, R. and {Blyth}, S. and {Brown}, M.~L. and {Bruggen}, M. and {Cluver}, M. and {Colafrancesco}, S. and {Cotter}, G. and {Cress}, C. and {Dav{\'e}}, R. and {Ferrari}, C. and {Hardcastle}, M.~J. and {Hale}, C.~L. and {Harrison}, I. and {Hatfield}, P.~W. and {Klockner}, H.~R. and {Kolwa}, S. and {Malefahlo}, E. and {Marubini}, T. and {Mauch}, T. and {Moodley}, K. and {Morganti}, R. and {Norris}, R.~P. and {Peters}, J.~A. and {Prandoni}, I. and {Prescott}, M. and {Oliver}, S. and {Oozeer}, N. and {Rottgering}, H.~J.~A. and {Seymour}, N. and {Simpson}, C. and {Smirnov}, O. and {Smith}, D.~J.~B.},
        title = "{The MeerKAT International GHz Tiered Extragalactic Exploration (MIGHTEE) Survey}",
    booktitle = {MeerKAT Science: On the Pathway to the SKA},
         year = 2016,
        month = jan,
          eid = {6},
        pages = {6},
archivePrefix = {arXiv},
       eprint = {1709.01901},
 primaryClass = {astro-ph.GA},
       adsurl = {https://ui.adsabs.harvard.edu/abs/2016mks..confE...6J}
}

@INPROCEEDINGS{2017MS&E..198a2014T,
       author = {{Taylor}, A. Russ and {Jarvis}, Matt},
        title = "{MIGHTEE: The MeerKAT International GHz Tiered Extragalactic Exploration}",
    booktitle = {Materials Science and Engineering Conference Series},
         year = 2017,
       series = {Materials Science and Engineering Conference Series},
       volume = {198},
        month = may,
        pages = {012014},
          doi = {10.1088/1757-899X/198/1/012014},
       adsurl = {https://ui.adsabs.harvard.edu/abs/2017MS&E..198a2014T}
}

@ARTICLE{2016A&A...585A..29B,
       author = {{Brienza}, M. and {Godfrey}, L. and {Morganti}, R. and {Vilchez}, N. and {Maddox}, N. and {Murgia}, M. and {Orru}, E. and {Shulevski}, A. and {Best}, P.~N. and {Br{\"u}ggen}, M. and {Harwood}, J.~J. and {Jamrozy}, M. and {Jarvis}, M.~J. and {Mahony}, E.~K. and {McKean}, J. and {R{\"o}ttgering}, H.~J.~A.},
        title = "{LOFAR discovery of a 700-kpc remnant radio galaxy at low redshift}",
      journal = {\aap},
         year = 2016,
        month = jan,
       volume = {585},
          eid = {A29},
        pages = {A29},
          doi = {10.1051/0004-6361/201526754},
archivePrefix = {arXiv},
       eprint = {1508.07239},
 primaryClass = {astro-ph.GA},
       adsurl = {https://ui.adsabs.harvard.edu/abs/2016A&A...585A..29B}
}

@ARTICLE{2000MNRAS.315..371S,
       author = {{Schoenmakers}, Arno P. and {de Bruyn}, A.~G. and {R{\"o}ttgering}, H.~J.~A. and {van der Laan}, H. and {Kaiser}, C.~R.},
        title = "{Radio galaxies with a `double-double morphology' - I. Analysis of the radio properties and evidence for interrupted activity in active galactic nuclei}",
      journal = {\mnras},
         year = 2000,
        month = jun,
       volume = {315},
       number = {2},
        pages = {371-380},
          doi = {10.1046/j.1365-8711.2000.03430.x},
archivePrefix = {arXiv},
       eprint = {astro-ph/9912141},
 primaryClass = {astro-ph},
       adsurl = {https://ui.adsabs.harvard.edu/abs/2000MNRAS.315..371S}
}

@ARTICLE{2019A&A...622A..13M,
       author = {{Mahatma}, V.~H. and {Hardcastle}, M.~J. and {Williams}, W.~L. and {Best}, P.~N. and {Croston}, J.~H. and {Duncan}, K. and {Mingo}, B. and {Morganti}, R. and {Brienza}, M. and {Cochrane}, R.~K. and {G{\"u}rkan}, G. and {Harwood}, J.~J. and {Jarvis}, M.~J. and {Jamrozy}, M. and {Jurlin}, N. and {Morabito}, L.~K. and {R{\"o}ttgering}, H.~J.~A. and {Sabater}, J. and {Shimwell}, T.~W. and {Smith}, D.~J.~B. and {Shulevski}, A. and {Tasse}, C.},
        title = "{LoTSS DR1: Double-double radio galaxies in the HETDEX field}",
      journal = {\aap},
         year = 2019,
        month = feb,
       volume = {622},
          eid = {A13},
        pages = {A13},
          doi = {10.1051/0004-6361/201833973},
archivePrefix = {arXiv},
       eprint = {1811.08194},
 primaryClass = {astro-ph.GA},
       adsurl = {https://ui.adsabs.harvard.edu/abs/2019A&A...622A..13M}
}

@ARTICLE{2007A&A...470..875P,
       author = {{Parma}, P. and {Murgia}, M. and {de Ruiter}, H.~R. and {Fanti}, R. and {Mack}, K. -H. and {Govoni}, F.},
        title = "{In search of dying radio sources in the local universe}",
      journal = {\aap},
         year = 2007,
        month = aug,
       volume = {470},
       number = {3},
        pages = {875-888},
          doi = {10.1051/0004-6361:20077592},
archivePrefix = {arXiv},
       eprint = {0705.3209},
 primaryClass = {astro-ph},
       adsurl = {https://ui.adsabs.harvard.edu/abs/2007A&A...470..875P}
}

@ARTICLE{2020MNRAS.496.3381R,
       author = {{Randriamanakoto}, Z. and {Ishwara-Chandra}, C.~H. and {Taylor}, A.~R.},
        title = "{J1615+5452: a remnant radio galaxy in the ELAIS-N1 field}",
      journal = {\mnras},
         year = 2020,
        month = aug,
       volume = {496},
       number = {3},
        pages = {3381-3389},
          doi = {10.1093/mnras/staa1782},
archivePrefix = {arXiv},
       eprint = {2006.10028},
 primaryClass = {astro-ph.GA},
       adsurl = {https://ui.adsabs.harvard.edu/abs/2020MNRAS.496.3381R}
}

@ARTICLE{2011A&A...526A.148M,
       author = {{Murgia}, M. and {Parma}, P. and {Mack}, K. -H. and {de Ruiter}, H.~R. and {Fanti}, R. and {Govoni}, F. and {Tarchi}, A. and {Giacintucci}, S. and {Markevitch}, M.},
        title = "{Dying radio galaxies in clusters}",
      journal = {\aap},
         year = 2011,
        month = feb,
       volume = {526},
          eid = {A148},
        pages = {A148},
          doi = {10.1051/0004-6361/201015302},
archivePrefix = {arXiv},
       eprint = {1011.0567},
 primaryClass = {astro-ph.CO},
       adsurl = {https://ui.adsabs.harvard.edu/abs/2011A&A...526A.148M}
}

@ARTICLE{2021MNRAS.501.3833D,
       author = {{Delhaize}, J. and {Heywood}, I. and {Prescott}, M. and {Jarvis}, M.~J. and {Delvecchio}, I. and {Whittam}, I.~H. and {White}, S.~V. and {Hardcastle}, M.~J. and {Hale}, C.~L. and {Afonso}, J. and {Ao}, Y. and {Brienza}, M. and {Br{\"u}ggen}, M. and {Collier}, J.~D. and {Daddi}, E. and {Glowacki}, M. and {Maddox}, N. and {Morabito}, L.~K. and {Prandoni}, I. and {Randriamanakoto}, Z. and {Sekhar}, S. and {An}, Fangxia and {Adams}, N.~J. and {Blyth}, S. and {Bowler}, R.~A.~A. and {Leeuw}, L. and {Marchetti}, L. and {Randriamampandry}, S.~M. and {Thorat}, K. and {Seymour}, N. and {Smirnov}, O. and {Taylor}, A.~R. and {Tasse}, C. and {Vaccari}, M.},
        title = "{MIGHTEE: are giant radio galaxies more common than we thought?}",
      journal = {\mnras},
         year = 2021,
        month = mar,
       volume = {501},
       number = {3},
        pages = {3833-3845},
          doi = {10.1093/mnras/staa3837},
archivePrefix = {arXiv},
       eprint = {2012.05759},
 primaryClass = {astro-ph.GA},
       adsurl = {https://ui.adsabs.harvard.edu/abs/2021MNRAS.501.3833D}
}

@ARTICLE{2020A&A...638A..29B,
       author = {{Brienza}, M. and {Morganti}, R. and {Harwood}, J. and {Duchet}, T. and {Rajpurohit}, K. and {Shulevski}, A. and {Hardcastle}, M.~J. and {Mahatma}, V. and {Godfrey}, L.~E.~H. and {Prandoni}, I. and {Shimwell}, T.~W. and {Intema}, H.},
        title = "{Radio spectral properties and jet duty cycle in the restarted radio galaxy 3C388}",
      journal = {\aap},
         year = 2020,
        month = jun,
       volume = {638},
          eid = {A29},
        pages = {A29},
          doi = {10.1051/0004-6361/202037457},
archivePrefix = {arXiv},
       eprint = {2003.13476},
 primaryClass = {astro-ph.GA},
       adsurl = {https://ui.adsabs.harvard.edu/abs/2020A&A...638A..29B}
}

@ARTICLE{2020PASP..132c5001L,
       author = {{Lacy}, M. and {Baum}, S.~A. and {Chandler}, C.~J. and {Chatterjee}, S. and {Clarke}, T.~E. and {Deustua}, S. and {English}, J. and {Farnes}, J. and {Gaensler}, B.~M. and {Gugliucci}, N. and {Hallinan}, G. and {Kent}, B.~R. and {Kimball}, A. and {Law}, C.~J. and {Lazio}, T.~J.~W. and {Marvil}, J. and {Mao}, S.~A. and {Medlin}, D. and {Mooley}, K. and {Murphy}, E.~J. and {Myers}, S. and {Osten}, R. and {Richards}, G.~T. and {Rosolowsky}, E. and {Rudnick}, L. and {Schinzel}, F. and {Sivakoff}, G.~R. and {Sjouwerman}, L.~O. and {Taylor}, R. and {White}, R.~L. and {Wrobel}, J. and {Andernach}, H. and {Beasley}, A.~J. and {Berger}, E. and {Bhatnager}, S. and {Birkinshaw}, M. and {Bower}, G.~C. and {Brandt}, W.~N. and {Brown}, S. and {Burke-Spolaor}, S. and {Butler}, B.~J. and {Comerford}, J. and {Demorest}, P.~B. and {Fu}, H. and {Giacintucci}, S. and {Golap}, K. and {G{\"u}th}, T. and {Hales}, C.~A. and {Hiriart}, R. and {Hodge}, J. and {Horesh}, A. and {Ivezi{\'c}}, {\v{Z}}. and {Jarvis}, M.~J. and {Kamble}, A. and {Kassim}, N. and {Liu}, X. and {Loinard}, L. and {Lyons}, D.~K. and {Masters}, J. and {Mezcua}, M. and {Moellenbrock}, G.~A. and {Mroczkowski}, T. and {Nyland}, K. and {O'Dea}, C.~P. and {O'Sullivan}, S.~P. and {Peters}, W.~M. and {Radford}, K. and {Rao}, U. and {Robnett}, J. and {Salcido}, J. and {Shen}, Y. and {Sobotka}, A. and {Witz}, S. and {Vaccari}, M. and {van Weeren}, R.~J. and {Vargas}, A. and {Williams}, P.~K.~G. and {Yoon}, I.},
        title = "{The Karl G. Jansky Very Large Array Sky Survey (VLASS). Science Case and Survey Design}",
      journal = {\pasp},
         year = 2020,
        month = mar,
       volume = {132},
       number = {1009},
          eid = {035001},
        pages = {035001},
          doi = {10.1088/1538-3873/ab63eb},
archivePrefix = {arXiv},
       eprint = {1907.01981},
 primaryClass = {astro-ph.IM},
       adsurl = {https://ui.adsabs.harvard.edu/abs/2020PASP..132c5001L}
}

@INPROCEEDINGS{2016mks..confE...1J,
       author = {{Jonas}, J. and {MeerKAT Team}},
        title = "{The MeerKAT Radio Telescope}",
    booktitle = {MeerKAT Science: On the Pathway to the SKA},
         year = 2016,
        month = jan,
          eid = {1},
        pages = {1},
       adsurl = {https://ui.adsabs.harvard.edu/abs/2016mks..confE...1J}
}

@ARTICLE{2007MNRAS.382.1019B,
       author = {{Brocksopp}, C. and {Kaiser}, C.~R. and {Schoenmakers}, A.~P. and {de Bruyn}, A.~G.},
        title = "{Three episodes of jet activity in the Fanaroff-Riley type II radio galaxy B0925+420}",
      journal = {\mnras},
         year = 2007,
        month = dec,
       volume = {382},
       number = {3},
        pages = {1019-1028},
          doi = {10.1111/j.1365-2966.2007.12483.x},
archivePrefix = {arXiv},
       eprint = {0709.4548},
 primaryClass = {astro-ph},
       adsurl = {https://ui.adsabs.harvard.edu/abs/2007MNRAS.382.1019B}
}

@ARTICLE{2023MNRAS.525L..87C,
       author = {{Chavan}, Kshitij and {Dabhade}, Pratik and {Saikia}, D.~J.},
        title = "{A giant radio galaxy with three cycles of episodic jet activity from LoTSS DR2}",
      journal = {\mnras},
         year = 2023,
        month = oct,
       volume = {525},
       number = {1},
        pages = {L87-L92},
          doi = {10.1093/mnrasl/slad100},
archivePrefix = {arXiv},
       eprint = {2307.08553},
 primaryClass = {astro-ph.GA},
       adsurl = {https://ui.adsabs.harvard.edu/abs/2023MNRAS.525L..87C}
}

@ARTICLE{2011MNRAS.417L..36H,
       author = {{Hota}, Ananda and {Sirothia}, S.~K. and {Ohyama}, Youichi and {Konar}, C. and {Kim}, Suk and {Rey}, Soo-Chang and {Saikia}, D.~J. and {Croston}, J.~H. and {Matsushita}, Satoki},
        title = "{Discovery of a spiral-host episodic radio galaxy}",
      journal = {\mnras},
         year = 2011,
        month = oct,
       volume = {417},
       number = {1},
        pages = {L36-L40},
          doi = {10.1111/j.1745-3933.2011.01115.x},
archivePrefix = {arXiv},
       eprint = {1107.4742},
 primaryClass = {astro-ph.CO},
       adsurl = {https://ui.adsabs.harvard.edu/abs/2011MNRAS.417L..36H}
}

@ARTICLE{2016ApJ...826..132S,
       author = {{Singh}, Veeresh and {Ishwara-Chandra}, C.~H. and {Kharb}, Preeti and {Srivastava}, Shweta and {Janardhan}, P.},
        title = "{J1216+0709: A Radio Galaxy with Three Episodes of AGN Jet Activity}",
      journal = {\apj},
         year = 2016,
        month = aug,
       volume = {826},
       number = {2},
          eid = {132},
        pages = {132},
          doi = {10.3847/0004-637X/826/2/132},
archivePrefix = {arXiv},
       eprint = {1605.05821},
 primaryClass = {astro-ph.GA},
       adsurl = {https://ui.adsabs.harvard.edu/abs/2016ApJ...826..132S}
}

@ARTICLE{1974MNRAS.167P..31F,
       author = {{Fanaroff}, B.~L. and {Riley}, J.~M.},
        title = "{The morphology of extragalactic radio sources of high and low luminosity}",
      journal = {\mnras},
         year = 1974,
        month = may,
       volume = {167},
        pages = {31P-36P},
          doi = {10.1093/mnras/167.1.31P},
       adsurl = {https://ui.adsabs.harvard.edu/abs/1974MNRAS.167P..31F}
}

@ARTICLE{2000MNRAS.315..381K,
       author = {{Kaiser}, Christian R. and {Schoenmakers}, Arno P. and {R{\"o}ttgering}, Huub J.~A.},
        title = "{Radio galaxies with a `double-double' morphology - II. The evolution of double-double radio galaxies and implications for the alignment effect in FRII sources}",
      journal = {\mnras},
         year = 2000,
        month = jun,
       volume = {315},
       number = {2},
        pages = {381-394},
          doi = {10.1046/j.1365-8711.2000.03431.x},
archivePrefix = {arXiv},
       eprint = {astro-ph/9912142},
 primaryClass = {astro-ph},
       adsurl = {https://ui.adsabs.harvard.edu/abs/2000MNRAS.315..381K}
}

@ARTICLE{2000MNRAS.315..395S,
       author = {{Schoenmakers}, Arno P. and {de Bruyn}, A.~G. and {R{\"o}ttgering}, H.~J.~A. and {van der Laan}, H.},
        title = "{Radio galaxies with a `double-double' morphology - III. The case of B1834+620}",
      journal = {\mnras},
         year = 2000,
        month = jun,
       volume = {315},
       number = {2},
        pages = {395-406},
          doi = {10.1046/j.1365-8711.2000.03432.x},
archivePrefix = {arXiv},
       eprint = {astro-ph/9912143},
 primaryClass = {astro-ph},
       adsurl = {https://ui.adsabs.harvard.edu/abs/2000MNRAS.315..395S}
}

@ARTICLE{2006MNRAS.372..693K,
       author = {{Konar}, C. and {Saikia}, D.~J. and {Jamrozy}, M. and {Machalski}, J.},
        title = "{Spectral ageing analysis of the double-double radio galaxy J1453+3308}",
      journal = {\mnras},
         year = 2006,
        month = oct,
       volume = {372},
       number = {2},
        pages = {693-702},
          doi = {10.1111/j.1365-2966.2006.10874.x},
archivePrefix = {arXiv},
       eprint = {astro-ph/0607660},
 primaryClass = {astro-ph},
       adsurl = {https://ui.adsabs.harvard.edu/abs/2006MNRAS.372..693K}
}

@ARTICLE{2010A&A...510A..84M,
       author = {{Machalski}, J. and {Jamrozy}, M. and {Konar}, C.},
        title = "{Spectral ageing analysis and dynamical analysis of the double-double radio galaxy J1548-3216}",
      journal = {\aap},
         year = 2010,
        month = feb,
       volume = {510},
          eid = {A84},
        pages = {A84},
          doi = {10.1051/0004-6361/200912797},
archivePrefix = {arXiv},
       eprint = {0912.1484},
 primaryClass = {astro-ph.CO},
       adsurl = {https://ui.adsabs.harvard.edu/abs/2010A&A...510A..84M}
}

@ARTICLE{2025MNRAS.536.2187H,
       author = {{Hale}, C.~L. and {Heywood}, I. and {Jarvis}, M.~J. and {Whittam}, I.~H. and {Best}, P.~N. and {An}, Fangxia and {Bowler}, R.~A.~A. and {Harrison}, I. and {Matthews}, A. and {Smith}, D.~J.~B. and {Taylor}, A.~R. and {Vaccari}, M.},
        title = "{MIGHTEE: the continuum survey Data Release 1}",
      journal = {\mnras},
         year = 2025,
        month = jan,
       volume = {536},
       number = {3},
        pages = {2187-2211},
          doi = {10.1093/mnras/stae2528},
archivePrefix = {arXiv},
       eprint = {2411.04958},
 primaryClass = {astro-ph.GA},
       adsurl = {https://ui.adsabs.harvard.edu/abs/2025MNRAS.536.2187H}
}

@ARTICLE{2011A&A...535A..65D,
       author = {{Durret}, F. and {Adami}, C. and {Cappi}, A. and {Maurogordato}, S. and {M{\'a}rquez}, I. and {Ilbert}, O. and {Coupon}, J. and {Arnouts}, S. and {Benoist}, C. and {Blaizot}, J. and {Edorh}, T.~M. and {Garilli}, B. and {Guennou}, L. and {Le Brun}, V. and {Le F{\`e}vre}, O. and {Mazure}, A. and {McCracken}, H.~J. and {Mellier}, Y. and {Mezrag}, C. and {Slezak}, E. and {Tresse}, L. and {Ulmer}, M.~P.},
        title = "{Galaxy cluster searches based on photometric redshifts in the four CFHTLS Wide fields}",
      journal = {\aap},
         year = 2011,
        month = nov,
       volume = {535},
          eid = {A65},
        pages = {A65},
          doi = {10.1051/0004-6361/201116985},
archivePrefix = {arXiv},
       eprint = {1109.0850},
 primaryClass = {astro-ph.CO},
       adsurl = {https://ui.adsabs.harvard.edu/abs/2011A&A...535A..65D}
}

@ARTICLE{2018PASJ...70S..20O,
       author = {{Oguri}, Masamune and {Lin}, Yen-Ting and {Lin}, Sheng-Chieh and {Nishizawa}, Atsushi J. and {More}, Anupreeta and {More}, Surhud and {Hsieh}, Bau-Ching and {Medezinski}, Elinor and {Miyatake}, Hironao and {Jian}, Hung-Yu and {Lin}, Lihwai and {Takada}, Masahiro and {Okabe}, Nobuhiro and {Speagle}, Joshua S. and {Coupon}, Jean and {Leauthaud}, Alexie and {Lupton}, Robert H. and {Miyazaki}, Satoshi and {Price}, Paul A. and {Tanaka}, Masayuki and {Chiu}, I. -Non and {Komiyama}, Yutaka and {Okura}, Yuki and {Tanaka}, Manobu M. and {Usuda}, Tomonori},
        title = "{An optically-selected cluster catalog at redshift 0.1 < z < 1.1 from the Hyper Suprime-Cam Subaru Strategic Program S16A data}",
      journal = {\pasj},
         year = 2018,
        month = jan,
       volume = {70},
          eid = {S20},
        pages = {S20},
          doi = {10.1093/pasj/psx042},
archivePrefix = {arXiv},
       eprint = {1701.00818},
 primaryClass = {astro-ph.CO},
       adsurl = {https://ui.adsabs.harvard.edu/abs/2018PASJ...70S..20O}
}

@ARTICLE{2018MNRAS.475..343W,
       author = {{Wen}, Z.~L. and {Han}, J.~L. and {Yang}, F.},
        title = "{A catalogue of clusters of galaxies identified from all sky surveys of 2MASS, WISE, and SuperCOSMOS}",
      journal = {\mnras},
         year = 2018,
        month = mar,
       volume = {475},
       number = {1},
        pages = {343-352},
          doi = {10.1093/mnras/stx3189},
archivePrefix = {arXiv},
       eprint = {1712.02491},
 primaryClass = {astro-ph.GA},
       adsurl = {https://ui.adsabs.harvard.edu/abs/2018MNRAS.475..343W}
}

@ARTICLE{2018A&A...620A..12C,
       author = {{Chiappetti}, L. and {Fotopoulou}, S. and {Lidman}, C. and {Faccioli}, L. and {Pacaud}, F. and {Elyiv}, A. and {Paltani}, S. and {Pierre}, M. and {Plionis}, M. and {Adami}, C. and {Alis}, S. and {Altieri}, B. and {Baldry}, I. and {Bolzonella}, M. and {Bongiorno}, A. and {Brown}, M. and {Driver}, S. and {Elmer}, E. and {Franzetti}, P. and {Grootes}, M. and {Guglielmo}, V. and {Iovino}, A. and {Koulouridis}, E. and {Lef{\`e}vre}, J.~P. and {Liske}, J. and {Maurogordato}, S. and {Melnyk}, O. and {Owers}, M. and {Poggianti}, B. and {Polletta}, M. and {Pompei}, E. and {Ponman}, T. and {Robotham}, A. and {Sadibekova}, T. and {Tuffs}, R. and {Valtchanov}, I. and {Vignali}, C. and {Wagner}, G.},
        title = "{The XXL Survey: XXVII. The 3XLSS point source catalogue}",
      journal = {\aap},
         year = 2018,
        month = dec,
       volume = {620},
          eid = {A12},
        pages = {A12},
          doi = {10.1051/0004-6361/201731880},
archivePrefix = {arXiv},
       eprint = {1810.03929},
 primaryClass = {astro-ph.GA},
       adsurl = {https://ui.adsabs.harvard.edu/abs/2018A&A...620A..12C}
}

@ARTICLE{2021ApJS..253...56Z,
       author = {{Zou}, Hu and {Gao}, Jinghua and {Xu}, Xin and {Zhou}, Xu and {Ma}, Jun and {Zhou}, Zhimin and {Zhang}, Tianmeng and {Nie}, Jundan and {Wang}, Jiali and {Xue}, Suijian},
        title = "{Galaxy Clusters from the DESI Legacy Imaging Surveys. I. Cluster Detection}",
      journal = {\apjs},
         year = 2021,
        month = apr,
       volume = {253},
       number = {2},
          eid = {56},
        pages = {56},
          doi = {10.3847/1538-4365/abe5b0},
archivePrefix = {arXiv},
       eprint = {2101.12340},
 primaryClass = {astro-ph.GA},
       adsurl = {https://ui.adsabs.harvard.edu/abs/2021ApJS..253...56Z}
}

@ARTICLE{2022RAA....22f5001Z,
       author = {{Zou}, Hu and {Sui}, Jipeng and {Xue}, Suijian and {Zhou}, Xu and {Ma}, Jun and {Zhou}, Zhimin and {Nie}, Jundan and {Zhang}, Tianmeng and {Feng}, Lu and {Shen}, Zhixia and {Wang}, Jiali},
        title = "{Photometric Redshifts and Galaxy Clusters for DES DR2, DESI DR9, and HSC-SSP PDR3 Data}",
      journal = {Research in Astronomy and Astrophysics},
         year = 2022,
        month = jun,
       volume = {22},
       number = {6},
          eid = {065001},
        pages = {065001},
          doi = {10.1088/1674-4527/ac6416},
archivePrefix = {arXiv},
       eprint = {2203.17035},
 primaryClass = {astro-ph.GA},
       adsurl = {https://ui.adsabs.harvard.edu/abs/2022RAA....22f5001Z}
}

@ARTICLE{1973A&A....26..423J,
       author = {{Jaffe}, W.~J. and {Perola}, G.~C.},
        title = "{Dynamical Models of Tailed Radio Sources in Clusters of Galaxies}",
      journal = {\aap},
         year = 1973,
        month = aug,
       volume = {26},
        pages = {423},
       adsurl = {https://ui.adsabs.harvard.edu/abs/1973A&A....26..423J}
}

@ARTICLE{1991MNRAS.253..147T,
       author = {{Tribble}, Peter C.},
        title = "{Radio emission in random magnetic field : radio haloes and the structure of the magnetic field in the Coma cluster.}",
      journal = {\mnras},
         year = 1991,
        month = nov,
       volume = {253},
        pages = {147},
          doi = {10.1093/mnras/253.1.147},
       adsurl = {https://ui.adsabs.harvard.edu/abs/1991MNRAS.253..147T}
}

@ARTICLE{1993MNRAS.261...57T,
       author = {{Tribble}, Peter C.},
        title = "{Radio spectral ageing in a random magnetic field.}",
      journal = {\mnras},
         year = 1993,
        month = mar,
       volume = {261},
        pages = {57-62},
          doi = {10.1093/mnras/261.1.57},
       adsurl = {https://ui.adsabs.harvard.edu/abs/1993MNRAS.261...57T}
}

@ARTICLE{2013MNRAS.435.3353H,
       author = {{Harwood}, Jeremy J. and {Hardcastle}, Martin J. and {Croston}, Judith H. and {Goodger}, Joanna L.},
        title = "{Spectral ageing in the lobes of FR-II radio galaxies: new methods of analysis for broad-band radio data}",
      journal = {\mnras},
         year = 2013,
        month = nov,
       volume = {435},
       number = {4},
        pages = {3353-3375},
          doi = {10.1093/mnras/stt1526},
archivePrefix = {arXiv},
       eprint = {1308.4137},
 primaryClass = {astro-ph.CO},
       adsurl = {https://ui.adsabs.harvard.edu/abs/2013MNRAS.435.3353H}
}

@ARTICLE{2004IJMPD..13.1549G,
       author = {{Govoni}, Federica and {Feretti}, Luigina},
        title = "{Magnetic Fields in Clusters of Galaxies}",
      journal = {International Journal of Modern Physics D},
         year = 2004,
        month = jan,
       volume = {13},
       number = {8},
        pages = {1549-1594},
          doi = {10.1142/S0218271804005080},
archivePrefix = {arXiv},
       eprint = {astro-ph/0410182},
 primaryClass = {astro-ph},
       adsurl = {https://ui.adsabs.harvard.edu/abs/2004IJMPD..13.1549G}
}

@ARTICLE{2018A&A...618A..45B,
       author = {{Brienza}, M. and {Morganti}, R. and {Murgia}, M. and {Vilchez}, N. and {Adebahr}, B. and {Carretti}, E. and {Concu}, R. and {Govoni}, F. and {Harwood}, J. and {Intema}, H. and {Loi}, F. and {Melis}, A. and {Paladino}, R. and {Poppi}, S. and {Shulevski}, A. and {Vacca}, V. and {Valente}, G.},
        title = "{Duty cycle of the radio galaxy B2 0258+35}",
      journal = {\aap},
         year = 2018,
        month = oct,
       volume = {618},
          eid = {A45},
        pages = {A45},
          doi = {10.1051/0004-6361/201832846},
archivePrefix = {arXiv},
       eprint = {1807.07280},
 primaryClass = {astro-ph.GA},
       adsurl = {https://ui.adsabs.harvard.edu/abs/2018A&A...618A..45B}
}

@ARTICLE{2021Galax...9...88M,
       author = {{Morganti}, Raffaella and {Jurlin}, Nika and {Oosterloo}, Tom and {Brienza}, Marisa and {Orr{\'u}}, Emanuela and {Kutkin}, Alexander and {Prandoni}, Isabella and {Adams}, Elizabeth A.~K. and {D{\'e}nes}, Helga and {Hess}, Kelley M. and {Shulevski}, Aleksandar and {van der Hulst}, Thijs and {Ziemke}, Jacob},
        title = "{Combining LOFAR and Apertif Data for Understanding the Life Cycle of Radio Galaxies}",
      journal = {Galaxies},
         year = 2021,
        month = nov,
       volume = {9},
       number = {4},
          eid = {88},
        pages = {88},
          doi = {10.3390/galaxies9040088},
archivePrefix = {arXiv},
       eprint = {2111.04776},
 primaryClass = {astro-ph.GA},
       adsurl = {https://ui.adsabs.harvard.edu/abs/2021Galax...9...88M}
}

@ARTICLE{1995ApJ...450..559B,
       author = {{Becker}, Robert H. and {White}, Richard L. and {Helfand}, David J.},
        title = "{The FIRST Survey: Faint Images of the Radio Sky at Twenty Centimeters}",
      journal = {\apj},
         year = 1995,
        month = sep,
       volume = {450},
        pages = {559},
          doi = {10.1086/176166},
       adsurl = {https://ui.adsabs.harvard.edu/abs/1995ApJ...450..559B}
}

@ARTICLE{1998AJ....115.1693C,
       author = {{Condon}, J.~J. and {Cotton}, W.~D. and {Greisen}, E.~W. and {Yin}, Q.~F. and {Perley}, R.~A. and {Taylor}, G.~B. and {Broderick}, J.~J.},
        title = "{The NRAO VLA Sky Survey}",
      journal = {\aj},
         year = 1998,
        month = may,
       volume = {115},
       number = {5},
        pages = {1693-1716},
          doi = {10.1086/300337},
       adsurl = {https://ui.adsabs.harvard.edu/abs/1998AJ....115.1693C}
}

@ARTICLE{2016A&A...594A..13P,
       author = {{Planck Collaboration} and {Ade}, P.~A.~R. and {Aghanim}, N. and {Arnaud}, M. and {Ashdown}, M. and {Aumont}, J. and {Baccigalupi}, C. and {Banday}, A.~J. and {Barreiro}, R.~B. and {Bartlett}, J.~G. and {Bartolo}, N. and {Battaner}, E. and {Battye}, R. and {Benabed}, K. and {Beno{\^\i}t}, A. and {Benoit-L{\'e}vy}, A. and {Bernard}, J. -P. and {Bersanelli}, M. and {Bielewicz}, P. and {Bock}, J.~J. and {Bonaldi}, A. and {Bonavera}, L. and {Bond}, J.~R. and {Borrill}, J. and {Bouchet}, F.~R. and {Boulanger}, F. and {Bucher}, M. and {Burigana}, C. and {Butler}, R.~C. and {Calabrese}, E. and {Cardoso}, J. -F. and {Catalano}, A. and {Challinor}, A. and {Chamballu}, A. and {Chary}, R. -R. and {Chiang}, H.~C. and {Chluba}, J. and {Christensen}, P.~R. and {Church}, S. and {Clements}, D.~L. and {Colombi}, S. and {Colombo}, L.~P.~L. and {Combet}, C. and {Coulais}, A. and {Crill}, B.~P. and {Curto}, A. and {Cuttaia}, F. and {Danese}, L. and {Davies}, R.~D. and {Davis}, R.~J. and {de Bernardis}, P. and {de Rosa}, A. and {de Zotti}, G. and {Delabrouille}, J. and {D{\'e}sert}, F. -X. and {Di Valentino}, E. and {Dickinson}, C. and {Diego}, J.~M. and {Dolag}, K. and {Dole}, H. and {Donzelli}, S. and {Dor{\'e}}, O. and {Douspis}, M. and {Ducout}, A. and {Dunkley}, J. and {Dupac}, X. and {Efstathiou}, G. and {Elsner}, F. and {En{\ss}lin}, T.~A. and {Eriksen}, H.~K. and {Farhang}, M. and {Fergusson}, J. and {Finelli}, F. and {Forni}, O. and {Frailis}, M. and {Fraisse}, A.~A. and {Franceschi}, E. and {Frejsel}, A. and {Galeotta}, S. and {Galli}, S. and {Ganga}, K. and {Gauthier}, C. and {Gerbino}, M. and {Ghosh}, T. and {Giard}, M. and {Giraud-H{\'e}raud}, Y. and {Giusarma}, E. and {Gjerl{\o}w}, E. and {Gonz{\'a}lez-Nuevo}, J. and {G{\'o}rski}, K.~M. and {Gratton}, S. and {Gregorio}, A. and {Gruppuso}, A. and {Gudmundsson}, J.~E. and {Hamann}, J. and {Hansen}, F.~K. and {Hanson}, D. and {Harrison}, D.~L. and {Helou}, G. and {Henrot-Versill{\'e}}, S. and {Hern{\'a}ndez-Monteagudo}, C. and {Herranz}, D. and {Hildebrandt}, S.~R. and {Hivon}, E. and {Hobson}, M. and {Holmes}, W.~A. and {Hornstrup}, A. and {Hovest}, W. and {Huang}, Z. and {Huffenberger}, K.~M. and {Hurier}, G. and {Jaffe}, A.~H. and {Jaffe}, T.~R. and {Jones}, W.~C. and {Juvela}, M. and {Keih{\"a}nen}, E. and {Keskitalo}, R. and {Kisner}, T.~S. and {Kneissl}, R. and {Knoche}, J. and {Knox}, L. and {Kunz}, M. and {Kurki-Suonio}, H. and {Lagache}, G. and {L{\"a}hteenm{\"a}ki}, A. and {Lamarre}, J. -M. and {Lasenby}, A. and {Lattanzi}, M. and {Lawrence}, C.~R. and {Leahy}, J.~P. and {Leonardi}, R. and {Lesgourgues}, J. and {Levrier}, F. and {Lewis}, A. and {Liguori}, M. and {Lilje}, P.~B. and {Linden-V{\o}rnle}, M. and {L{\'o}pez-Caniego}, M. and {Lubin}, P.~M. and {Mac{\'\i}as-P{\'e}rez}, J.~F. and {Maggio}, G. and {Maino}, D. and {Mandolesi}, N. and {Mangilli}, A. and {Marchini}, A. and {Maris}, M. and {Martin}, P.~G. and {Martinelli}, M. and {Mart{\'\i}nez-Gonz{\'a}lez}, E. and {Masi}, S. and {Matarrese}, S. and {McGehee}, P. and {Meinhold}, P.~R. and {Melchiorri}, A. and {Melin}, J. -B. and {Mendes}, L. and {Mennella}, A. and {Migliaccio}, M. and {Millea}, M. and {Mitra}, S. and {Miville-Desch{\^e}nes}, M. -A. and {Moneti}, A. and {Montier}, L. and {Morgante}, G. and {Mortlock}, D. and {Moss}, A. and {Munshi}, D. and {Murphy}, J.~A. and {Naselsky}, P. and {Nati}, F. and {Natoli}, P. and {Netterfield}, C.~B. and {N{\o}rgaard-Nielsen}, H.~U. and {Noviello}, F. and {Novikov}, D. and {Novikov}, I. and {Oxborrow}, C.~A. and {Paci}, F. and {Pagano}, L. and {Pajot}, F. and {Paladini}, R. and {Paoletti}, D. and {Partridge}, B. and {Pasian}, F. and {Patanchon}, G. and {Pearson}, T.~J. and {Perdereau}, O. and {Perotto}, L. and {Perrotta}, F. and {Pettorino}, V. and {Piacentini}, F. and {Piat}, M. and {Pierpaoli}, E. and {Pietrobon}, D. and {Plaszczynski}, S. and {Pointecouteau}, E. and {Polenta}, G. and {Popa}, L. and {Pratt}, G.~W. and {Pr{\'e}zeau}, G.},
        title = "{Planck 2015 results. XIII. Cosmological parameters}",
      journal = {\aap},
         year = 2016,
        month = sep,
       volume = {594},
          eid = {A13},
        pages = {A13},
          doi = {10.1051/0004-6361/201525830},
archivePrefix = {arXiv},
       eprint = {1502.01589},
 primaryClass = {astro-ph.CO},
       adsurl = {https://ui.adsabs.harvard.edu/abs/2016A&A...594A..13P}
}

@ARTICLE{2014A&A...569A..52S,
       author = {{Singh}, V. and {Beelen}, A. and {Wadadekar}, Y. and {Sirothia}, S. and {Ishwara-Chandra}, C.~H. and {Basu}, A. and {Omont}, A. and {McAlpine}, K. and {Ivison}, R.~J. and {Oliver}, S. and {Farrah}, D. and {Lacy}, M.},
        title = "{Multiwavelength characterization of faint ultra steep spectrum radio sources: A search for high-redshift radio galaxies}",
      journal = {\aap},
         year = 2014,
        month = sep,
       volume = {569},
          eid = {A52},
        pages = {A52},
          doi = {10.1051/0004-6361/201423644},
archivePrefix = {arXiv},
       eprint = {1405.1737},
 primaryClass = {astro-ph.GA},
       adsurl = {https://ui.adsabs.harvard.edu/abs/2014A&A...569A..52S}
}

@ARTICLE{2024MNRAS.528.2511T,
       author = {{Taylor}, A.~R. and {Sekhar}, S. and {Heino}, L. and {Scaife}, A.~M.~M. and {Stil}, J. and {Bowles}, M. and {Jarvis}, M. and {Heywood}, I. and {Collier}, J.~D.},
        title = "{MIGHTEE polarization early science fields: the deep polarized sky}",
      journal = {\mnras},
         year = 2024,
        month = feb,
       volume = {528},
       number = {2},
        pages = {2511-2522},
          doi = {10.1093/mnras/stae169},
archivePrefix = {arXiv},
       eprint = {2312.13230},
 primaryClass = {astro-ph.GA},
       adsurl = {https://ui.adsabs.harvard.edu/abs/2024MNRAS.528.2511T}
}

@ARTICLE{2015MNRAS.454.3403H,
       author = {{Harwood}, Jeremy J. and {Hardcastle}, Martin J. and {Croston}, Judith H.},
        title = "{Spectral ageing in the lobes of cluster-centre FR II radio galaxies}",
      journal = {\mnras},
         year = 2015,
        month = dec,
       volume = {454},
       number = {4},
        pages = {3403-3422},
          doi = {10.1093/mnras/stv2194},
archivePrefix = {arXiv},
       eprint = {1509.06757},
 primaryClass = {astro-ph.GA},
       adsurl = {https://ui.adsabs.harvard.edu/abs/2015MNRAS.454.3403H}
}

@ARTICLE{2017A&A...606A..98B,
       author = {{Brienza}, M. and {Godfrey}, L. and {Morganti}, R. and {Prandoni}, I. and {Harwood}, J. and {Mahony}, E.~K. and {Hardcastle}, M.~J. and {Murgia}, M. and {R{\"o}ttgering}, H.~J.~A. and {Shimwell}, T.~W. and {Shulevski}, A.},
        title = "{Search and modelling of remnant radio galaxies in the LOFAR Lockman Hole field}",
      journal = {\aap},
         year = 2017,
        month = oct,
       volume = {606},
          eid = {A98},
        pages = {A98},
          doi = {10.1051/0004-6361/201730932},
archivePrefix = {arXiv},
       eprint = {1708.01904},
 primaryClass = {astro-ph.GA},
       adsurl = {https://ui.adsabs.harvard.edu/abs/2017A&A...606A..98B}
}

@ARTICLE{1991ApJ...381...63F,
       author = {{Fernini}, Ilias and {Leahy}, J. Patrick and {Burns}, Jack O. and {Basart}, John P.},
        title = "{Depolarization Asymmetry in the Quasar 3C 47}",
      journal = {\apj},
         year = 1991,
        month = nov,
       volume = {381},
        pages = {63},
          doi = {10.1086/170628},
       adsurl = {https://ui.adsabs.harvard.edu/abs/1991ApJ...381...63F}
}

@ARTICLE{1998PASP..110..493O,
       author = {{O'Dea}, Christopher P.},
        title = "{The Compact Steep-Spectrum and Gigahertz Peaked-Spectrum Radio Sources}",
      journal = {\pasp},
         year = 1998,
        month = may,
       volume = {110},
       number = {747},
        pages = {493-532},
          doi = {10.1086/316162},
       adsurl = {https://ui.adsabs.harvard.edu/abs/1998PASP..110..493O}
}

@ARTICLE{2003MNRAS.342..399G,
       author = {{Gizani}, Nectaria A.~B. and {Leahy}, J.~P.},
        title = "{A multiband study of Hercules A - II. Multifrequency Very Large Array imaging}",
      journal = {\mnras},
         year = 2003,
        month = jun,
       volume = {342},
       number = {2},
        pages = {399-421},
          doi = {10.1046/j.1365-8711.2003.06469.x},
archivePrefix = {arXiv},
       eprint = {astro-ph/0305600},
 primaryClass = {astro-ph},
       adsurl = {https://ui.adsabs.harvard.edu/abs/2003MNRAS.342..399G}
}

@ARTICLE{2017A&A...600A..65S,
       author = {{Shulevski}, A. and {Morganti}, R. and {Harwood}, J.~J. and {Barthel}, P.~D. and {Jamrozy}, M. and {Brienza}, M. and {Brunetti}, G. and {R{\"o}ttgering}, H.~J.~A. and {Murgia}, M. and {White}, G.~J. and {Croston}, J.~H. and {Br{\"u}ggen}, M.},
        title = "{Radiative age mapping of the remnant radio galaxy B2 0924+30: the LOFAR perspective}",
      journal = {\aap},
         year = 2017,
        month = apr,
       volume = {600},
          eid = {A65},
        pages = {A65},
          doi = {10.1051/0004-6361/201630008},
archivePrefix = {arXiv},
       eprint = {1701.06903},
 primaryClass = {astro-ph.GA},
       adsurl = {https://ui.adsabs.harvard.edu/abs/2017A&A...600A..65S}
}

@ARTICLE{2025A&A...696A..97D,
       author = {{Dabhade}, P. and {Chavan}, K. and {Saikia}, D.~J. and {Oei}, M.~S.~S.~L. and {R{\"o}ttgering}, H.~J.~A.},
        title = "{Search and analysis of giant radio galaxies with associated nuclei (SAGAN): V. Study of giant double-double radio galaxies from LoTSS DR2}",
      journal = {\aap},
         year = 2025,
        month = apr,
       volume = {696},
          eid = {A97},
        pages = {A97},
          doi = {10.1051/0004-6361/202451918},
archivePrefix = {arXiv},
       eprint = {2408.13607},
 primaryClass = {astro-ph.GA},
       adsurl = {https://ui.adsabs.harvard.edu/abs/2025A&A...696A..97D}
}

@ARTICLE{2023JApA...44...13D,
       author = {{Dabhade}, Pratik and {Saikia}, D.~J. and {Mahato}, Mousumi},
        title = "{Decoding the giant extragalactic radio sources}",
      journal = {Journal of Astrophysics and Astronomy},
         year = 2023,
        month = jun,
       volume = {44},
       number = {1},
          eid = {13},
        pages = {13},
          doi = {10.1007/s12036-022-09898-5},
archivePrefix = {arXiv},
       eprint = {2208.02130},
 primaryClass = {astro-ph.GA},
       adsurl = {https://ui.adsabs.harvard.edu/abs/2023JApA...44...13D}
}

@ARTICLE{2014MNRAS.442.3192V,
       author = {{Vantyghem}, A.~N. and {McNamara}, B.~R. and {Russell}, H.~R. and {Main}, R.~A. and {Nulsen}, P.~E.~J. and {Wise}, M.~W. and {Hoekstra}, H. and {Gitti}, M.},
        title = "{Cycling of the powerful AGN in MS 0735.6+7421 and the duty cycle of radio AGN in clusters}",
      journal = {\mnras},
         year = 2014,
        month = aug,
       volume = {442},
       number = {4},
        pages = {3192-3205},
          doi = {10.1093/mnras/stu1030},
archivePrefix = {arXiv},
       eprint = {1405.6208},
 primaryClass = {astro-ph.GA},
       adsurl = {https://ui.adsabs.harvard.edu/abs/2014MNRAS.442.3192V}
}

@ARTICLE{2006MNRAS.366.1391S,
       author = {{Saikia}, D.~J. and {Konar}, C. and {Kulkarni}, V.~K.},
        title = "{J0041+3224: a new double-double radio galaxy}",
      journal = {\mnras},
         year = 2006,
        month = mar,
       volume = {366},
       number = {4},
        pages = {1391-1398},
          doi = {10.1111/j.1365-2966.2005.09926.x},
archivePrefix = {arXiv},
       eprint = {astro-ph/0512113},
 primaryClass = {astro-ph},
       adsurl = {https://ui.adsabs.harvard.edu/abs/2006MNRAS.366.1391S}
}

@ARTICLE{2012MNRAS.424.1061K,
       author = {{Konar}, C. and {Hardcastle}, M.~J. and {Jamrozy}, M. and {Croston}, J.~H. and {Nandi}, S.},
        title = "{Rejuvenated radio galaxies J0041+3224 and J1835+6204: how long can the quiescent phase of nuclear activity last?}",
      journal = {\mnras},
         year = 2012,
        month = aug,
       volume = {424},
       number = {2},
        pages = {1061-1076},
          doi = {10.1111/j.1365-2966.2012.21279.x},
       adsurl = {https://ui.adsabs.harvard.edu/abs/2012MNRAS.424.1061K}
}

@ARTICLE{2015A&A...579A..27S,
       author = {{Shulevski}, A. and {Morganti}, R. and {Barthel}, P.~D. and {Murgia}, M. and {van Weeren}, R.~J. and {White}, G.~J. and {Br{\"u}ggen}, M. and {Kunert-Bajraszewska}, M. and {Jamrozy}, M. and {Best}, P.~N. and {R{\"o}ttgering}, H.~J.~A. and {Chyzy}, K.~T. and {de Gasperin}, F. and {B{\^\i}rzan}, L. and {Brunetti}, G. and {Brienza}, M. and {Rafferty}, D.~A. and {Anderson}, J. and {Beck}, R. and {Deller}, A. and {Zarka}, P. and {Schwarz}, D. and {Mahony}, E. and {Orr{\'u}}, E. and {Bell}, M.~E. and {Bentum}, M.~J. and {Bernardi}, G. and {Bonafede}, A. and {Breitling}, F. and {Broderick}, J.~W. and {Butcher}, H.~R. and {Carbone}, D. and {Ciardi}, B. and {de Geus}, E. and {Duscha}, S. and {Eisl{\"o}ffel}, J. and {Engels}, D. and {Falcke}, H. and {Fallows}, R.~A. and {Fender}, R. and {Ferrari}, C. and {Frieswijk}, W. and {Garrett}, M.~A. and {Grie{\ss}meier}, J. and {Gunst}, A.~W. and {Heald}, G. and {Hoeft}, M. and {H{\"o}randel}, J. and {Horneffer}, A. and {van der Horst}, A.~J. and {Intema}, H. and {Juette}, E. and {Karastergiou}, A. and {Kondratiev}, V.~I. and {Kramer}, M. and {Kuniyoshi}, M. and {Kuper}, G. and {Maat}, P. and {Mann}, G. and {McFadden}, R. and {McKay-Bukowski}, D. and {McKean}, J.~P. and {Meulman}, H. and {Mulcahy}, D.~D. and {Munk}, H. and {Norden}, M.~J. and {Paas}, H. and {Pandey-Pommier}, M. and {Pizzo}, R. and {Polatidis}, A.~G. and {Reich}, W. and {Rowlinson}, A. and {Scaife}, A.~M.~M. and {Serylak}, M. and {Sluman}, J. and {Smirnov}, O. and {Steinmetz}, M. and {Swinbank}, J. and {Tagger}, M. and {Tang}, Y. and {Tasse}, C. and {Thoudam}, S. and {Toribio}, M.~C. and {Vermeulen}, R. and {Vocks}, C. and {Wijers}, R.~A.~M.~J. and {Wise}, M.~W. and {Wucknitz}, O.},
        title = "{The peculiar radio galaxy 4C 35.06: a case for recurrent AGN activity?}",
      journal = {\aap},
         year = 2015,
        month = jul,
       volume = {579},
          eid = {A27},
        pages = {A27},
          doi = {10.1051/0004-6361/201425416},
archivePrefix = {arXiv},
       eprint = {1504.06642},
 primaryClass = {astro-ph.GA},
       adsurl = {https://ui.adsabs.harvard.edu/abs/2015A&A...579A..27S}
}

@ARTICLE{2020MNRAS.497.3638W,
       author = {{Walg}, S. and {Achterberg}, A. and {Markoff}, S. and {Keppens}, R. and {Porth}, O.},
        title = "{Relativistic AGN jets - III. Synthesis of synchrotron emission from double-double radio galaxies}",
      journal = {\mnras},
         year = 2020,
        month = sep,
       volume = {497},
       number = {3},
        pages = {3638-3657},
          doi = {10.1093/mnras/staa2195},
archivePrefix = {arXiv},
       eprint = {2007.14815},
 primaryClass = {astro-ph.HE},
       adsurl = {https://ui.adsabs.harvard.edu/abs/2020MNRAS.497.3638W}
}

@ARTICLE{2023Galax..11...74M,
       author = {{Mahatma}, Vijay H.},
        title = "{The Dynamics and Energetics of Remnant and Restarting RLAGN}",
      journal = {Galaxies},
         year = 2023,
        month = jun,
       volume = {11},
       number = {3},
          eid = {74},
        pages = {74},
          doi = {10.3390/galaxies11030074},
archivePrefix = {arXiv},
       eprint = {2306.08471},
 primaryClass = {astro-ph.GA},
       adsurl = {https://ui.adsabs.harvard.edu/abs/2023Galax..11...74M}
}

@ARTICLE{1991ApJ...369..308C,
       author = {{Clarke}, David A. and {Burns}, Jack O.},
        title = "{Numerical Simulations of a Restarting Jet}",
      journal = {\apj},
         year = 1991,
        month = mar,
       volume = {369},
        pages = {308},
          doi = {10.1086/169762},
       adsurl = {https://ui.adsabs.harvard.edu/abs/1991ApJ...369..308C}
}

@ARTICLE{2013MNRAS.433.1453W,
       author = {{Walg}, S. and {Achterberg}, A. and {Markoff}, S. and {Keppens}, R. and {Meliani}, Z.},
        title = "{Relativistic AGN jets I. The delicate interplay between jet structure, cocoon morphology and jet-head propagation}",
      journal = {\mnras},
         year = 2013,
        month = aug,
       volume = {433},
       number = {2},
        pages = {1453-1478},
          doi = {10.1093/mnras/stt823},
archivePrefix = {arXiv},
       eprint = {1305.2157},
 primaryClass = {astro-ph.HE},
       adsurl = {https://ui.adsabs.harvard.edu/abs/2013MNRAS.433.1453W}
}

@ARTICLE{2014MNRAS.439.3969W,
       author = {{Walg}, S. and {Achterberg}, A. and {Markoff}, S. and {Keppens}, R. and {Porth}, O.},
        title = "{Relativistic AGN jets - II. Jet properties and mixing effects for episodic jet activity}",
      journal = {\mnras},
         year = 2014,
        month = apr,
       volume = {439},
       number = {4},
        pages = {3969-3985},
          doi = {10.1093/mnras/stu253},
archivePrefix = {arXiv},
       eprint = {1311.4234},
 primaryClass = {astro-ph.GA},
       adsurl = {https://ui.adsabs.harvard.edu/abs/2014MNRAS.439.3969W}
}

@ARTICLE{2008MNRAS.385.2117S,
       author = {{Safouris}, V. and {Subrahmanyan}, R. and {Bicknell}, G.~V. and {Saripalli}, L.},
        title = "{PKS B1545-321: bow shocks of a relativistic jet?}",
      journal = {\mnras},
         year = 2008,
        month = apr,
       volume = {385},
       number = {4},
        pages = {2117-2135},
          doi = {10.1111/j.1365-2966.2008.12975.x},
archivePrefix = {arXiv},
       eprint = {0802.0538},
 primaryClass = {astro-ph},
       adsurl = {https://ui.adsabs.harvard.edu/abs/2008MNRAS.385.2117S}
}

@ARTICLE{2021A&A...650A.170B,
       author = {{Biava}, Nadia and {Brienza}, Marisa and {Bonafede}, Annalisa and {Gitti}, Myriam and {Bonnassieux}, Etienne and {Harwood}, Jeremy and {Edge}, Alastair C. and {Riseley}, Christopher J. and {Vantyghem}, Adrian},
        title = "{Constraining the AGN duty cycle in the cool-core cluster MS 0735.6+7421 with LOFAR data}",
      journal = {\aap},
         year = 2021,
        month = jun,
       volume = {650},
          eid = {A170},
        pages = {A170},
          doi = {10.1051/0004-6361/202040063},
archivePrefix = {arXiv},
       eprint = {2104.08294},
 primaryClass = {astro-ph.HE},
       adsurl = {https://ui.adsabs.harvard.edu/abs/2021A&A...650A.170B}
}

@ARTICLE{2020MNRAS.499.5719R,
       author = {{Ruffa}, Ilaria and {Laing}, Robert A. and {Prandoni}, Isabella and {Paladino}, Rosita and {Parma}, Paola and {Davis}, Timothy A. and {Bureau}, Martin},
        title = "{The AGN fuelling/feedback cycle in nearby radio galaxies - III. 3D relative orientations of radio jets and CO discs and their interaction}",
      journal = {\mnras},
         year = 2020,
        month = dec,
       volume = {499},
       number = {4},
        pages = {5719-5731},
          doi = {10.1093/mnras/staa3166},
archivePrefix = {arXiv},
       eprint = {2010.04685},
 primaryClass = {astro-ph.GA},
       adsurl = {https://ui.adsabs.harvard.edu/abs/2020MNRAS.499.5719R}
}

@ARTICLE{2009ApJ...702.1230T,
       author = {{Taylor}, A.~R. and {Stil}, J.~M. and {Sunstrum}, C.},
        title = "{A Rotation Measure Image of the Sky}",
      journal = {\apj},
         year = 2009,
        month = sep,
       volume = {702},
       number = {2},
        pages = {1230-1236},
          doi = {10.1088/0004-637X/702/2/1230},
       adsurl = {https://ui.adsabs.harvard.edu/abs/2009ApJ...702.1230T}
}

@BOOK{1970ranp.book.....P,
       author = {{Pacholczyk}, A.~G.},
        title = "{Radio astrophysics. Nonthermal processes in galactic and extragalactic sources}",
         year = 1970,
       adsurl = {https://ui.adsabs.harvard.edu/abs/1970ranp.book.....P}
}

@ARTICLE{2012ApJ...760...77A,
       author = {{An}, Tao and {Baan}, Willem A.},
        title = "{The Dynamic Evolution of Young Extragalactic Radio Sources}",
      journal = {\apj},
         year = 2012,
        month = nov,
       volume = {760},
       number = {1},
          eid = {77},
        pages = {77},
          doi = {10.1088/0004-637X/760/1/77},
archivePrefix = {arXiv},
       eprint = {1211.1760},
 primaryClass = {astro-ph.CO},
       adsurl = {https://ui.adsabs.harvard.edu/abs/2012ApJ...760...77A}
}

@ARTICLE{2009AN....330..120F,
       author = {{Fanti}, C.},
        title = "{Radio properties of CSSs and GPSs}",
      journal = {Astronomische Nachrichten},
         year = 2009,
        month = feb,
       volume = {330},
       number = {2},
        pages = {120-127},
          doi = {10.1002/asna.200811137},
       adsurl = {https://ui.adsabs.harvard.edu/abs/2009AN....330..120F}
}

@ARTICLE{1982ApJ...257..538B,
       author = {{Burns}, J.~O. and {Christiansen}, W.~A. and {Hough}, D.~H.},
        title = "{Multifrequency VLA observations of 3C 388: evidence for an intermittent jet ?}",
      journal = {\apj},
         year = 1982,
        month = jun,
       volume = {257},
        pages = {538-558},
          doi = {10.1086/160010},
       adsurl = {https://ui.adsabs.harvard.edu/abs/1982ApJ...257..538B}
}

@ARTICLE{2016arXiv161205560C,
       author = {{Chambers}, K.~C. and {Magnier}, E.~A. and {Metcalfe}, N. and {Flewelling}, H.~A. and {Huber}, M.~E. and {Waters}, C.~Z. and {Denneau}, L. and {Draper}, P.~W. and {Farrow}, D. and {Finkbeiner}, D.~P. and {Holmberg}, C. and {Koppenhoefer}, J. and {Price}, P.~A. and {Rest}, A. and {Saglia}, R.~P. and {Schlafly}, E.~F. and {Smartt}, S.~J. and {Sweeney}, W. and {Wainscoat}, R.~J. and {Burgett}, W.~S. and {Chastel}, S. and {Grav}, T. and {Heasley}, J.~N. and {Hodapp}, K.~W. and {Jedicke}, R. and {Kaiser}, N. and {Kudritzki}, R. -P. and {Luppino}, G.~A. and {Lupton}, R.~H. and {Monet}, D.~G. and {Morgan}, J.~S. and {Onaka}, P.~M. and {Shiao}, B. and {Stubbs}, C.~W. and {Tonry}, J.~L. and {White}, R. and {Ba{\~n}ados}, E. and {Bell}, E.~F. and {Bender}, R. and {Bernard}, E.~J. and {Boegner}, M. and {Boffi}, F. and {Botticella}, M.~T. and {Calamida}, A. and {Casertano}, S. and {Chen}, W. -P. and {Chen}, X. and {Cole}, S. and {Deacon}, N. and {Frenk}, C. and {Fitzsimmons}, A. and {Gezari}, S. and {Gibbs}, V. and {Goessl}, C. and {Goggia}, T. and {Gourgue}, R. and {Goldman}, B. and {Grant}, P. and {Grebel}, E.~K. and {Hambly}, N.~C. and {Hasinger}, G. and {Heavens}, A.~F. and {Heckman}, T.~M. and {Henderson}, R. and {Henning}, T. and {Holman}, M. and {Hopp}, U. and {Ip}, W. -H. and {Isani}, S. and {Jackson}, M. and {Keyes}, C.~D. and {Koekemoer}, A.~M. and {Kotak}, R. and {Le}, D. and {Liska}, D. and {Long}, K.~S. and {Lucey}, J.~R. and {Liu}, M. and {Martin}, N.~F. and {Masci}, G. and {McLean}, B. and {Mindel}, E. and {Misra}, P. and {Morganson}, E. and {Murphy}, D.~N.~A. and {Obaika}, A. and {Narayan}, G. and {Nieto-Santisteban}, M.~A. and {Norberg}, P. and {Peacock}, J.~A. and {Pier}, E.~A. and {Postman}, M. and {Primak}, N. and {Rae}, C. and {Rai}, A. and {Riess}, A. and {Riffeser}, A. and {Rix}, H.~W. and {R{\"o}ser}, S. and {Russel}, R. and {Rutz}, L. and {Schilbach}, E. and {Schultz}, A.~S.~B. and {Scolnic}, D. and {Strolger}, L. and {Szalay}, A. and {Seitz}, S. and {Small}, E. and {Smith}, K.~W. and {Soderblom}, D.~R. and {Taylor}, P. and {Thomson}, R. and {Taylor}, A.~N. and {Thakar}, A.~R. and {Thiel}, J. and {Thilker}, D. and {Unger}, D. and {Urata}, Y. and {Valenti}, J. and {Wagner}, J. and {Walder}, T. and {Walter}, F. and {Watters}, S.~P. and {Werner}, S. and {Wood-Vasey}, W.~M. and {Wyse}, R.},
        title = "{The Pan-STARRS1 Surveys}",
      journal = {arXiv e-prints},
         year = 2016,
        month = dec,
          eid = {arXiv:1612.05560},
        pages = {arXiv:1612.05560},
          doi = {10.48550/arXiv.1612.05560},
archivePrefix = {arXiv},
       eprint = {1612.05560},
 primaryClass = {astro-ph.IM},
       adsurl = {https://ui.adsabs.harvard.edu/abs/2016arXiv161205560C}
}

@ARTICLE{2025A&A...696A.239B,
       author = {{Brienza}, M. and {Rajpurohit}, K. and {Churazov}, E. and {Heywood}, I. and {Br{\"u}ggen}, M. and {Hoeft}, M. and {Vazza}, F. and {Bonafede}, A. and {Botteon}, A. and {Brunetti}, G. and {Gastaldello}, F. and {Khabibullin}, I. and {Lyskova}, N. and {Majumder}, A. and {R{\"o}ttgering}, H.~J.~A. and {Shimwell}, T.~W. and {Simionescu}, A. and {van Weeren}, R.~J.},
        title = "{Non-thermal filaments and AGN recurrent activity in the galaxy group Nest200047: A LOFAR, uGMRT, MeerKAT, and VLA radio spectral analysis}",
      journal = {\aap},
         year = 2025,
        month = apr,
       volume = {696},
          eid = {A239},
        pages = {A239},
          doi = {10.1051/0004-6361/202553676},
archivePrefix = {arXiv},
       eprint = {2502.18244},
 primaryClass = {astro-ph.GA},
       adsurl = {https://ui.adsabs.harvard.edu/abs/2025A&A...696A.239B}
}

@ARTICLE{2020PASA...37...18W,
       author = {{White}, Sarah V. and {Franzen}, Thomas M.~O. and {Riseley}, Chris J. and {Wong}, O. Ivy and {Kapi{\'n}ska}, Anna D. and {Hurley-Walker}, Natasha and {Callingham}, Joseph R. and {Thorat}, Kshitij and {Wu}, Chen and {Hancock}, Paul and {Hunstead}, Richard W. and {Seymour}, Nick and {Swan}, Jesse and {Wayth}, Randall and {Morgan}, John and {Chhetri}, Rajan and {Jackson}, Carole and {Weston}, Stuart and {Bell}, Martin and {For}, Bi-Qing and {Gaensler}, B.~M. and {Johnston-Hollitt}, Melanie and {Offringa}, Andr{\'e} and {Staveley-Smith}, Lister},
        title = "{The GLEAM 4-Jy (G4Jy) Sample: I. Definition and the catalogue}",
      journal = {\pasa},
         year = 2020,
        month = jun,
       volume = {37},
          eid = {e018},
        pages = {e018},
          doi = {10.1017/pasa.2020.9},
archivePrefix = {arXiv},
       eprint = {2004.13125},
 primaryClass = {astro-ph.GA},
       adsurl = {https://ui.adsabs.harvard.edu/abs/2020PASA...37...18W}
}

@ARTICLE{2020PASA...37...17W,
       author = {{White}, Sarah V. and {Franzen}, Thomas M.~O. and {Riseley}, Chris J. and {Wong}, O. Ivy and {Kapi{\'n}ska}, Anna D. and {Hurley-Walker}, Natasha and {Callingham}, Joseph R. and {Thorat}, Kshitij and {Wu}, Chen and {Hancock}, Paul and {Hunstead}, Richard W. and {Seymour}, Nick and {Swan}, Jesse and {Wayth}, Randall and {Morgan}, John and {Chhetri}, Rajan and {Jackson}, Carole and {Weston}, Stuart and {Bell}, Martin and {Gaensler}, B.~M. and {Johnston-Hollitt}, Melanie and {Offringa}, Andr{\'e} and {Staveley-Smith}, Lister},
        title = "{The GLEAM 4-Jy (G4Jy) Sample: II. Host galaxy identification for individual sources}",
      journal = {\pasa},
         year = 2020,
        month = jun,
       volume = {37},
          eid = {e017},
        pages = {e017},
          doi = {10.1017/pasa.2020.10},
archivePrefix = {arXiv},
       eprint = {2004.13025},
 primaryClass = {astro-ph.GA},
       adsurl = {https://ui.adsabs.harvard.edu/abs/2020PASA...37...17W}
}

@ARTICLE{2023ApJ...944..176D,
       author = {{Dutta}, Sushant and {Singh}, Veeresh and {Chandra}, C.~H. Ishwara and {Wadadekar}, Yogesh and {Kayal}, Abhijit and {Heywood}, Ian},
        title = "{Search and Characterization of Remnant Radio Galaxies in the XMM-LSS Deep Field}",
      journal = {\apj},
         year = 2023,
        month = feb,
       volume = {944},
       number = {2},
          eid = {176},
        pages = {176},
          doi = {10.3847/1538-4357/acaf01},
archivePrefix = {arXiv},
       eprint = {2212.10133},
 primaryClass = {astro-ph.GA},
       adsurl = {https://ui.adsabs.harvard.edu/abs/2023ApJ...944..176D}
}

@ARTICLE{2025PASA...42..124N,
       author = {{Norris}, Ray P. and {Yew}, Miranda and {Crawford}, Evan J. and {Gupta}, Nikhel and {Rudnick}, Lawrence and {Andernach}, Heinz and {Filipovi{\'c}}, Miroslav D. and {Gordon}, Yjan and {Hopkins}, Andrew and {Park}, Laurence and {Brown}, Michael and {Jimenez Gallardo}, Ana Maria and {Shabala}, Stanislav},
        title = "{EMU and the DRAGNs I: A catalogue of DRAGNs}",
      journal = {\pasa},
         year = 2025,
        month = sep,
       volume = {42},
          eid = {e124},
        pages = {e124},
          doi = {10.1017/pasa.2025.10076},
archivePrefix = {arXiv},
       eprint = {2507.23337},
 primaryClass = {astro-ph.GA},
       adsurl = {https://ui.adsabs.harvard.edu/abs/2025PASA...42..124N}
}

@dataset{vaccari_2023_8192777,
  author       = {Vaccari, Mattia},
  title        = {The Spitzer Data Fusion},
  month        = jul,
  year         = 2023,
  publisher    = {Zenodo},
  version      = {DR1},
  doi          = {10.5281/zenodo.8192777},
  url          = {https://doi.org/10.5281/zenodo.8192777},
}

@dataset{vaccari_2025_15176245,
  author       = {Vaccari, Mattia},
  title        = {The Spitzer Spectroscopic Data Fusion - Merged
                   Spectroscopic Redshift Catalogs in Spitzer Fields
                  },
  month        = mar,
  year         = 2025,
  publisher    = {Zenodo},
  version      = {20 March 2025},
  doi          = {10.5281/zenodo.15176245},
  url          = {https://doi.org/10.5281/zenodo.15176245},
}

@INPROCEEDINGS{2009pra..confE...4J,
       author = {{Jonas}, J.},
        title = "{The MeerKAT SKA precursor telescope}",
    booktitle = {Panoramic Radio Astronomy: Wide-field 1-2 GHz Research on Galaxy Evolution},
         year = 2009,
        month = jan,
          eid = {4},
        pages = {4},
       adsurl = {https://ui.adsabs.harvard.edu/abs/2009pra..confE...4J}
}

@ARTICLE{2015ApJ...809..168C,
       author = {{Callingham}, J.~R. and {Gaensler}, B.~M. and {Ekers}, R.~D. and {Tingay}, S.~J. and {Wayth}, R.~B. and {Morgan}, J. and {Bernardi}, G. and {Bell}, M.~E. and {Bhat}, R. and {Bowman}, J.~D. and {Briggs}, F. and {Cappallo}, R.~J. and {Deshpande}, A.~A. and {Ewall-Wice}, A. and {Feng}, L. and {Greenhill}, L.~J. and {Hazelton}, B.~J. and {Hindson}, L. and {Hurley-Walker}, N. and {Jacobs}, D.~C. and {Johnston-Hollitt}, M. and {Kaplan}, D.~L. and {Kudrayvtseva}, N. and {Lenc}, E. and {Lonsdale}, C.~J. and {McKinley}, B. and {McWhirter}, S.~R. and {Mitchell}, D.~A. and {Morales}, M.~F. and {Morgan}, E. and {Oberoi}, D. and {Offringa}, A.~R. and {Ord}, S.~M. and {Pindor}, B. and {Prabu}, T. and {Procopio}, P. and {Riding}, J. and {Srivani}, K.~S. and {Subrahmanyan}, R. and {Udaya Shankar}, N. and {Webster}, R.~L. and {Williams}, A. and {Williams}, C.~L.},
        title = "{Broadband Spectral Modeling of the Extreme Gigahertz-peaked Spectrum Radio Source PKS B0008-421}",
      journal = {\apj},
         year = 2015,
        month = aug,
       volume = {809},
       number = {2},
          eid = {168},
        pages = {168},
          doi = {10.1088/0004-637X/809/2/168},
archivePrefix = {arXiv},
       eprint = {1507.04819},
 primaryClass = {astro-ph.GA},
       adsurl = {https://ui.adsabs.harvard.edu/abs/2015ApJ...809..168C}
}

@ARTICLE{2015AJ....149...74T,
       author = {{Tingay}, S.~J. and {Macquart}, J. -P. and {Collier}, J.~D. and {Rees}, G. and {Callingham}, J.~R. and {Stevens}, J. and {Carretti}, E. and {Wayth}, R.~B. and {Wong}, G.~F. and {Trott}, C.~M. and {McKinley}, B. and {Bernardi}, G. and {Bowman}, J.~D. and {Briggs}, F. and {Cappallo}, R.~J. and {Corey}, B.~E. and {Deshpande}, A.~A. and {Emrich}, D. and {Gaensler}, B.~M. and {Goeke}, R. and {Greenhill}, L.~J. and {Hazelton}, B.~J. and {Johnston-Hollitt}, M. and {Kaplan}, D.~L. and {Kasper}, J.~C. and {Kratzenberg}, E. and {Lonsdale}, C.~J. and {Lynch}, M.~J. and {McWhirter}, S.~R. and {Mitchell}, D.~A. and {Morales}, M.~F. and {Morgan}, E. and {Oberoi}, D. and {Ord}, S.~M. and {Prabu}, T. and {Rogers}, A.~E.~E. and {Roshi}, A. and {Udaya Shankar}, N. and {Srivani}, K.~S. and {Subrahmanyan}, R. and {Waterson}, M. and {Webster}, R.~L. and {Whitney}, A.~R. and {Williams}, A. and {Williams}, C.~L.},
        title = "{The Spectral Variability of the GHz-Peaked Spectrum Radio Source PKS 1718-649 and a Comparison of Absorption Models}",
      journal = {\aj},
         year = 2015,
        month = feb,
       volume = {149},
       number = {2},
          eid = {74},
        pages = {74},
          doi = {10.1088/0004-6256/149/2/74},
archivePrefix = {arXiv},
       eprint = {1412.4216},
 primaryClass = {astro-ph.GA},
       adsurl = {https://ui.adsabs.harvard.edu/abs/2015AJ....149...74T}
}

@ARTICLE{2018MNRAS.477..578C,
       author = {{Collier}, J.~D. and {Tingay}, S.~J. and {Callingham}, J.~R. and {Norris}, R.~P. and {Filipovi{\'c}}, M.~D. and {Galvin}, T.~J. and {Huynh}, M.~T. and {Intema}, H.~T. and {Marvil}, J. and {O'Brien}, A.~N. and {Roper}, Q. and {Sirothia}, S. and {Tothill}, N.~F.~H. and {Bell}, M.~E. and {For}, B.-Q. and {Gaensler}, B.~M. and {Hancock}, P.~J. and {Hindson}, L. and {Hurley-Walker}, N. and {Johnston-Hollitt}, M. and {Kapi{\'n}ska}, A.~D. and {Lenc}, E. and {Morgan}, J. and {Procopio}, P. and {Staveley-Smith}, L. and {Wayth}, R.~B. and {Wu}, C. and {Zheng}, Q. and {Heywood}, I. and {Popping}, A.},
        title = "{High-resolution observations of low-luminosity gigahertz-peaked spectrum and compact steep-spectrum sources}",
      journal = {\mnras},
         year = 2018,
        month = jun,
       volume = {477},
       number = {1},
        pages = {578-592},
          doi = {10.1093/mnras/sty564},
archivePrefix = {arXiv},
       eprint = {1802.09666},
 primaryClass = {astro-ph.GA},
       adsurl = {https://ui.adsabs.harvard.edu/abs/2018MNRAS.477..578C}
}

@INPROCEEDINGS{2008ASPC..386...46H,
       author = {{Hardcastle}, M.~J.},
        title = "{Hotspots, Jets and Environments}",
    booktitle = {Extragalactic Jets: Theory and Observation from Radio to Gamma Ray},
         year = 2008,
       editor = {{Rector}, T.~A. and {De Young}, D.~S.},
       series = {Astronomical Society of the Pacific Conference Series},
       volume = {386},
        month = jun,
        pages = {46},
          doi = {10.48550/arXiv.0707.1866},
archivePrefix = {arXiv},
       eprint = {0707.1866},
 primaryClass = {astro-ph},
       adsurl = {https://ui.adsabs.harvard.edu/abs/2008ASPC..386...46H}
}

@ARTICLE{1994AJ....108..766B,
       author = {{Bridle}, Alan H. and {Hough}, David H. and {Lonsdale}, Colin J. and {Burns}, Jack O. and {Laing}, Robert A.},
        title = "{Deep VLA Imaging of Twelve Extended 3CR Quasars}",
      journal = {\aj},
         year = 1994,
        month = sep,
       volume = {108},
        pages = {766},
          doi = {10.1086/117112},
       adsurl = {https://ui.adsabs.harvard.edu/abs/1994AJ....108..766B}
}

@ARTICLE{1997MNRAS.291...20L,
       author = {{Leahy}, J.~P. and {Black}, A.~R.~S. and {Dennett-Thorpe}, J. and {Hardcastle}, M.~J. and {Komissarov}, S. and {Perley}, R.~A. and {Riley}, J.~M. and {Scheuer}, P.~A.~G.},
        title = "{A study of FRII radio galaxies with z<0.15 - II. High-resolution maps of 11 sources at 3.6 CM}",
      journal = {\mnras},
         year = 1997,
        month = oct,
       volume = {291},
       number = {1},
        pages = {20-53},
          doi = {10.1093/mnras/291.1.20},
       adsurl = {https://ui.adsabs.harvard.edu/abs/1997MNRAS.291...20L}
}

@ARTICLE{2000A&A...363..507G,
       author = {{Gopal-Krishna} and {Wiita}, P.~J.},
        title = "{Extragalactic radio sources with hybrid morphology: implications for the Fanaroff-Riley dichotomy}",
      journal = {\aap},
         year = 2000,
        month = nov,
       volume = {363},
        pages = {507-516},
          doi = {10.48550/arXiv.astro-ph/0009441},
archivePrefix = {arXiv},
       eprint = {astro-ph/0009441},
 primaryClass = {astro-ph},
       adsurl = {https://ui.adsabs.harvard.edu/abs/2000A&A...363..507G}
}

@ARTICLE{2020MNRAS.491..803H,
       author = {{Harwood}, Jeremy J. and {Vernstrom}, Tessa and {Stroe}, Andra},
        title = "{Unveiling the cause of hybrid morphology radio sources (HyMoRS)}",
      journal = {\mnras},
         year = 2020,
        month = jan,
       volume = {491},
       number = {1},
        pages = {803-822},
          doi = {10.1093/mnras/stz3069},
archivePrefix = {arXiv},
       eprint = {1910.12857},
 primaryClass = {astro-ph.GA},
       adsurl = {https://ui.adsabs.harvard.edu/abs/2020MNRAS.491..803H}
}

@ARTICLE{2025arXiv250314745D,
       author = {{DESI Collaboration} and {Abdul-Karim}, M. and {Adame}, A.~G. and {Aguado}, D. and {Aguilar}, J. and {Ahlen}, S. and {Alam}, S. and {Aldering}, G. and {Alexander}, D.~M. and {Alfarsy}, R. and {Allen}, L. and {Allende Prieto}, C. and {Alves}, O. and {Anand}, A. and {Andrade}, U. and {Armengaud}, E. and {Avila}, S. and {Aviles}, A. and {Awan}, H. and {Bailey}, S. and {Baleato Lizancos}, A. and {Ballester}, O. and {Bault}, A. and {Bautista}, J. and {BenZvi}, S. and {Beraldo e Silva}, L. and {Bermejo-Climent}, J.~R. and {Beutler}, F. and {Bianchi}, D. and {Blake}, C. and {Blum}, R. and {Bolton}, A.~S. and {Bonici}, M. and {Brieden}, S. and {Brodzeller}, A. and {Brooks}, D. and {Buckley-Geer}, E. and {Burtin}, E. and {Canning}, R. and {Carnero Rosell}, A. and {Carr}, A. and {Carrilho}, P. and {Casas}, L. and {Castander}, F.~J. and {Cereskaite}, R. and {Cervantes-Cota}, J.~L. and {Chaussidon}, E. and {Chaves-Montero}, J. and {Chen}, S. and {Chen}, X. and {Claybaugh}, T. and {Cole}, S. and {Cooper}, A.~P. and {Cousinou}, M.-C. and {Cuceu}, A. and {Davis}, T.~M. and {Dawson}, K.~S. and {de Belsunce}, R. and {de la Cruz}, R. and {de la Macorra}, A. and {de Mattia}, A. and {Deiosso}, N. and {Della Costa}, J. and {Demina}, R. and {Demirbozan}, U. and {DeRose}, J. and {Dey}, A. and {Dey}, B. and {Ding}, J. and {Ding}, Z. and {Doel}, P. and {Douglass}, K. and {Dowicz}, M. and {Ebina}, H. and {Edelstein}, J. and {Eisenstein}, D.~J. and {Elbers}, W. and {Emas}, N. and {Escoffier}, S. and {Fagrelius}, P. and {Fan}, X. and {Fanning}, K. and {Fawcett}, V.~A. and {Fern\textbackslash'andez-Garc\textbackslash'ia}, E. and {Ferraro}, S. and {Findlay}, N. and {Font-Ribera}, A. and {Forero-Romero}, J.~E. and {Forero-S\textbackslash'anchez}, D. and {Frenk}, C.~S. and {G\textbackslash''ansicke}, B.~T. and {Galbany}, L. and {Garc\textbackslash'ia-Bellido}, J. and {Garcia-Quintero}, C. and {Garrison}, L.~H. and {Gazta\textbackslash\raisebox{-0.5ex}\textasciitildenaga}, E. and {Gil-Mar\textbackslash'in}, H. and {Gnedin}, O.~Y. and {Gontcho}, S. Gontcho A and {Gonzalez-Morales}, A.~X. and {Gonzalez-Perez}, V. and {Gordon}, C. and {Graur}, O. and {Green}, D. and {Gruen}, D. and {Gsponer}, R. and {Guandalin}, C. and {Gutierrez}, G. and {Guy}, J. and {Hahn}, C. and {Han}, J.~J. and {Han}, J. and {He}, S. and {Herrera-Alcantar}, H.~K. and {Honscheid}, K. and {Hou}, J. and {Howlett}, C. and {Huterer}, D. and {Ir\textbackslashv\{s\}i\textbackslashv\{c\}}, V. and {Ishak}, M. and {Jacques}, A. and {Jimenez}, J. and {Jing}, Y.~P. and {Joachimi}, B. and {Joudaki}, S. and {Joyce}, R. and {Jullo}, E. and {Juneau}, S. and {Kara\textbackslashc\{c\}ayl\{\textbackslashi\}}, N.~G. and {Karim}, T. and {Kehoe}, R. and {Kent}, S. and {Khederlarian}, A. and {Kirkby}, D. and {Kisner}, T. and {Kitaura}, F.-S. and {Kizhuprakkat}, N. and {Kong}, H. and {Koposov}, S.~E. and {Kremin}, A. and {Krolewski}, A. and {Lahav}, O. and {Lai}, Y. and {Lamman}, C. and {Lan}, T.-W. and {Landriau}, M. and {Lang}, D. and {Lange}, J.~U. and {Lasker}, J. and {Le Goff}, J.~M. and {Le Guillou}, L. and {Leauthaud}, A. and {Levi}, M.~E. and {Li}, S. and {Li}, T.~S. and {Lodha}, K. and {Lokken}, M. and {Luo}, Y. and {Magneville}, C. and {Manera}, M. and {Manser}, C.~J. and {Margala}, D. and {Martini}, P. and {Maus}, M. and {McCullough}, J. and {McDonald}, P. and {Medina}, G.~E. and {Medina-Varela}, L. and {Meisner}, A. and {Mena-Fern\textbackslash'andez}, J. and {Menegas}, A. and {Mezcua}, M. and {Miquel}, R. and {Montero-Camacho}, P. and {Moon}, J. and {Moustakas}, J. and {Mu\textbackslash\raisebox{-0.5ex}\textasciitildenoz-Guti\textbackslash'errez}, A. and {Mu\textbackslash\raisebox{-0.5ex}\textasciitildenoz-Santos}, D. and {Myers}, A.~D. and {Myles}, J. and {Nadathur}, S. and {Najita}, J. and {Napolitano}, L. and {Newman}, J.~A. and {Nikakhtar}, F. and {Nikutta}, R. and {Niz}, G. and {Noriega}, H.~E. and {Padmanabhan}, N. and {Paillas}, E. and {Palanque-Delabrouille}, N. and {Palmese}, A. and {Pan}, J. and {Pan}, Z. and {Parkinson}, D. and {Peacock}, J. and {Percival}, W.~J. and {P\textbackslash'erez-Fern\textbackslash'andez}, A. and {P\textbackslash'erez-R\textbackslash`afols}, I. and {Peterson}, P.},
        title = "{Data Release 1 of the Dark Energy Spectroscopic Instrument}",
      journal = {arXiv e-prints},
         year = 2025,
        month = mar,
          eid = {arXiv:2503.14745},
        pages = {arXiv:2503.14745},
          doi = {10.48550/arXiv.2503.14745},
archivePrefix = {arXiv},
       eprint = {2503.14745},
 primaryClass = {astro-ph.CO},
       adsurl = {https://ui.adsabs.harvard.edu/abs/2025arXiv250314745D}
}

@ARTICLE{2021A&ARv..29....3O,
       author = {{O'Dea}, Christopher P. and {Saikia}, D.~J.},
        title = "{Compact steep-spectrum and peaked-spectrum radio sources}",
      journal = {\aapr},
         year = 2021,
        month = dec,
       volume = {29},
       number = {1},
          eid = {3},
        pages = {3},
          doi = {10.1007/s00159-021-00131-w},
archivePrefix = {arXiv},
       eprint = {2009.02750},
 primaryClass = {astro-ph.GA},
       adsurl = {https://ui.adsabs.harvard.edu/abs/2021A&ARv..29....3O}
}

@ARTICLE{1995A&A...295..629S,
       author = {{Sanghera}, H.~S. and {Saikia}, D.~J. and {Luedke}, E. and {Spencer}, R.~E. and {Foulsham}, P.~A. and {Akujor}, C.~E. and {Tzioumis}, A.~K.},
        title = "{High-resolution radio observations of compact steep spectrum sources.}",
      journal = {\aap},
         year = 1995,
        month = mar,
       volume = {295},
        pages = {629},
       adsurl = {https://ui.adsabs.harvard.edu/abs/1995A&A...295..629S}
}

@ARTICLE{2005A&A...441...89G,
       author = {{Giroletti}, M. and {Giovannini}, G. and {Taylor}, G.~B.},
        title = "{Low power compact radio galaxies at high angular resolution}",
      journal = {\aap},
         year = 2005,
        month = oct,
       volume = {441},
       number = {1},
        pages = {89-101},
          doi = {10.1051/0004-6361:20053347},
archivePrefix = {arXiv},
       eprint = {astro-ph/0506497},
 primaryClass = {astro-ph},
       adsurl = {https://ui.adsabs.harvard.edu/abs/2005A&A...441...89G}
}

@ARTICLE{2012A&A...545A..91S,
       author = {{Shulevski}, A. and {Morganti}, R. and {Oosterloo}, T. and {Struve}, C.},
        title = "{Recurrent radio emission and gas supply: the radio galaxy B2 0258+35}",
      journal = {\aap},
         year = 2012,
        month = sep,
       volume = {545},
          eid = {A91},
        pages = {A91},
          doi = {10.1051/0004-6361/201219869},
archivePrefix = {arXiv},
       eprint = {1207.4348},
 primaryClass = {astro-ph.CO},
       adsurl = {https://ui.adsabs.harvard.edu/abs/2012A&A...545A..91S}
}
%\begin{thebibliography}{99}
%\end{thebibliography}
%%%%%%%%%%%%%%%%%%%%%%%%%%%%%%%%%%%%%%%%%%%%%%%%%%
%%%%%%%%%%%%%%%%% APPENDICES %%%%%%%%%%%%%%%%%%%%%
\appendix
\section{Additional information on J022248-060934}
\begin{figure*}
		\centering
		\begin{tabular}{c}
			{\resizebox{1\hsize}{!}{\includegraphics[trim={0.5cm 0.1cm 0.1cm 0.2cm},clip]{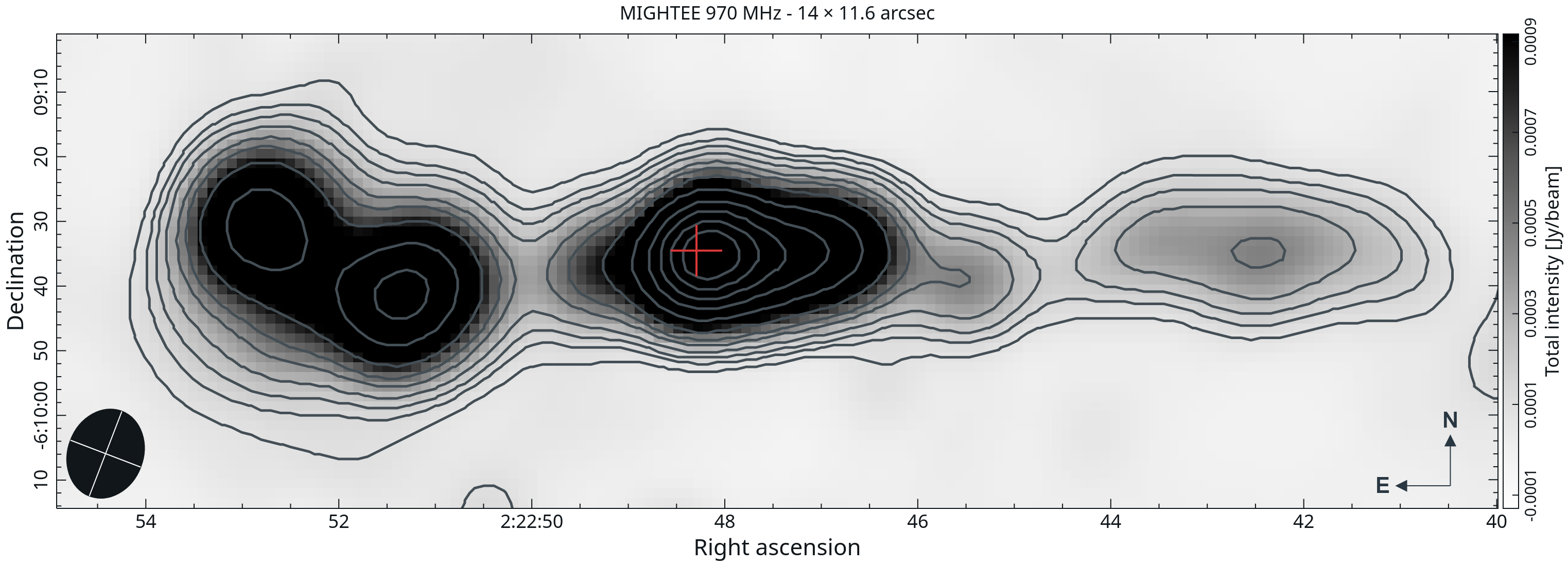}}}\\
            \\
            \vspace{0.4cm}
            {\resizebox{1\hsize}{!}{\includegraphics[trim={0.5cm 0.1cm 0.1cm 0.2cm},clip]{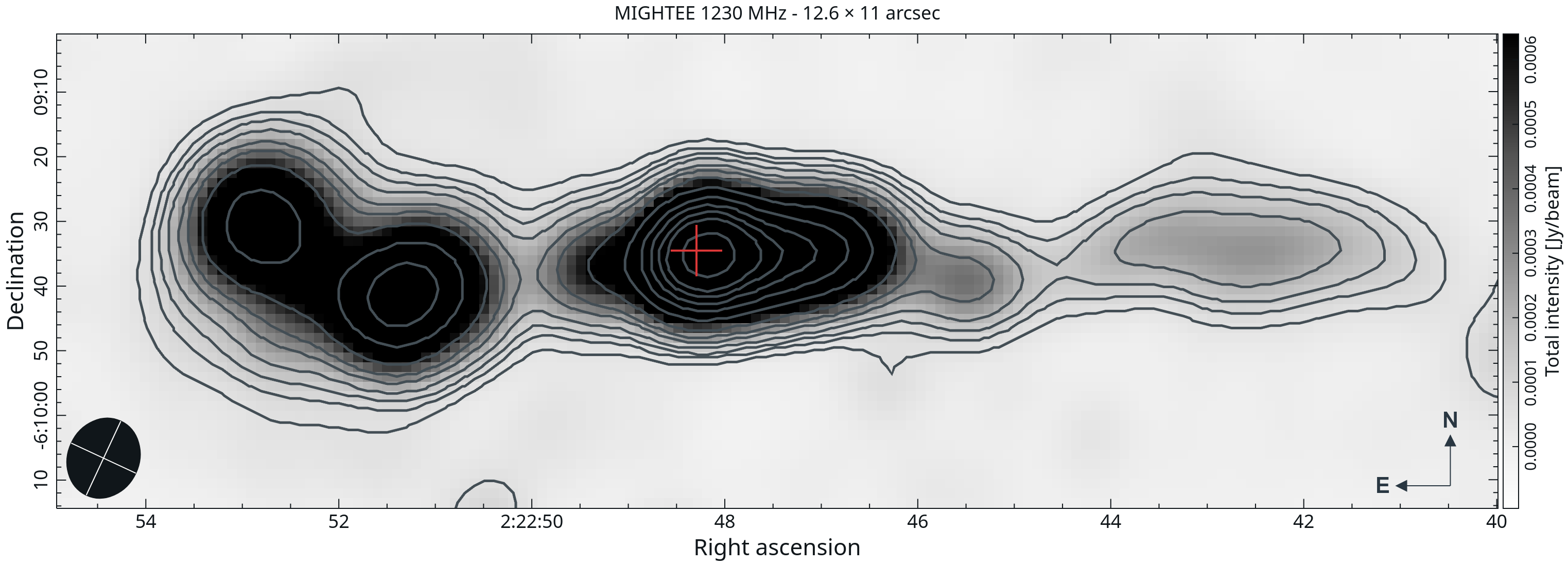}}}\\
            %\vspace{0.5cm}
            {\resizebox{1\hsize}{!}{\includegraphics[trim={0.5cm 0.1cm 0.1cm 0.2cm},clip]{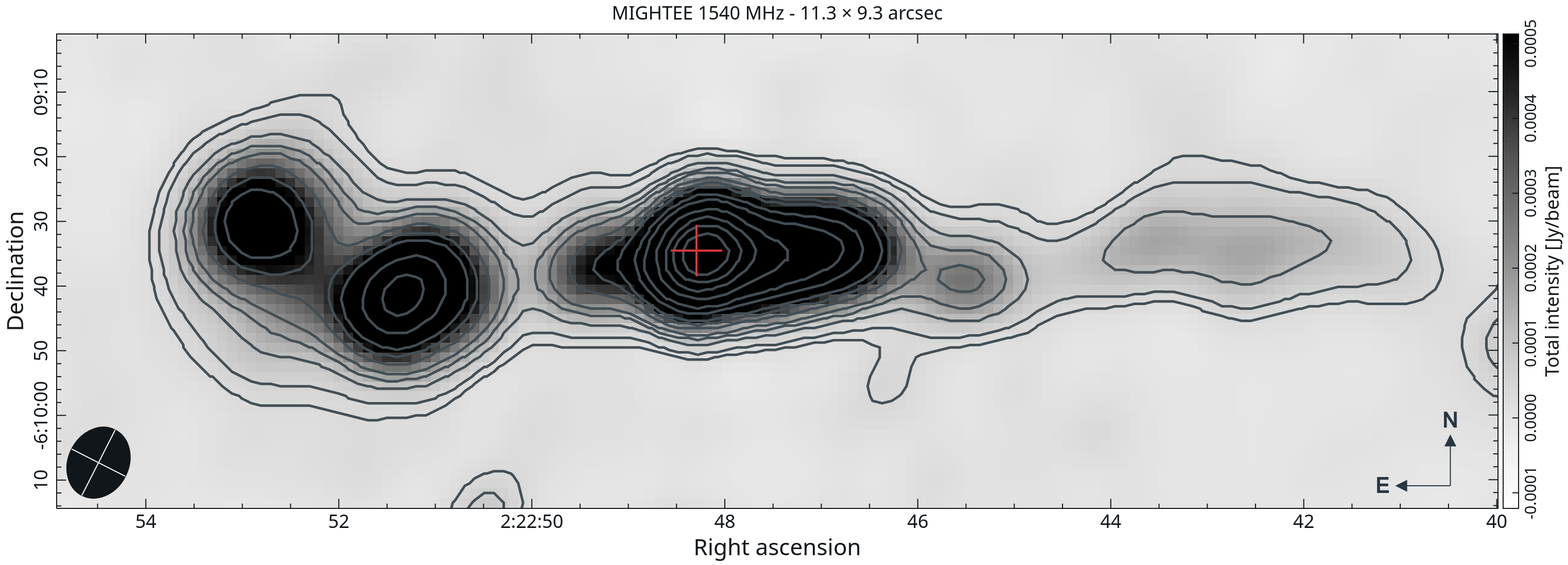}}}\\
		\end{tabular}
		\caption{ Radio morphologies of the TDRG shown in MIGHTEE sub-band images. \textit{Top panel:} MIGHTEE 970 MHz  (contour levels: $-$3, 3, 5, 7, 10, 15, 20, 30, 50, 75, 100, 125, 150 $\times \sigma_{local}$ = 28.4 $\mu$Jy/beam); \textit{ Middle panel:} MIGHTEE 1230 MHz (contour levels: $-$3, 3, 5, 7, 10, 15, 20, 30, 50, 75, 100, 125, 150, 200 $\times \sigma_{local}$ = 19.1 $\mu$Jy/beam); \textit{Bottom panel:} MIGHTEE 1540 MHz (contour levels: $-$3, 3, 5, 10, 15, 20, 30, 40, 50, 75, 100, 150, 200, 250, 300 $\times \sigma_{local}$ = 11.1 $\mu$Jy/beam). The beams of the images are shown in the bottom left corner. The red cross marks the host location.}  
		
		\label{fig: source_morpho_appendix}
	\end{figure*}

The MIGHTEE sub-band radio images of the TDRG are shown in Figure \ref{fig: source_morpho_appendix}. The respective resolutions are 14$\times$11.6 arcsec, 12.6$\times$11 arcsec and 11.3$\times$9.3 arcsec at 970, 1230 and 1540 MHz. Although at lower resolution compared to the 5 arcsec MIGHTEE-hi, these images still depict the structures of the source.

The archival radio images of the source from the complementary data are shown in Figure \ref{fig: source_morpho_archival}. From the top to the bottom panel, we display images at 150 MHz from LOFAR, at 325 MHz from GMRT Legacy and from VLASS at 3 GHz. The source is only resolved into a DDRG at low frequencies while the core is the only component recovered at high frequency.

Table \ref{tab: flux_appendix} lists additional flux densities of the source. The measurements were taken from the radio images in Figure \ref{fig: source_morpho_appendix} and Figure \ref{fig: source_morpho_archival}.

      \begin{table}  
 \caption{Flux densities of the components of the source at 150 MHz from LOFAR (2), 325 MHz from GMRT (3) and from MIGHTEE sub-band images respectively at 970, 1230 and 1540 MHz (4 - 6) and  at 3000 MHz from VLASS (7).}
 
 \label{tab: flux_appendix}
		\centering
        {\resizebox{0.9\hsize}{!}{
		\begin{tabular}{lccc}
  
	\hline
 Components & $S_{150}$  & $S_{325}$ & $S_{970}$\\%& $S_{1230}$  & $S_{1540}$ & $S_{3000}$ \\
    (1)& (2)& (3) & (4)\\
    & [mJy]& [mJy] & [mJy]\\
    \hline
    \hline
    
    Core & 10.53$\pm$1.05 & 11.90$\pm$1.19& 5.72$\pm$0.57\\
    Core$_{peak}$ & 6.96$\pm0.69$ & 8.72$\pm$0.87& 5.64$\pm$0.56\\
    Lobe $_{out (E)}$ & 53.85$\pm$5.38& 23.07 $\pm$2.30& 4.71$\pm$0.47 \\
    Lobe $_{out (W)}$ & 10.76$\pm$1.07& 6.30$\pm$ 0.63& 1.72$\pm$0.17 \\
    Lobe $_{ mid (E)}$ & 31.11$\pm$3.11& 17.93$\pm$ 1.79& 4.86$\pm$0.48\\
    Lobe $_{mid (W)}$ & $-$ & $-$& 0.59$\pm$0.05\\
    Lobe $_{in (E)}$ & $-$ & $-$& 0.87$\pm$ 0.08\\ 
    Lobe $_{in (W)}$ &8.66$\pm$0.86 & 7.04$\pm$0.70& 3.29$\pm$0.32\\
    \hline
    \hline
    Components & $S_{1230}$  & $S_{1540}$ & $S_{3000}$ \\
    (1)& (5)&(6)&(7)\\
    &[mJy]&[mJy]& [mJy]\\
    \hline
    \hline
   Core & 5.59$\pm$ 0.55& 5.53$\pm$0.55 &3.29$\pm$0.25\\
   Core$_{peak}$ & 5.18$\pm$0.51& 4.78$\pm$0.47&3.20$\pm$0.32 \\
   Lobe $_{out (E)}$ & 3.35$\pm$0.33& 2.31$\pm$0.23&$-$\\
    Lobe $_{out (W)}$ & 1.15$\pm$0.11& 0.81$\pm$0.08&$-$\\
    Lobe $_{ mid (E)}$ & 3.87$\pm$0.38& 3.02$\pm$0.30&$-$\\
    Lobe $_{mid (W)}$& 0.49$\pm$ 0.04& 0.40$\pm$0.04&$-$\\
    Lobe $_{in (E)}$ &0.75$\pm$ 0.07& 0.65$\pm$0.06& $-$\\
     Lobe $_{in (W)}$ & 2.89$\pm$0.28& 2.50$\pm$0.25 &$-$\\
     \hline
		\end{tabular}
        
        }}
	\end{table}

Figure \ref{fig: alphamap_appendix} shows additional spectral index maps of the source. These maps show the spectral features across the source from 150 to 1540 MHz but computed between narrower wavelength. 

Table \ref{tab: Tribble} summarises the age fitting results using the Tribble model. In general, these results are consistent, within the margin of errors, with the ones using the JP model which are presented in Section \ref{sec: results_agemap} and discussed in Section \ref{subsec: disc_duty_cycle}. 

 \begin{figure*}
		\centering
		\begin{tabular}{c}
{\resizebox{1\hsize}{!}{\includegraphics[trim={0.5cm 0.2cm 0.2cm 0.1cm},clip]{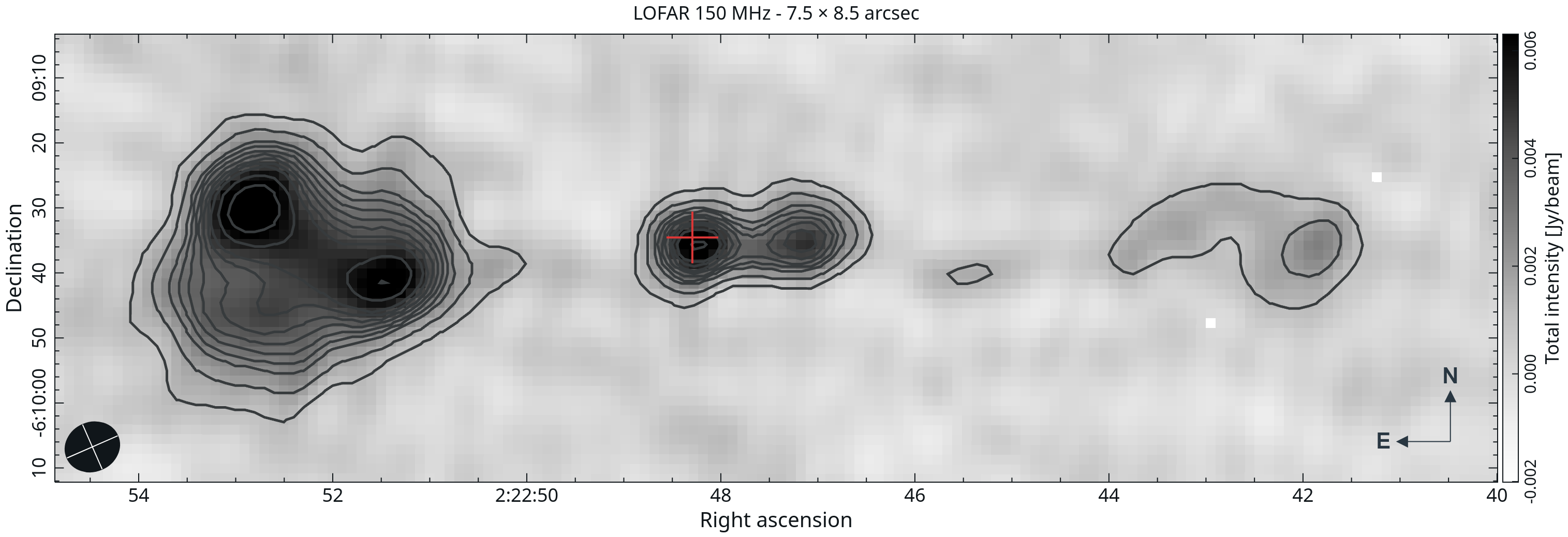}}}\\
\\
{\resizebox{1\hsize}{!}{\includegraphics[trim={0.5cm 0.2cm 0.2cm 0.1cm},clip]{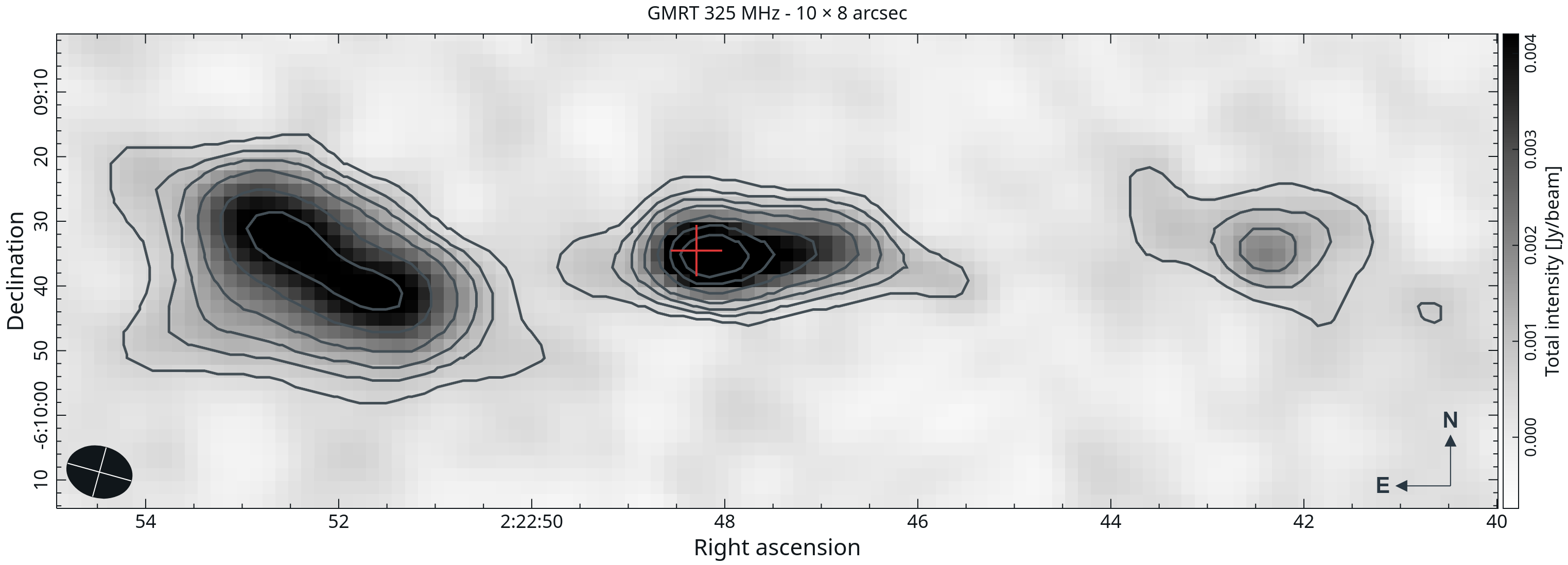}}}\\
\\
{\resizebox{1\hsize}{!}{\includegraphics[trim={0.5cm 0.2cm 0.2cm 0.1cm},clip]{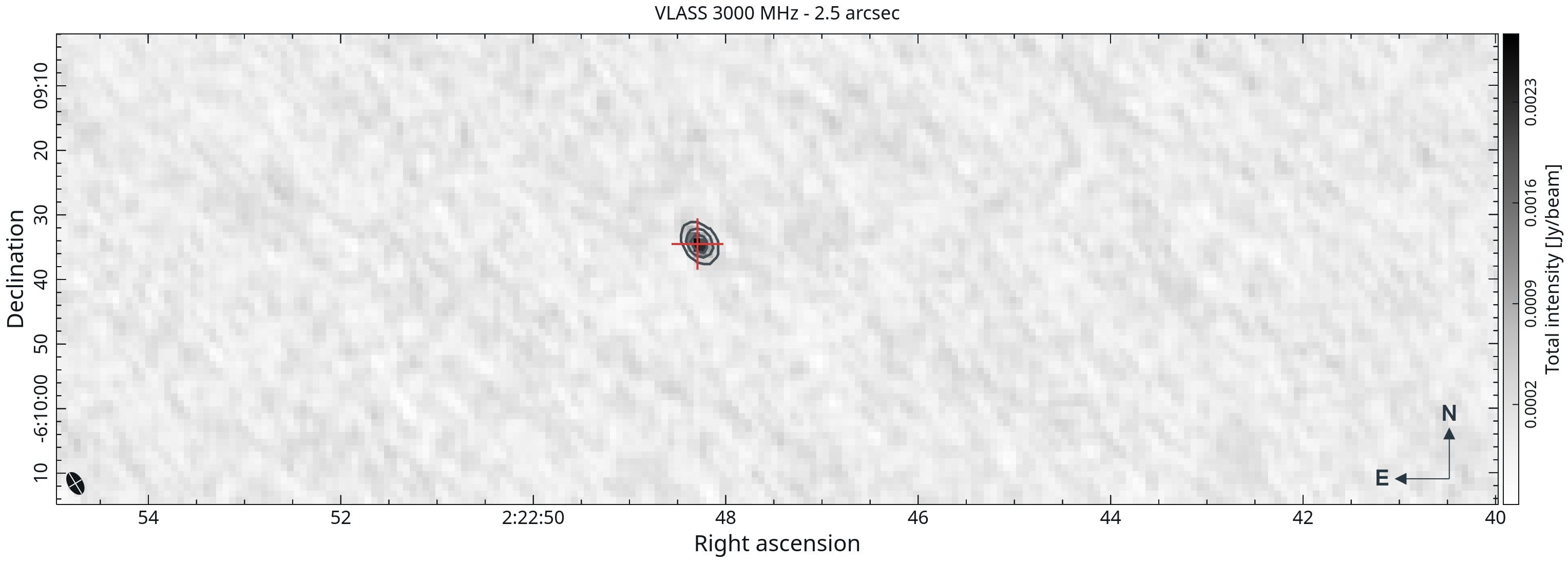}}}
\makebox[-2.5pt][r]{% Similar to \llap
    \raisebox{9em}{%
      \includegraphics[width=.2\linewidth]{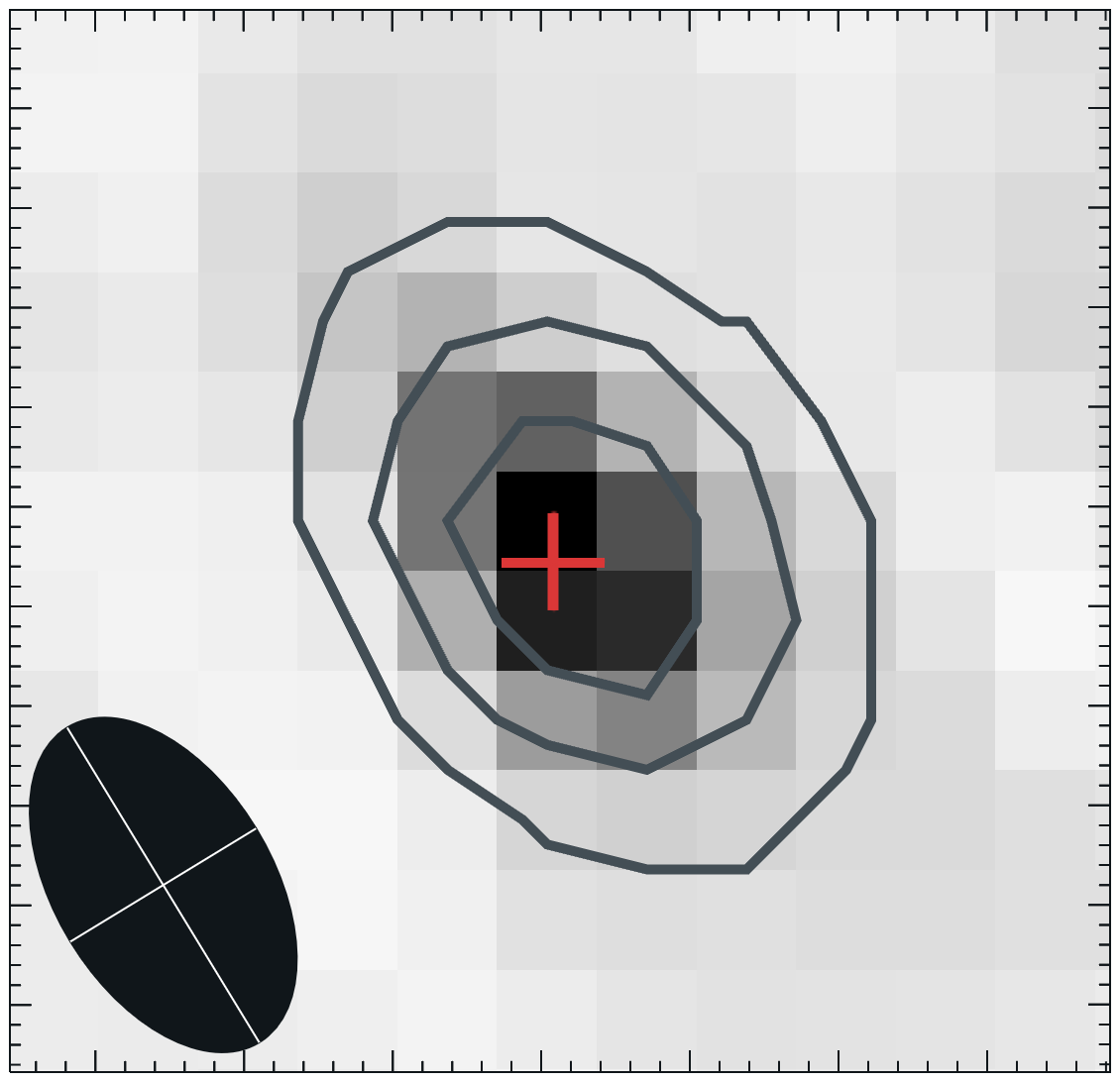}% Inserted image/inset
    }\hspace*{2.5em}%
  }\\
		\end{tabular}
		\caption{ Radio morphologies of the TDRG shown in the 150 MHz LOFAR image (\textit{top panel}, contour levels: $-$3, 3, 5, 7, 8, 9, 10, 11, 12, 13, 15, 17 $\times \sigma_{local}$ = 366.1 $\mu$Jy/beam) and the 325 MHz GMRT (\textit{middle panel}, contour levels: $-$3, 3, 5, 7, 10, 15, 20, 25 $\times \sigma_{local}$ = 194.4 $\mu$Jy/beam) and the radio core in the 3000 MHz VLASS (\textit{bottom panel} with a zoomed-in image in the top right inset, contour levels: $-$3, 3, 5, 7 $\times \sigma_{local}$ = 133.1 $\mu$Jy/beam)). The solid black circle in the bottom left corner denotes the synthesized beam. The red cross indicates the location of the host galaxy. }
		
		\label{fig: source_morpho_archival}
	\end{figure*}

 \begin{figure*}
		\centering
		\begin{tabular}{c}		
  {\resizebox{0.9\hsize}{!}{\includegraphics[trim={0.5cm 0 0.5cm 0},clip]{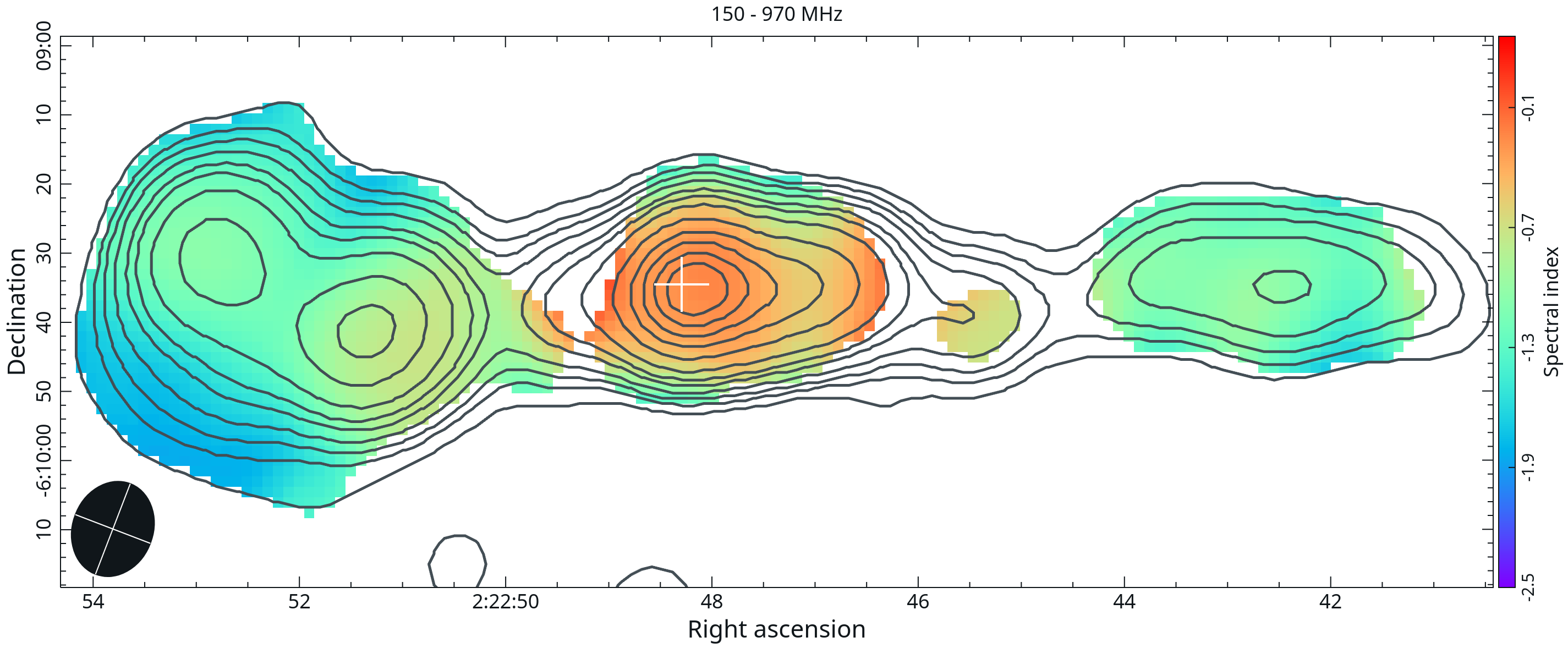}}}\\
  \\
  {\resizebox{0.9\hsize}{!}{\includegraphics[trim={0.5cm 0 0.5cm 0},clip]{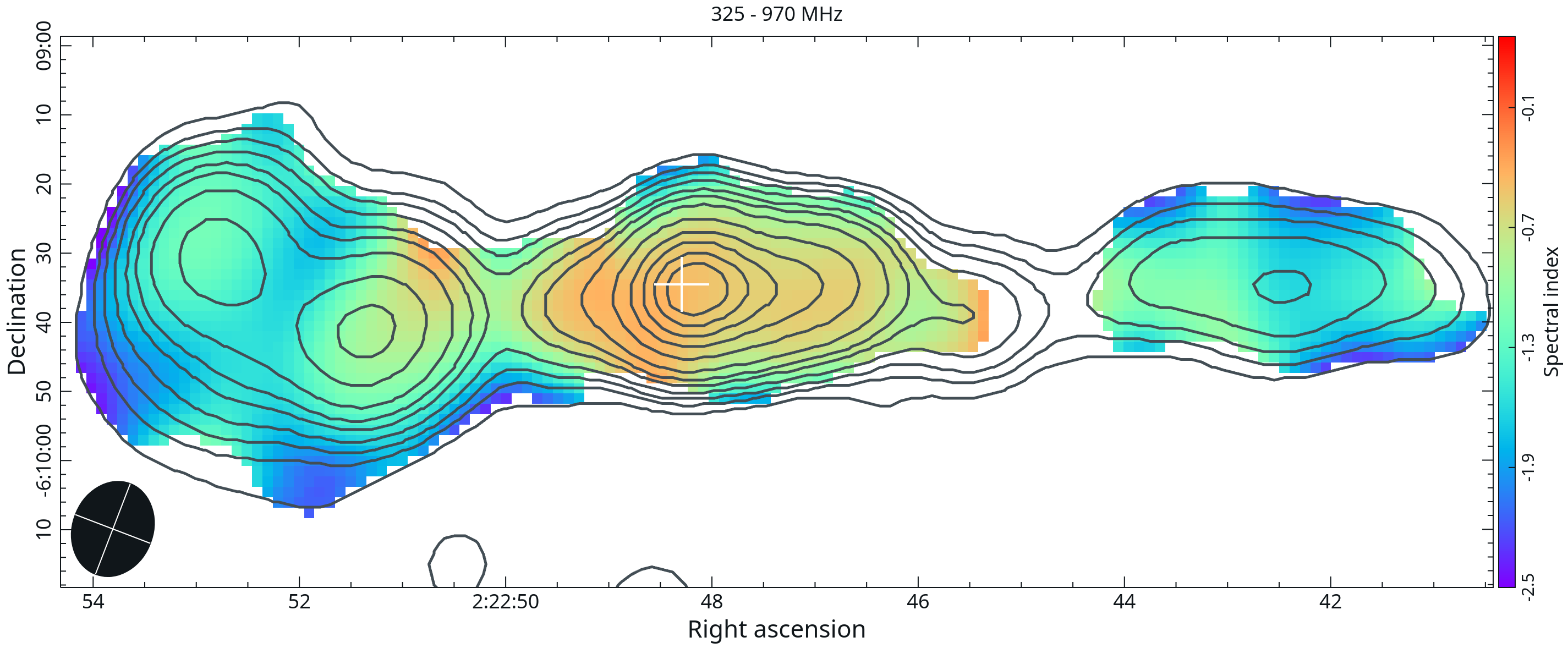}}}\\
  \\
  {\resizebox{0.9\hsize}{!}{\includegraphics[trim={0.5cm 0 0.5cm 0},clip]{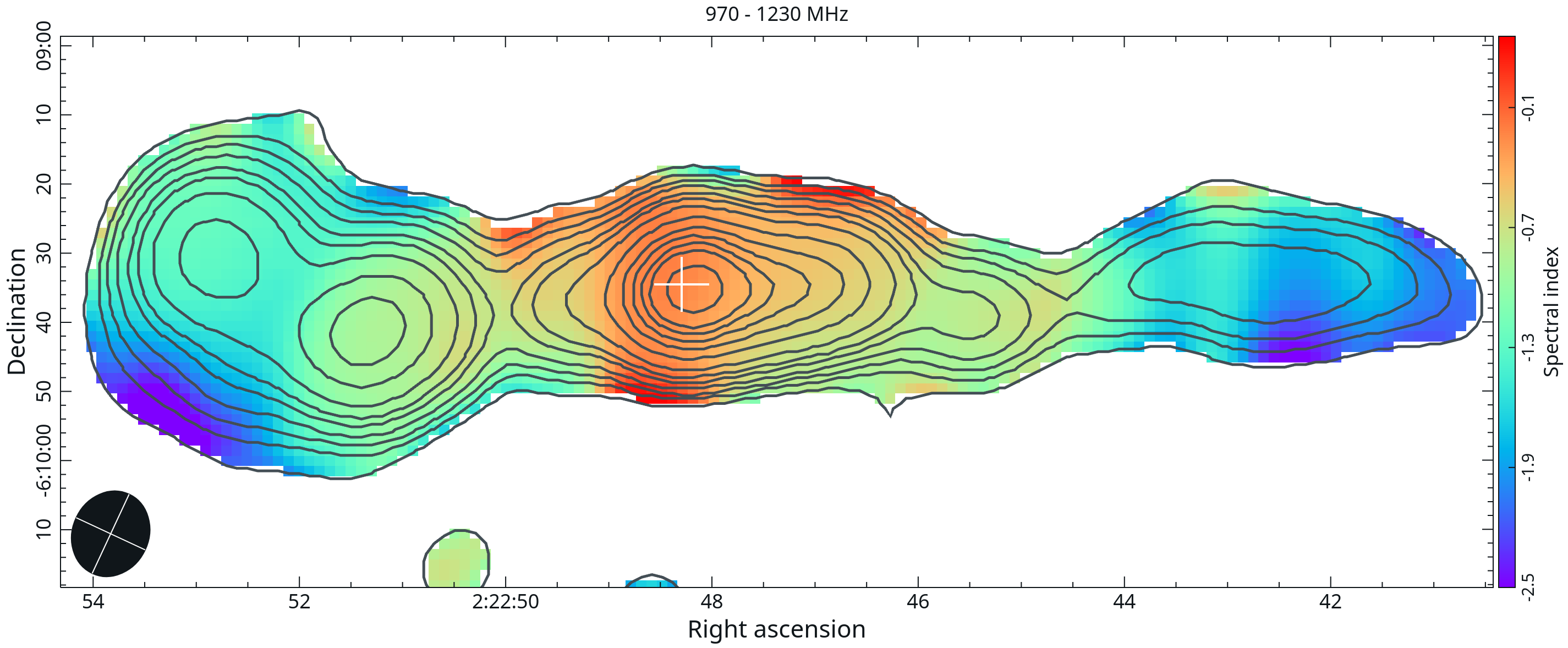}}}\\
    \end{tabular}
\caption{Additional spectral index maps above 3$\sigma_{local}$ level of the TDRG. \textit{From top to bottom panel:} maps computed between 150 to 970 MHz (14 $\times$ 11.6 arcsec), 325 to 970 MHz (14 $\times$ 11.6 arcsec), 970 to 1230 MHz (14 $\times$ 11.6 arcsec), 970 to 1540 MHz (14 $\times$ 11.6 arcsec) and 1230 to 1540 MHz (12.6 $\times$ 11 arcsec). The contours are from the higher frequency images used to compute the maps and their levels are the same as in Figure \ref{fig: source_morpho_appendix}.  The solid black circle in the bottom left corner denotes the synthesized beam. The white cross indicates the location of the host galaxy} 
		
		\label{fig: alphamap_appendix}
        \addtocounter{figure}{-1}
	\end{figure*}
    
    	\begin{figure*}
		\centering
		\begin{tabular}{c}	 
          {\resizebox{0.9\hsize}{!}{\includegraphics[trim={0.5cm 0 0.5cm 0},clip]{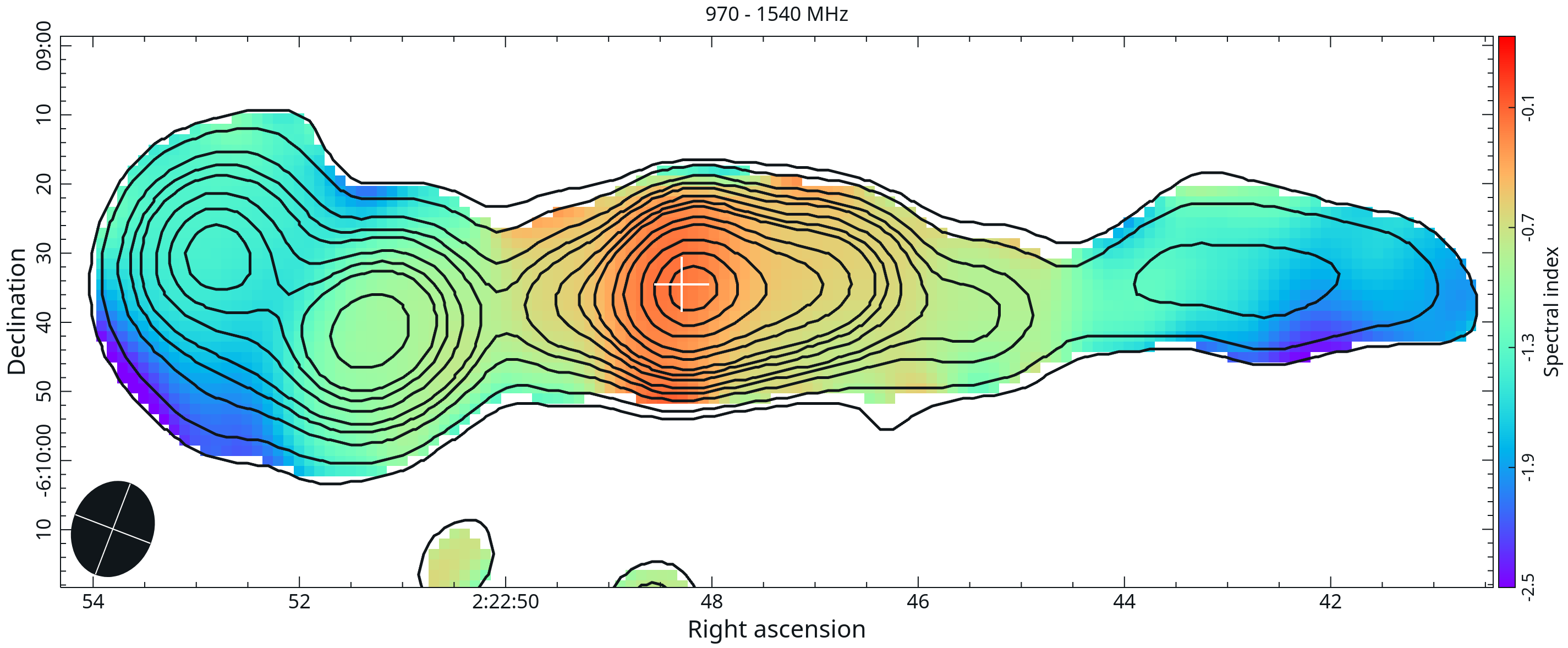}}}\\
    \\
  {\resizebox{0.9\hsize}{!}{\includegraphics[trim={0.5cm 0 0.5cm 0},clip]{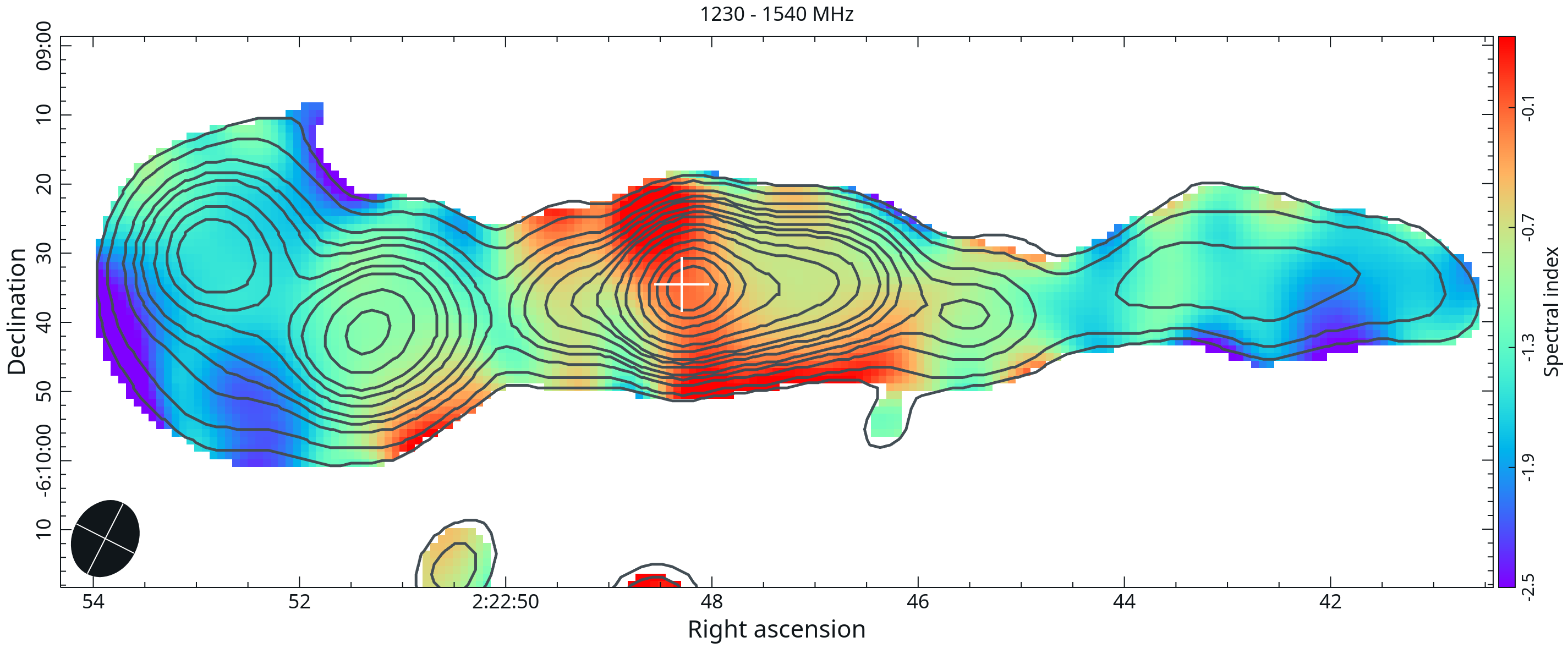}}}\\
   \end{tabular}
\caption{Cont.} 
	\end{figure*}

    \begin{table*}
 \caption{Spectral age fitting results from BRATS using the Tribble model. Column lists: (1): indicates the fitted component where $^\dagger$ indicates the components for which the spectral age fit was done using the MIGHTEE sub-band data only due to low detection in LOFAR and GMRT, (2): estimated equipartition magnetic field, (3): resulting injection index, (4): average $\chi_{red}^2$ over the fitted components, (5): fit number of degrees of freedom, (6): indicates whether the model can be rejected or not, (7): confidence level at which the model can be rejected or not, (8 $-$ 12): number of pixels with the corresponding level of confidence bin, (13): the minimum age of a pixel within a component and (14): the maximum age of a pixel within a component.} 
 \label{tab: Tribble}
		\centering
        {\resizebox{1\hsize}{!}{
        
		  \begin{tabular}{lccccccccccccc}
          
        \hline
         Component & B$_{eq}$ & $\alpha_{inj}$& Average & dof & Rejected& Confidence & &&Confidence bins&&& Min age& Max age\\
         \cline{8-12}
        & [$\mu$G]& &$\chi_{red}^2$& &&level& $<$68&68$-$90&90$-$95&95$-$99&$>$99& [Myr]& [Myr]\\
        (1)&(2)&(3)&(4)&(5)&(6)&(7)&&&(8 $-$ 12)&&&(13)&(14)\\
        \hline
        \hline
         Lobe$_{out (E)}$&15.9&$-$0.77&0.35&3&No& 68 & 394 & 19 & 0& 0 & 0  & 6.78 $^{+0.64}_{-0.85}$ & 15.18$^{+1.39}_{-1.28}$\\
         &&&&&&&&&&&&\\
         
         Lobe$_{mid (E)}$& 11.4&$-$0.68&0.48&3&No& 68 & 273 & 19 & 0& 0 & 0  & 2.78 $^{+1.89}_{-2.78}$ & 14.38$^{+1.11}_{-0.95}$\\
         &&&&&&&&&&&&\\
         $^\dagger$Lobe$_{in (E)}$& 7.4&$-$0.51&0.01&1&No& 68 & 173 & 0 & 0& 0 & 0  & 0.00 $^{+5.56}_{-0.00}$ & 9.18$^{+14.28}_{-9.18}$\\
         &&&&&&&&&&&&\\
        Lobe$_{in (W)}$& 8.8&$-$0.50&0.20&3&No& 68 & 150 & 0 & 0& 0 & 0  & 0.00 $^{+2.94}_{-0.00}$ & 9.58$^{+3.15}_{-7.24}$\\
         &&&&&&&&&&&&&\\
         $^\dagger$Lobe$_{mid (W)}$& 6.45&$-$0.51&0.005&1&No& 68 & 133 & 0 & 0& 0 & 0  & 4.38 $^{+13.57}_{-4.38}$ & 10.02$^{+13.31}_{-10.02}$\\
         &&&&&&&&&&&&\\
        Lobe$_{out (W)}$ & 9.0&$-$0.51&0.18&3&No& 68 & 280 & 0 & 0& 0 & 0  & 12.42 $^{+3.39}_{-4.70}$ & 20.78$^{+1.72}_{-2.35}$\\
        \hline
        
        \end{tabular}{}
        }}
        
\end{table*}

% Don't change these lines
\bsp	% typesetting comment
\label{lastpage}
\end{document}